\documentclass[a4paper, twocolumn, 10pt, superscriptaddress, amsmath,amssymb,pra, aps]{revtex4-1}

\usepackage[T1]{fontenc} 
\usepackage{amsmath, amsthm, amsfonts, amssymb,amstext}
\usepackage{mathptmx}
\usepackage[mathscr]{eucal}
\usepackage{bm}
\usepackage{xfrac}
\usepackage{mathrsfs}
\usepackage{latexsym}
\usepackage{array}
\usepackage{verbatim}
\usepackage{graphicx}
\usepackage{bbold}
\usepackage{ulem}
\usepackage[toc,page]{appendix}
\usepackage{color}
\usepackage{tikz}
\usepackage{textcomp} 
\usepackage{subfigure}
\usepackage{float}
\usepackage{soul}

\usepackage{mathptmx}
\usepackage{amssymb}
\usepackage{amsmath}
\usepackage{amsfonts}
\usepackage{array}

\usepackage{bm}
\usepackage{xcolor}

\usepackage{braket}

\definecolor{mygrey}{gray}{0.35}
\definecolor{myblue}{rgb}{0.2,0.2,0.8}
\definecolor{myzard}{cmyk}{0,0,0.05,0}
\definecolor{mywhite}{rgb}{1,1,1}
\definecolor{myred}{rgb}{1,0.,0.3}

\usepackage[colorlinks=true,citecolor=myblue,linkcolor=myred]{hyperref}

\def\beq{{\begin{equation}}}
\def\eeq{{\end{equation}}}

\def\a{{\rm a}}

\def\i{{\rm i}}

 \def\ee{\mathord{\rm e}}
 
 \def\ii{\mathord{\rm i}}

\def\half{\textstyle\frac{1}{2}}

\def\fourth{\textstyle\frac{1}{4}}

\renewcommand{\ii}{{\rm i}}

\def\beq{\begin{equation}}
\def\eeq{\end{equation}}
\def\barray{\begin{eqnarray}}
\def\earray{\end{eqnarray}}

\DeclareFontEncoding{LS1}{}{}
\DeclareFontSubstitution{LS1}{stix}{m}{n}
\DeclareSymbolFont{symbols4}{LS1}{stixbb}{m}{it}
\DeclareMathSymbol{\varhexagonblack}{\mathord}{symbols4}{"DD}
\DeclareMathSymbol{\hexagonblack}  {\mathord}{symbols4}{"DE}

\graphicspath{{./gfx/}}

\usepackage{natbib}

\usepackage{color}   
\usepackage{xcolor}
\definecolor{celeste}{cmyk}{1.00, 0.00, 0.00, 0.4}
\usepackage{hyperref}
\hypersetup{
	colorlinks=true, 
	linktoc=all,     
	breaklinks=true, 
  linkcolor=celeste, 
  citecolor=celeste,  
  urlcolor=celeste
}

\def\beq{{\begin{equation}}}
\def\eeq{{\end{equation}}}

\def\a{{\rm a}}

\def\i{{\rm i}}

 \def\ee{\mathord{\rm e}}
 
 \def\ii{\mathord{\rm i}}

\def\half{\textstyle\frac{1}{2}}

\def\fourth{\textstyle\frac{1}{4}}

\renewcommand{\ii}{{\rm i}}

\def\beq{\begin{equation}}
\def\eeq{\end{equation}}
\def\barray{\begin{eqnarray}}
\def\earray{\end{eqnarray}}

\newenvironment{lcase}
{\left\lbrace \begin{aligned}}
	{\end{aligned} \right.}

\newcommand\sbgs[0]{\Phi_0} 
\newcommand\Hb[0]{H_\textrm{b}}

\begin{document}

\title{Real-time collisions of fractional charges in a trapped-ion Jackiw-Rebbi field theory}

\author{A. Kahan}
\affiliation{Instituto de F\'isica Te\'orica UAM-CSIC, Universidad Aut\'onoma de Madrid, Cantoblanco, 28049, Madrid, Spain} 
\author{P. Viñas}
\affiliation{Instituto de F\'isica Te\'orica UAM-CSIC, Universidad Aut\'onoma de Madrid, Cantoblanco, 28049, Madrid, Spain}
\author{T. V. Zache}
\affiliation{Institute for Theoretical Physics, University of Innsbruck, Innsbruck, 6020, Austria}
\affiliation{Institute for Quantum Optics and Quantum Information of the Austrian Academy of Sciences,
Innsbruck, 6020, Austria}
\author{A. Bermudez}
\affiliation{Instituto de F\'isica Te\'orica UAM-CSIC, Universidad Aut\'onoma de Madrid, Cantoblanco, 28049, Madrid, Spain} 

\begin{abstract}
We propose and analyze a trapped-ion quantum simulator of the Jackiw–Rebbi model, a paradigmatic quantum field theory in (1+1) dimensions where solitonic excitations of a scalar field can bind fermionic zero modes leading to fractionally-charged excitations. In our approach, the scalar field is a coarse-grained description of the planar zigzag ion displacements in the vicinity of a structural phase transition. The internal electronic states of the ions encode spins with interactions mediated by the transverse phonons and in-plane spin-phonon couplings with a zigzag pattern,  which together correspond to a Yukawa-coupled Dirac field. Instead of assuming a fixed soliton background, we study the effect of back-reaction and quantum fluctuations on the coupled dynamics of the full fermion–boson system. We start by applying a Born–Oppenheimer approximation to obtain an effective Peierls–Nabarro potential for the topological kink, unveiling how the fermionic back-reaction can lead to localization of the kink. Beyond this limit, a truncated Wigner approximation combined with fermionic Gaussian states captures the quantum spreading and localization of a kink and kink-antikink scattering. Our results reveal how back-reaction and quantum fluctuations modify the stability and real-time evolution of fractionalized fermions,  predicting experimentally accessible signatures in current trapped-ion architectures.
\end{abstract}

\maketitle

\setcounter{tocdepth}{2}
\begingroup
\hypersetup{linkcolor=black}
\tableofcontents
\endgroup

\section{\bf Introduction}

Understanding real-time dynamics of strongly-coupled quantum field theories (QFTs) remains an open problem of relevance across very different scales~\cite{Calzetta_Hu_2008}, ranging from non-equilibrium processes in the early universe~\cite{PhysRevD.56.3258,annurev:/content/journals/10.1146/annurev.nucl.49.1.35,PhysRevLett.91.111601}, over the quark–gluon plasma dynamics in high-energy collisions~\cite{KRASNITZ1999237, BAIER200151,kolb2003hydrodynamicdescriptionultrarelativisticheavyion} to transport in strongly correlated quantum materials~\cite{RevModPhys.58.323,PhysRevLett.68.2512,RevModPhys.86.779}. The advent of quantum lattice simulators (Q$\ell$Ss)~\cite{Feynman1982,RevModPhys.86.153,Cirac2012}, i.e. highly-controllable isolated quantum many-body systems on a physical lattice that evolve under a  tunable Hamiltonian, opens new possibilities in this direction. From this perspective, Q$\ell$Ss  can be understood as experimental regularizations of the target QFT under study with the potential to overcome the limitations of classical numerical methods for the simulation of real-time quantum many-body  dynamics~\cite{Zohar_2016,Banuls2020,doi:10.1098/rsta.2021.0064,Klco_2022,PRXQuantum.4.027001,Menu2024,Halimeh2025}. 

 Even if the final goal is the accurate quantum simulation of the relativistic QFTs relevant to the Standard Model of particle physics~\cite{Peskin:1995ev}, or those that emerge as coarse-grained descriptions of concrete lattice models in quantum materials~\cite{Fradkin2013}, current technologies are currently targeting simplified models~\cite{Martinez_2016,PhysRevA.98.032331,Schweizer_2019, Kokail_2019,Mil_2020,Yang_2020,Zhou_2021,Mildenberger_2022,Meth_2023, De_2024,Cochran2025Visualizing,Gonzalez-Cuadra2025,cobos2025realtimedynamics21dgauge,saner2025realtimeobservationaharonovbohminterference,than2025observationquantumfieldtheorydynamicsspinphonon}. In spite of the simplifications,  some of these minimal models capture the main ingredients of more complex QFTs and, moreover,  can lead to exotic phenomena that have not been observed in any experiment to date. In this work, we focus on the Jackiw-Rebbi (JR) model~\cite{PhysRevD.13.3398},  a cornerstone in the study of QFTs that unveiled the interplay of spontaneous symmetry breaking and non-trivial topology. In particular, the JR model allows to explore how solitonic excitations of a real $\lambda\phi^4$ QFT in $D=(1+1)$ dimensions, which interpolate between the two possible symmetry-broken vacua~\cite{PhysRevD.10.4130,PismaZhETF.20.430,PhysRevD.11.1486}, can polarize the fermionic Dirac-field vacuum in a very exotic manner~\cite{PhysRevLett.47.986}. As noted above, these low-dimensional solitons share some of the key properties with those found in higher-$D$ gauge theories~\cite{Manton:2004}. In particular, the JR mechanism of binding fermionic zero modes bound to topological defects generalizes 
to vortices~\cite{NIELSEN197345,JACKIW1981681} and monopoles~\cite{HOOFT1974276,PhysRevD.16.3068}, where the existence and number 
of zero modes follow from a corresponding index theorem in open space~\cite{Callias1978}. A paradigmatic result of the interplay of spontaneous symmetry breaking and topology is that the fermions bound to these zero modes can actually carry a fractional charge or fermion number~\cite{PhysRevD.13.3398,NIEMI198699}.

  \begin{figure*}
    \centering
    \includegraphics[width=0.65\linewidth]{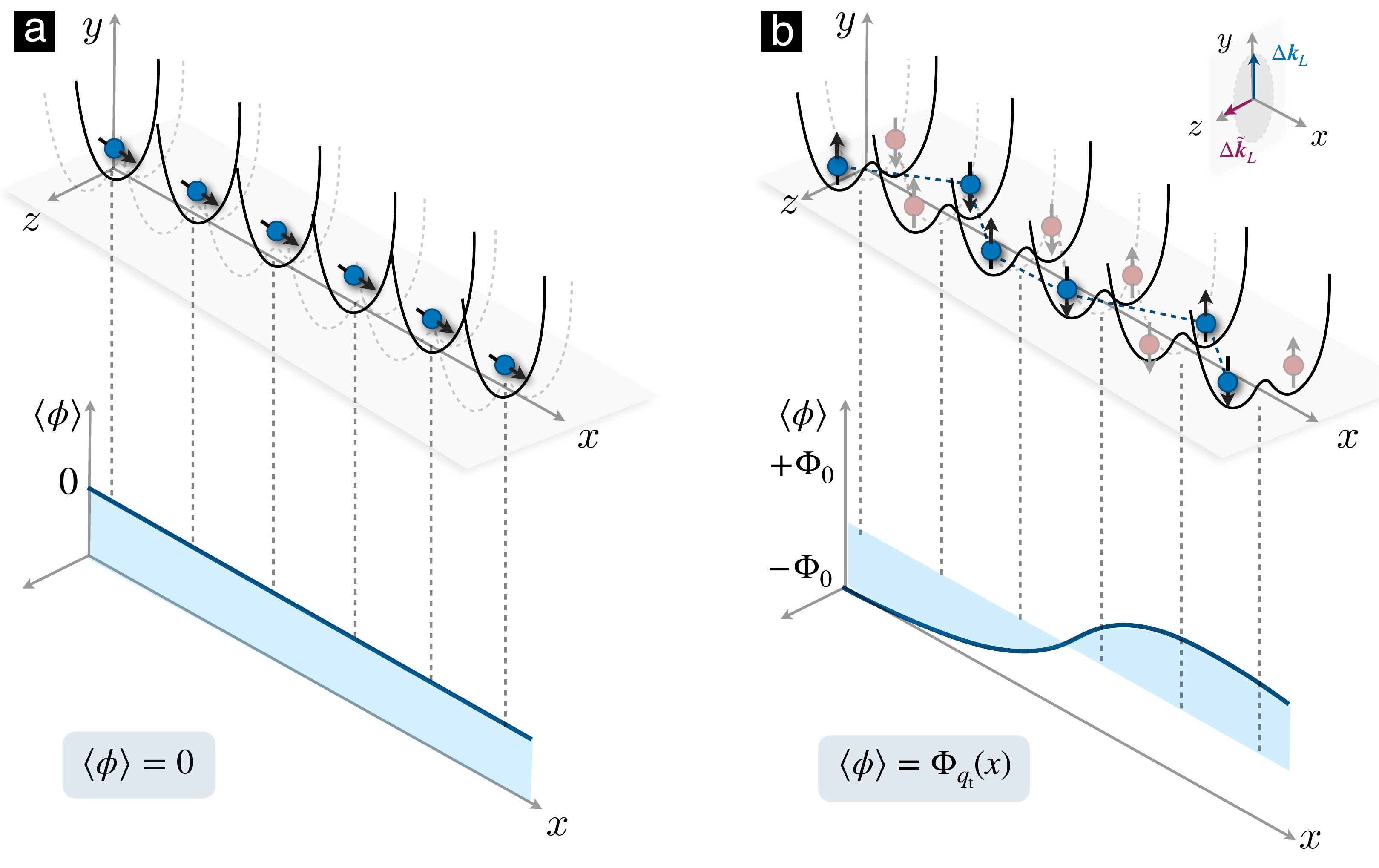}
    \caption{{\bf Trapped-ion Jackiw-Rebbi model:}
A linear chain of ions is tuned across the linear {\bf (a)} to zigzag {\bf (b)} structural phase transition, so that transverse displacements along the $z$ axis act as a coarse-grained scalar field $\phi(x)$ supporting topological kink configurations that interpolate between the ``zig-zag''  and ``zag-zig'' configurations $\langle \phi \rangle = \pm \Phi_0$, as depicted in {\bf (b)}.
Two internal electronic states of each ion encode a staggered fermionic field via a Jordan--Wigner transformation, and phonon-mediated spin-spin interactions induced by a M\o lmer-S\o rensen laser scheme acting along $\Delta\boldsymbol{k}_L$ yield an effective Dirac kinetic term (see inset of {\bf (b)}).
A state-dependent optical dipole force along the crystal plane  $\Delta\tilde{\boldsymbol{k}}_L$ (inset) couples spin and motion, implementing the Yukawa-type interaction.
Together, these elements realize an analog trapped-ion Q$\ell$S of the Jackiw-Rebbi model that would allow to explore real-time dynamics of scalar solitons and fractional fermion bound states, particularly looking into characteristic effects due to back-reaction and quantum fluctuations that go beyond typical approximations.}  
    \label{fig:trapped_ion_scheme}
\end{figure*}
The most common approach to the JR model adopts a simplified approximation: the soliton is treated as a prescribed classical field,   focusing on the resulting properties of the quantum Dirac field in this background, such as the appearance of the fractional charge on the zero mode bound to the soliton~\cite{PhysRevD.13.3398,JACKIW1981253,PhysRevB.25.6447,RAJARAMAN1982151,PhysRevD.30.809,PhysRevD.30.2136,PhysRevD.30.2194,PhysRevB.31.6112,PhysRevB.25.6447}. 
The stability of the solitons, which appear in this classical approximation as localized lump-like solutions of nonlinear field equations, relies on the fact that they cannot be locally unwound at a finite energy cost~\cite{PhysRevD.10.4130}. This robustness can also be understood from a topological perspective~\cite{RevModPhys.51.591,Manton:2004} in which the solitons are understood as maps from an asymptotic spatial manifold $S^{D-2}$ to the target vacuum manifold of the theory $\mathscr{T}^{\rm f}_{\rm g}$. In the  JR model, these correspond to two points sufficiently separated from the soliton center $S^{0}=\{x_0-d,x_0+d\}$, and by the two asymptotic values of the scalar field $\mathcal{T}^{\rm f}_{\rm g}=\{-\Phi_0,\Phi_0\}$. These maps fall into distinct homotopy classes~\cite{Nakahara:2003nw} which, in this case, can be labeled by an integer or discrete topological charge. The homotopy group in the JR model counts the number of connected components of the vacuum manifold $\pi_0(\mathcal{T}^{\rm f}_{\rm g})=\mathbb{Z}_2$, and determines if the classical field asymptotes to the same or to a different SSB groundstate. 
 As discussed below, one can define a topological charge $q_{\rm t}=\pm 1$ for the soliton, which is conserved under local perturbations and underlies the robustness of the solitons, typically referred to as kinks and anti-kinks.

 When this classical soliton interacts with the Dirac field, it leads to trapped fermion  modes~\cite{PhysRevD.10.4130}, including a zero-energy mode that is exponentially localized about the soliton center.  As discussed in~\cite{PhysRevD.13.3398, JACKIW1981253, PhysRevB.25.6447, RAJARAMAN1982151, PhysRevD.30.809, PhysRevD.30.2136, PhysRevD.30.2194, PhysRevB.31.6112, PhysRevB.25.6447}, this zero mode leads to a spectrum of excitations with fractional charges $q_{\rm f}\in\frac{1}{2}(2\mathbb{Z}+1)$. In addition to the standard bosonic and fermionic particles/anti-particles that are excited from one of the SSB vacua, one can find composite excitations in a different sector localised around the soliton and characterized by a  topological and a half-integer charge   $(q_{\rm t},q_{\rm f})$. These excitations can be Lorentz boosted to any velocity below the speed of sound and propagate without dispersion. In contrast to other integrable models~\cite{PhysRevD.11.3424}, however, these composite excitations do not display perfectly elastic collisions.  A more contemporary account of the topology of this composite particle arises in the context of symmetry-protected topological phases~\cite{PhysRevB.82.115120, RevModPhys.88.035005}, in which solitons are called defects. Their topology is classified by the jump of a Chern–Simons invariant ${\rm CS}_1=\tfrac{1}{2}\,{\rm mod}\mathbb{Z}$, which is computed by the integral~\cite{PhysRevLett.62.2747} of the Berry connection~\cite{doi:10.1098/rspa.1984.0023,RevModPhys.82.1959} at two asymptotic regions far from the soliton center. The occupied fermionic bands can be understood as a vector bundle over a base space ${\rm BZ}^{D-1}\times S^{D-2}$,  assuming periodic boundary conditions leading to a Brillouin zone, and also a smooth background soliton leading to a manifold wrapping around the defect, here  $k\in[-\pi,\pi]$ and $S^0=\{x_0-d,x_0+d\}$.  Note that, even if the integrals on each side of the soliton are not quantized without a specific regularization~\cite{Ryu_2010}, their difference is regularization independent, and detects the localized zero-energy mode at the topological soliton.

Note that the discussion so far neglects any quantum fluctuations of the scalar field from the outset, as well as the back-reaction from the Dirac fermions onto the scalar solitons. The goal of the present work is to go beyond this limit, exploring dynamical effects that are genuinely induced by either quantum fluctuations or back-reaction. We further address how such effects might be observed in an analog Q$\ell$Ss based on trapped-ion crystals~\cite{Blatt2012}. Following the seminal trapped-ion proposal~\cite{PhysRevLett.92.207901}, the main focus of experimental progress in recent years~\cite{RevModPhys.93.025001} has focused on the quantum simulation of 
 effective spin models with long-range interactions mediated by the quantized vibrational oscillations of the crystal, i.e., the phonons. We note, however, that there is also a remarkable progress in using the phonons not as passive mediators of a dispersive interaction, but also as active degrees of freedom in the quantum simulator~\cite{Gerritsma2010, PhysRevLett.106.060503, PhysRevA.85.031401, Ulm2013, Pyka2013,  PhysRevLett.110.133004, Toyoda:2015ama, PhysRevLett.117.170401, Zhang:2016lyo, PhysRevLett.120.073001, PhysRevLett.123.180502, PhysRevLett.123.213605, D3SC02453A, Whitlow2023,doi:10.1126/sciadv.ads8011, Sun:2024obv, so2025quantumsimulationchargeexciton, saner2025realtimeobservationaharonovbohminterference,  than2025observationquantumfieldtheorydynamicsspinphonon}. Particularly relevant to the current work, Than et {\it al.} have recently reported on the progress towards the quantum simulation of Yukawa-coupled QFTs of Dirac fields coupled to a free scalar field following the scheme in~\cite{PhysRevResearch.3.043072}, which combines digital and analog quantum-simulation tools~\cite{PhysRevLett.109.200501}. We should also mention a previous theoretical proposal to explore the JR model with slow light in an atomic ensemble~\cite{Angelakis2014}, which can, in principle, encode a Dirac spinor under a fixed solitonic background controlled by shaping the detuning of the light-matter interaction.

 In the present work, we instead focus on a fully-analog Q$\ell$S based on trapped ions in a zigzag ladder (see Fig.~\ref{fig:trapped_ion_scheme}). The effective $\lambda\phi^4$   QFT emerges in the vicinity of the structural phase transition between the linear chain and the zigzag ladder, and allows to identify the scalar field with the coarse-grained alternating displacements in the direction of the zigzag ladder~\cite{PhysRevB.77.064111,  PhysRevLett.105.075701, PhysRevX.7.041012}. We note that solitonic excitations interpolating between the two possible zigzag patterns have already been observed in experiments~\cite{Ulm2013, Pyka2013, PhysRevLett.110.133004}. We envision that the phonons transverse to the plane of the ladder serve to mediate long-range spin-spin interactions, which can lead to an effective massless Dirac QFT under a Jordan-Wigner transformation~\cite{Jordan1928}.  We note that the light-cone-like propagation of excitations in the spin model has been measured in~\cite{Jurcevic2014, Richerme2014}, which, as we argue below, is consistent with this massless Dirac QFT with an effective speed of light that depends on the long-range tail of the spin-spin couplings. In addition, we study how a state-dependent dipole potential within the ladder plane can be exploited to mimic the required Yukawa fermion-boson vertex, leading to the JR model.

In order to identify possibly new dynamical effects brought by quantum fluctuations and back-reaction, we start by considering an adiabatic  Born-Oppenheimer-type approximation~\cite{https://doi.org/10.1002/andp.19273892002}, in which the Dirac fermions adapt instantaneously to the scalar field. Considering in particular the displacement of the center of the topological soliton, henceforth referred to as kink $q_{\rm t}=+1$ or anti-kink $q_{\rm t}=-1$,  and assuming that the corresponding fermionic zero mode reacts instantaneously to this displacement, we can calculate the periodic potential experienced by the composite $(q_{\rm t},q_{\rm f})$ excitation. We find that this potential has two main contributions: {\it (i)} a so-called Peierls-Nabarro (PN) term~\cite{RPeierls_1940, Nabarro_1947, braun2004frenkel} stemming from the discreteness of the lattice regularization of the JR QFT, i.e. the explicit breaking of continuous translation invariance, and {\it (ii)} a back-reaction potential, which can be qualitatively understood as a consequence of the inertia of the fermionic charge bound by the kink.  We find that this second contribution dominates as the Yukawa coupling grows, and that it can considerably alter the kink localization dynamics in a lattice JR model. 

In the absence of back-reaction, a classical kink that is initially at rest will remain localized at the same position, provided that this position coincides with a minimum of the PN potential. When including quantum fluctuations, such a kink will spread as a function of time, leading to a diffusive behavior. As shown below, this behavior can be captured by a truncated Wigner approximation (TWA) designed to account for the leading order effects of quantum fluctuations in a semi-classical expansion~\cite{Carmichael1999, Polkovnikov2003, Polkovnikov2010}. As discussed below, quantum effects are incorporated by sampling initial phase-space fluctuations consistent with the small-displacement dispersion relation about the kink classical profile, and then evolving each of these initial conditions with the equations of motion for the lattice scalar field. This process leads to an ensemble of trajectories for the observable under study. Combining this method with the equations of motion for  Dirac fermions, which can be integrated exactly for each trajectory using fermionic Gaussian states~\cite{10.5555/2011637.2011640, Kraus_2010, Surace2022},   we can also explore genuine back-reaction in real-time dynamics, thus going beyond the previous Born-Oppenheimer approximation. We use this extended TWA+Gaussian approximation to test the above picture based on the adiabatic potential of the topological kink with the bound charge on the fermionic zero mode. As discussed in more detail below, we find that due to this back-reaction, the spread of the kink gets gradually lowered until, at some specific coupling strength, it gets completely localised within a potential well and shows breathing oscillations. We also investigate real-time collisions of kink–antikink pairs, showing how
their dynamics range from an almost-elastic bounce to the formation
of long-lived bound states, depending on their initial kinetic energy. The TWA is used to show that quantum fluctuations smear the sharp classical
boundaries, but the distinct collision regimes are still manifest, pointing to interesting possible experiments that explore such phenomenology using trapped-ion Q$\ell$Ss.

This article is organised as follows. In Sec.~\ref{sec:QLS}, we introduce a scheme for a trapped-ion Q$\ell$S 
 of scalar-field solitons coupled to Dirac fields, beginning with a review of the continuum
Jackiw-Rebbi QFT, its lattice formulation, and finally its mapping
to a trapped-ion system. In Sec.~\ref{sec:methods}, we develop the
adiabatic and phase-space methods used throughout this work, including a
Born-Oppenheimer treatment of the fermion--soliton back-reaction and a
semiclassical truncated-Wigner approach complemented by fermionic Gaussian
state methods. Sec.~\ref{sec:single_soliton} applies these methods to the real-time dynamics of a
fractionally-charged fermion bound to a topological soliton, focusing on the
diffusion, localisation, and drag of the fractional charge in both static and
moving-soliton scenarios. In Sec.~\ref{sec:soliton_collisions}, we explore the collisions of
half-charged fermions on soliton-antisoliton configurations, starting with the analysis of classical kink-antikink scattering in both the continuum and on the lattice. We then show that, even if quantum fluctuations modify soliton dynamics, several interesting regimes with persistent oscillating bound states and confined/deconfined half-charges can still be differentiated. Finally, Sec.~\ref{sec:conc_out}
presents our conclusions and outlook.

\section{\bf Trapped-ion scalar solitons and Dirac fields}
\label{sec:QLS}
\subsection{Continuum Jackiw-Rebbi  field theory }
Let us start by defining the  JR model in detail, a relativistic QFT of a massless Dirac field $\psi(t,x)$  with a  Yukawa coupling $g$ to a real scalar field $\phi(t,x)$, both defined in a flat $D=(1+1)$-dimensional Minkowski spacetime. The scalar field has a bare mass $m_0$ and  quartic self-interaction $\lambda$, and the corresponding  action in natural units $\hbar=c=1$ reads as follows
\begin{equation}
\label{eq:JR_action}
S=\!\!\int\!\!{\rm d}^2x\Big(\ii\overline{\psi}(\gamma^{\mu}\partial_\mu-g\phi)\psi+\half(\partial^\mu\phi\partial_\mu\phi-m_0^2\phi^2)-\frac{\lambda}{4}\phi^4\Big),
\end{equation}
where we have introduced the  adjoint $\overline{\psi}(t,x)=\psi^{\dagger}(t,x)\gamma^0$. The gamma matrices  fulfill a Clifford algebra $\{\gamma^\mu,\gamma^{\nu}\}=2\eta^{\mu\nu}$ with $\mu,\nu\in\{0,1\}$ labeling the spacetime coordinates with metric $\eta={\rm diag}(1,-1)$. For Q$\ell$Ss, it is customary to consider the Hamiltonian formulation of the field theory $H=\!\!\int\!\!{\rm d}x\,\mathcal{H}$, which has the following  Hamiltonian density 
\beq
\label{eq:H_JR}
\mathcal{H}=\overline{\psi}\big(-\ii\gamma^{1}\partial_x+g\phi\big)\psi+\half\big(\pi^2+(\partial_x\phi)^2+m_0^2\phi^2\big)+\frac{\lambda}{4}\phi^4.
\eeq
As customary, canonical quantization proceeds by considering equal-time   commutation $[\phi(t,x),\pi(t,y)]=\ii\delta(x-y)$ and  anti-commutation $\{\psi(t,x),\psi^{{\dagger}}\!(t,y)\}=\delta(x-y)$   relations for the bosonic and fermionic fields, respectively.   

\subsubsection{Vanishing interactions}

Let us start by setting $g=0$, and noting that the bosonic sector can undergo a second-order quantum phase transition where a global $\mathbb{Z}_2$ inversion symmetry on the scalar field $\phi\to -\phi$ gets spontaneously broken~\cite{PhysRevD.13.2778}. This can be characterized through the onset of a non-zero vacuum expectation value (VEV) of the scalar field $\langle\phi(x)\rangle=\Phi_0\neq 0,\,\,\forall x$ for $m_0^2<m_{\rm c}^2(\lambda)$. In the rest of this paper, we will use lower-case symbols for quantum field operators and the corresponding uppercase ones for classical fields and phase-space variables. The above VEV can already be captured at the classical level by inspecting   the classical potential 
\beq
\label{eq:classical_potential}
V_{\rm cl}(\Phi)=\!\int\!{\rm d}x\left(\tfrac{m_0^2}{2}\Phi^2(x)+\tfrac{\lambda}{4}\Phi^4(x)\right),
\eeq
which forms a double well for $m_0^2<0$ and $\lambda>0$ and leads to two possible spatially-homogeneous ground states 
\begin{equation}\label{sol_vacio}
 \Phi_0 = 
 \begin{lcase}
    & 0,\hspace{9ex}m_0^2>0, \\
    & \pm \sqrt{-\frac{m_0^2}{\lambda}},\hspace{1ex}m_0^2<0.
\end{lcase}
\end{equation}
 The second line of this equation describes a scalar vacuum with a broken $\mathbb{Z}_2$ symmetry as $\Phi_0\neq-\Phi_0$ for $\Phi_0\neq 0$. Even for these low spacetime dimensions, this transition persists when including quantum fluctuations, which lead to an effective potential with radiative corrections  $V_{\rm cl}(\Phi)\mapsto V_{\rm eff}(\Phi)$~\cite{PhysRevD.7.1888, PhysRevD.9.1686}. As a result, the critical point is shifted from $m_{0}^2=0$  to a negative value of the bare mass  $m_0^2<m_{\rm c}^2(\lambda)<0$ which, in contrast to higher-dimensional instances~\cite{PhysRevLett.28.240}, cannot be computed using perturbative renormalization group flows~\cite{WILSON197475, RevModPhys.66.129}. 
 
 As usual in QFT, the calculation of this shift, and thus the physical mass of the excitations, is afflicted by UV divergences. In  $D=(1+1)$ dimensions, all the UV divergences of this model can be accounted for through a one-loop Feynman diagram, the so-called tadpole~\cite{Peskin:1995ev}. On a  lattice of  spacing $a$, one can define a tadpole-renormalised mass $\mu$  by
  \beq
 \label{eq:tadpole_continum}
\mu^2=m_0^2+3\lambda\!\!\int\!\! \frac{{\rm d}^2k}{(2\pi)^2}\,\,\frac{1}{k^2-m_0^2+\ii\epsilon},
 \eeq
 where $\epsilon\to0^+$, and the integral is regularised with a UV cutoff.
  We note that for $D=1+1$ dimensions, this procedure removes all  UV divergences of the QFT, as any higher-order term with a propagator depending on this mass no longer diverges. Reproducing the classical potential analysis after this one-loop renormalization would predict the critical point at $\mu^2<0$ with a VEV given by $\Phi_0\mapsto\Phi=\sqrt{-\mu^2/\lambda}$. 
To date, the most accurate estimations of the shift of this UV-renormalised critical point due to quantum corrections beyond this limit predict $\lambda/\mu^2|_{\rm c}\approx 11.086(9)>0=\lambda/\mu^2|_{\rm c, MF}$, which have been obtained using matrix product state techniques~\cite{PhysRevD.88.085030, PhysRevResearch.2.033278, PhysRevD.106.L071501}.  These calculations confirm that the scaling of the VEV  in the vicinity of the phase transition changes from the above mean-field   scaling~\eqref{sol_vacio} with $m_0^2\mapsto\mu^2$ to a scaling law consistent with the $D=2$ Ising universality class.

\begin{figure}
    \centering
    \includegraphics[width=0.7\linewidth]{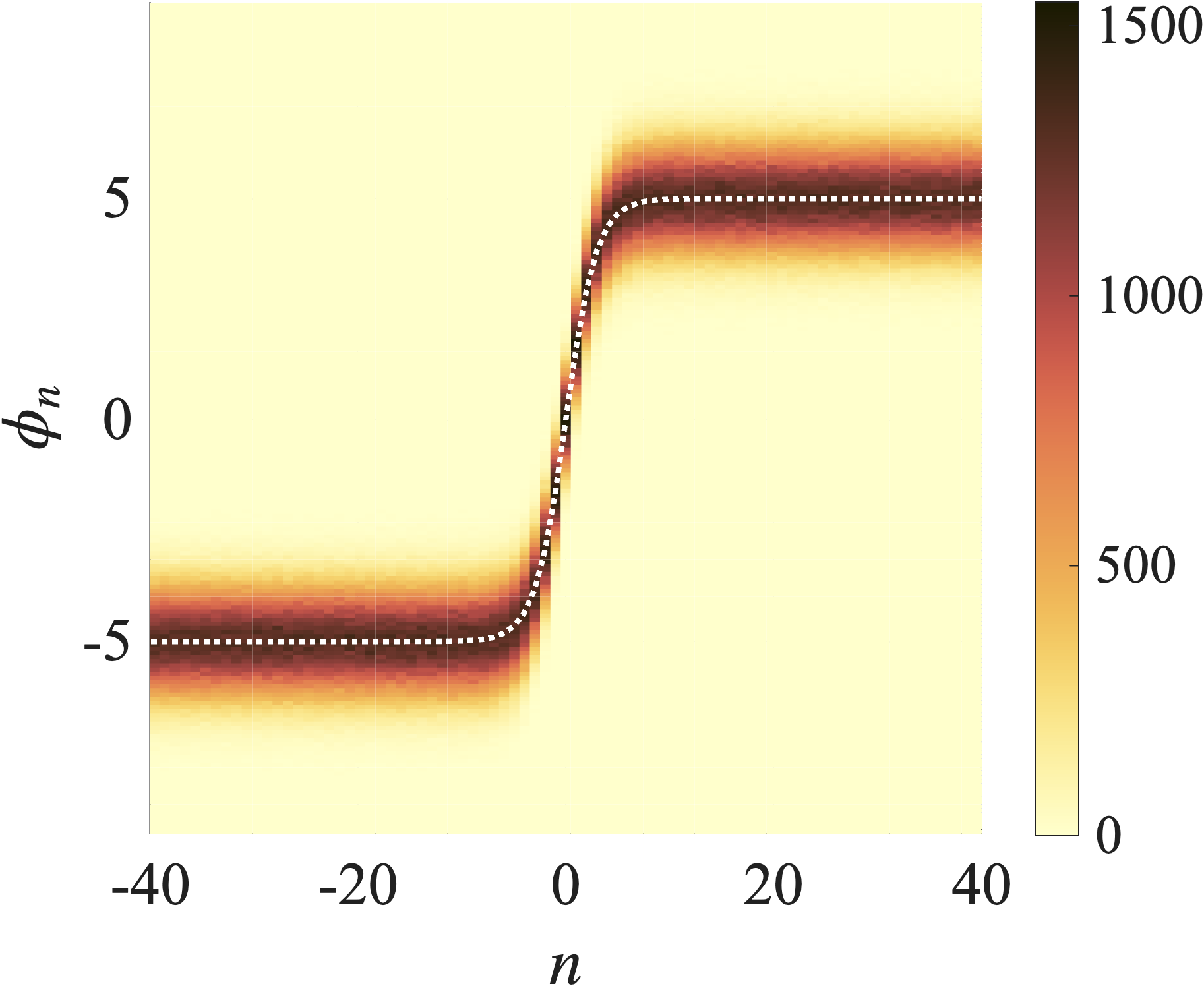}
    \caption{{\bf Classical solitons of a lattice scalar field:} The dotted white line shows the classical kink solution~\eqref{kink_cont} with $q_t=1$, plotted on top of a histogram of initial conditions for the truncated Wigner approach. This histogram is drawn from the Wigner distribution of the ground state of the model of coupled harmonic oscillators resulting from the linearization of the $\lambda\phi^4$ QFT around the kink solution, as defined in Eq.~\eqref{harmonic_chain_ground_state}.   Parameters used are $N=100$ lattice sites, lengthscale $\xi_0/a=3.0$, and amplitude $\Phi_0=5$, which uniquely identify the bare mass and quartic coupling.
    }
    \label{fig_TWA_initial}
\end{figure}

Within the spontaneous symmetry-broken (SSB) regime, in addition to standard particle-like excitations of the scalar field that can be produced from the vacuum by acting with a local perturbation, one also finds a different type of non-perturbative solitonic excitations~\cite{PhysRevD.10.4130, PismaZhETF.20.430, PhysRevD.11.1486}. These classical solitons can  be derived    analytically   in the classical limit
\begin{equation}\label{kink_cont}
        \Phi_{q_{\rm t}} (x) = q_{\rm t}\Phi_0 \tanh{\left( \frac{x-x_0}{\xi_0} \right)},\hspace{2ex} \xi_0^2 = - \frac{2}{m_0^2}
\end{equation}
where $\xi_0$ quantifies a width of the excitation, and we have included an arbitrary constant $x_0$ to describe the center of the kink $q_{\rm t}=+1$ or anti-kink $q_{\rm t}=-1$, reflecting the translational invariance of the continuum field theory. 
As advanced in the introduction, these solutions asymptotically approach the values of the two SSB ground states~\eqref{sol_vacio} (see the dashed line in Fig.~\ref{fig_TWA_initial}), and display an intermediate interpolating region leading to a finite excitation energy. We note that these excitations are stable and do not decay, as this would require
flipping the vacuum over an infinite spatial region, which requires an infinite energy. Moreover, they   can move freely   with momentum $\overline{P}_0$,   leading to a relativistic particle-like dispersion   $E_{q_{\rm t}}=(\overline{P}_0^2+M_0^2)^{1/2}$ with  the following rest mass  
\beq
\label{eq:soliton_mass}
M_0=\int\!{\rm d}x\left(\frac{1}{2}(\partial_x\Phi_{q_{\rm t}})^2+\frac{m_0^2}{2}\Phi_{q_{\rm t}}^2+\frac{\lambda}{4}\Phi_{q_{\rm t}}^4\right)=
\frac{2\sqrt{2}}{3}\,\frac{m_0^3}{\lambda}
\eeq
that is independent of the soliton center $x_0$. 

Interest in these solutions stems from the fact that they behave as extended particle solutions, but, nonetheless, they are not connected to the asymptotic free-field modes: they are the result of the inherent nonlinearities in the field equations and cannot be excited locally.  Moreover, these excitations have a conserved topological current $\partial_\mu J^{\mu}_{q_{\rm t}}=0$ that can be obtained from  $J^{\mu}_{q_{\rm t}}(x)=\frac{1}{2\Phi_0}\epsilon^{\mu\nu}\partial_{\nu}\Phi_{q_{\rm t}}(x)$. One can formally define a topological charge by  integrating this current, and identify it with the previous kink/anti-kink index
\beq
\label{eq:top_charge}
\!\!\int\!\!{\rm d}x\,\,J^0_{q_{\rm t}}=\frac{\Phi_{q_{\rm t}}(+\infty)-\Phi_{q_{\rm t}}(-\infty)}{2\Phi_0}=q_{\rm t}.
\eeq
Using the asymptotic values in Eq.~\eqref{kink_cont}, one clearly finds the $q_{\rm t}=\pm 1$ topological charges, and understands the robustness to local deformations that perturb the particular profile~\eqref{kink_cont} but do not flip the sign of the asymptotic values. In contrast,  the spatially-homogeneous SSB  vacuum has a vanishing $q_{\rm t} =0$.

When considering quantum fluctuations, a semiclassical functional method~\cite{PhysRevD.10.4114} shows that these particle-like extended states actually survive quantization~\cite{PhysRevD.10.4130}. In the regime of small fluctuations with respect to the classical solution, these quantum solitons are characterized by an energy spectrum with both discrete and continuous modes. In particular, the discrete solution at zero energy corresponds to a collective mode that can be identified with the center-of-mass motion of the soliton, confirming the relativistic dispersion introduced above. After subtraction of the tadpole divergence~\eqref{eq:tadpole} 
in all expressions, including the soliton amplitude $\Phi_0\mapsto \Phi=\sqrt{-\mu^2/\lambda}$ and width  $\xi^2_0\mapsto \xi^2=2/\mu^2$, the continuum and discrete vibrational modes of the soliton contribute together to the following shift of the soliton mass $M_0\mapsto M= 2^{3/2}\mu^3/3\lambda+2\mu(1/2\sqrt{3}-3/2\pi)$~\cite{PhysRevD.10.4130}. 

\subsubsection{Finite Yukawa coupling}

When switching back the Yukawa coupling $g\neq 0$,  we note that the $\mathbb{Z}_2$ symmetry $\phi\to-\phi$  must be supplemented with a discrete  axial rotation $\psi\to\gamma^5\psi$ that sends $\overline{\psi}\psi\to-\overline{\psi}\psi$. In a mean-field-type argument, the soliton  induces a  Dirac fermion mass $m_{q_{\rm t}}(x)=g\langle\phi(x)\rangle_{q_{\rm t}}=q_{\rm t}g\Phi\tanh{((x-x_0)/\xi)}$  that changes sign as one moves across the soliton. This can lead to various fermionic bound states depending on the ratio of the Yukawa coupling to the quartic interactions $g/\lambda^2$ ~\cite{PhysRevD.10.4130}. We consider the following irreducible representation of the gamma matrices $\gamma^0=\sigma^z$ and $\gamma^1=\ii\sigma^y$. As occurs for the continuum modes $\epsilon_{\pm}(k)=\pm({k^2+g^2\Phi^2})^{1/2}$, there is a spectral symmetry relating the discrete eigenmodes,  
which always come in pairs $\{\ket{\epsilon_n},\ket{-\epsilon_n}=\sigma^y\ket{\epsilon_n}\}$. An exception to this rule is a mode of zero-energy $\epsilon=0$ that fulfills 
\beq
\label{eq:zero_mode_ode}
\partial_x\Psi_{q_{\rm f}}(x)=\ii m_{\rm t}(x)\,\gamma^1\Psi_{q_{\rm f}}(x).
\eeq
The only normalizable solution to this differential equation is exponentially localised around the soliton
\beq\label{continuous_zero_mode}
\Psi_{q_{\rm f}}(x)=
\mathcal{N}\cosh^{-\sqrt{\frac{2g^2}{\lambda}}}\!\!\left(\frac{x-x_0}{\xi}\right)\begin{pmatrix}
1  \\
\ii q_{\rm t}  
\end{pmatrix},
\eeq
where $\mathcal{N}$ is a normalization factor. In contrast to the other modes, this one is self-conjugate  $\sigma^y\Psi_{q_{\rm f}}(x)=q_{\rm t}\Psi_{q_{\rm f}}(x)$ up to an irrelevant global phase, and is thus said to be unpaired. 
Within the same mean-field consideration, we can study the spectrum of the Dirac fermions in Eq.~\eqref{eq:JR_action} for a fixed scalar field configuration. This describes a Schrödinger equation in a Pöschl-Teller potential, which has both scattering and bound states, where the energies of the latter are \cite{PhysRevD.10.4130}

\begin{equation}\label{cont_limit_bound_states}
    \varepsilon_n= \pm \frac{1}{\xi}
    \sqrt{2 n 
    \sqrt{\frac{2g^2}{\lambda}} -n^2}, 
    \quad n=0,1,..., \lfloor \sqrt{2g^2/\lambda}\rfloor
\end{equation} 

\noindent which includes the zero-energy state.

 Although the zero-energy mode was initially argued to host a charge-neutral spin-1/2 excitation~\cite{PhysRevD.10.4130}, Jackiw and Rebbi later showed that its unpaired nature actually leads to a fermion with half-integer charge~\cite{PhysRevD.13.3398, NIEMI198699}. Paralleling the discussion of the topological charge above Eq.~\eqref{eq:top_charge}, the fermion charge in units $e=1$  is related to the conserved $U(1)$ current $j^{\mu}_{q_{\rm f}}(x)=\overline{\psi}(x)\gamma^\mu\psi(x)$ by means of $q_{\rm f}=\int{\rm d}x:j^0_{q_{\rm f}}:\,$. This expression needs to be regularized to subtract the diverging contribution of the filled Dirac sea.  Whereas  occupying (emptying)   positive (negative) energy  levels accounts for  fermions  and anti-fermions with opposite as excitations~\cite{Peskin:1995ev},  the  unpaired zero mode instead leads to a degenerate vacuum 
 \beq
 \label{eq:hs_zero_mode}
\ket{{\rm g}_-}=\prod_{\epsilon<0}\gamma^{\dagger}_{\epsilon}(k)\ket{0}, \hspace{2ex}\ket{{\rm g}_+}=\gamma^{\dagger}_{0}\prod_{\epsilon<0}\gamma^{\dagger}_{\epsilon}(k)\ket{0}.
 \eeq
 Using  a point splitting  regularization of the fermion current  $:j^0_{q_{\rm f}}:=\lim_{a\to 0}\half\big(\psi^\dagger(x+a)\psi(x-a)-\psi(x-\a)\psi^\dagger(x+a)\big)$  equivalent to normal ordering, one finds that all the contributions to the charge density from the paired finite-energy eigenmodes  $\pm \epsilon_n$ cancel, whereas the unpaired zero mode induces an spectral imbalance  that leads to a fractional charge
 \beq
 \label{eq:half_charge}
 q_{\rm f}\ket{ {\rm g}_{\pm}}=\!\!\int\!\!{\rm d}x:j^0_{q_{\rm f}}:\ket{ {\rm g}_{\pm}}=\!\!\int\!\!{\rm d}x\,\half|\Psi_{q_{\rm f}}(x)|^2\ket{{\rm g}_\pm}=\pm \half \ket{{\rm g}_\pm }.
 \eeq
Hence, the populated bound state at the kink $q_{\rm t}=+1$ or anti-kink $q_{\rm t}=-1$ carries a half-integer charge $q_{\rm f}=1/2$ 
  corresponding to a spin-1/2 particle that travels with the soliton, inheriting its stability from the previous topological arguments.  As noted in the introduction, the soliton is treated as a prescribed classical field in this approximation~\cite{PhysRevD.13.3398,JACKIW1981253,PhysRevB.25.6447,RAJARAMAN1982151,PhysRevD.30.809,PhysRevD.30.2136,PhysRevD.30.2194,PhysRevB.31.6112,PhysRevB.25.6447}. 

  Considering the Berry connection associated to the filled  Dirac sea of any of the two SSB groundstates $\mathcal{A}_{\pm}(k)=\bra{\epsilon_-(k)}\ii\partial_k\ket{\epsilon_-(k)}={\mp\ii g\Phi}/2(k^2+g^2\Phi^2)$~\cite{Ryu_2010}, one can assign a Chern-Simons invariant to the defect assuming that the soliton is a fixed slowly-varying background~\cite{RevModPhys.88.035005}. In this regime, the Chern-Simons invariant ${\rm CS}_1 =\tfrac{\ii}{2\pi}\int_{{\rm BZ}\times S^0}{\rm d}k{\rm d}x\,A(k,x)$  characterizes the topological nature of the fermionic defect through the difference of the Zak phases at two points $S_0=\{x_0-d,x_0+d\}$ far way from the kink/anti-kink, and reads
  \beq
  {\rm CS}_1 =\pm\ii\int_{\rm BZ}\tfrac{{\rm d}k }{2\pi}\big(A_+(k)-
A_-(k)\big)=\frac{q_{\rm t}}{2}{\rm mod}\mathbb{Z}.
  \eeq
  As already noted in the introduction,  the difference of the Zak phases at both sides of the soliton is regularization independent, and accounts for the topology of the localized zero-energy mode at the soliton.

 \subsection{Lattice Jackiw-Rebbi  field theory}
 
 Let us now consider the Hamiltonian field theory in Eq.~\eqref{eq:H_JR}, and discuss a standard discretization based on Kogut-Susskind staggered fermions~\cite{PhysRevD.11.395,PhysRevD.13.1043,PhysRevD.16.3031}.  We note that this approach not only regularizes the UV divergences by introducing a lattice spacing $a$, but also forms the basis to discuss the connection to trapped-ion Q$\ell$Ss in the following subsection. In the Kogut-Susskind approach, only the spatial spacetime coordinates are discretized, leading in this case to a simple chain
 \beq
 \Lambda_\ell=\left\{x_n=an-\tfrac{L}{2}:\hspace{1ex} n\in\{1,\cdots, N\}\right\},
 \eeq
where $N$ is an odd integer counting the number of sites. 

\subsubsection{Bosonic sector: Scalar field}

We start by analyzing the bosonic sector, considering that the scalar field and its canonical momentum are quantum operators defined on the individual sites  $\phi(x),\pi(x)\mapsto\phi_n,\pi_n$ and fulfilling the canonical algebra $[\phi_{n},\pi_{n'}]=\ii\delta_{n,n'}/a\to\ii\delta(x-y)$ as $a\to 0$ \cite{Jordan2014,Rothe:1992nt}. Substituting the spatial derivatives with finite differences, we get the following lattice Hamiltonian 
\begin{equation}\label{Hphi4}
    \Hb= a\!\sum_{n=1}^{N} \!\mathcal{H}_n=a\!\sum_{n} \! \left( \frac{\pi_n^2}{2} + \frac{1}{2} \big( \nabla_{\!a}^+
 \phi_n \big)^2 + \frac{m_0^2}{2} \phi_n^2  +  \frac{\lambda}{4} \phi_n^4 \right),
\end{equation}

\noindent where $\nabla^+_{\!a} \phi_n=(\phi_{n+1}-\phi_n)/a$ is the forward difference. Note that a simple gradient expansion for the field $\phi_{n+1}\approx\phi(x)+a\partial_x\phi$,  and equivalently for the canonical momentum, together with  $a\sum_{n}\mapsto\int{\rm d}x$ and $L=Na\to\infty$, clearly yield the original self-interacting scalar QFT with discretization errors scaling as $\mathcal{O}(a)$. We also note that a simple  change of the discretization to central differences $\nabla_{\!a}
 \phi_n=(\phi_{n+1}-\phi_{n-1})/2a$  would decrease these errors to $\mathcal{O}(a^2)$ but, on the other hand,  move from nearest-neighbor couplings to next-to-nearest ones, which cannot be easily mapped onto the trapped-ion Q$\ell$S of this work. We thus stick to the forward-difference choice.

 \begin{figure}
    \centering
   \includegraphics[width=0.7\linewidth]{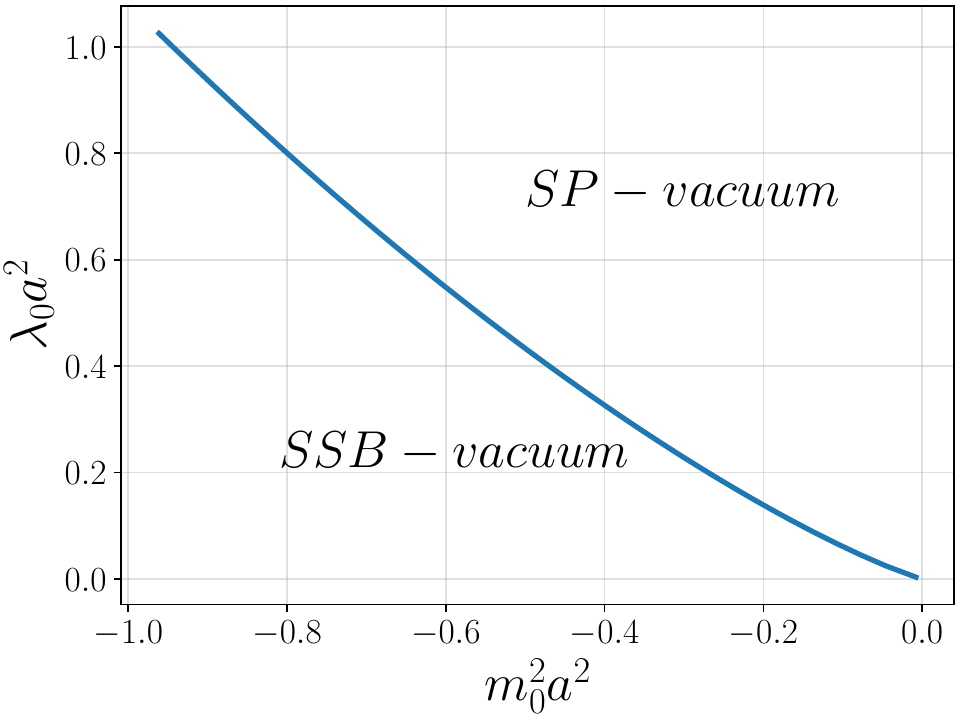}
    \caption{{\bf Critical line  Critical line for the parity-breaking phase transition: the parity-breaking phase transition:} we obtain critical points by means of the tadpole (\ref{eq:tadpole}) and sunrise (\ref{eq:sunrise}) mass shifts. After selecting a $\mu^2$ value, we first employ the sunrise contribution to obtain $\lambda$, after which the tadpole one is used to get $m_0^2$. The resulting line is represented in blue, and separates the spontaneous symmetry-broken vacuum (below the critical line) from the symmetry-preserved vacuum (above the critical line).}
    \label{fig_critical_line}
\end{figure}

The lattice serves as a UV regularization of the tadpole diagram discussed in Eq.~\eqref{eq:tadpole} by exchanging $k^2\mapsto k_0^2+\tfrac{2}{a^2}\sin^2(\tfrac{ka}{2})$ for  $k=-\frac{\pi}{a}+\frac{2\pi}{L}j$ with $j=1,\cdots,  N$  labeling the discrete momenta within the Brillouin zone. In fact, the tadpole diagrams can  be resummed
 to all orders in the quartic coupling, leading to the following self-consistent equation
 \beq
 \label{eq:tadpole}
m_0^2= \mu^2-\frac{3\lambda}{2\pi}\frac{1}{\sqrt{1+\fourth \mu^2a^2}}\,\mathsf{K}\!\left(\frac{1}{1+\fourth \mu^2a^2}\right),
 \eeq
 where we have introduced the  complete
elliptic integral of the first kind $\mathsf{K}(x) = \int^{\pi/2}_0{\rm d}\theta(1-x\sin^2\theta)^{-1/2}$. This approximation coincides with the Hartree
method of mean-field theory, and solving it for $\mu^2=0$ can give an estimate of how the critical point changes with coupling strength $m_{\rm c}^2(\lambda)<0$.
However, as noted in \cite{Vinas2023}, when trying to solve this non-linear equation in $D=1+1$ dimensions, one faces an infra-red (IR) divergence. To obtain a finite prediction, one must incorporate further quantum corrections going beyond the tadpole family. In particular, by considering all possible tadpole-like decorations to the so-called sunrise diagram---the lowest-order diagram of the perturbative expansion depending on external momenta---the condition for the symmetry breaking  $\mu^2=0$ changes to
\beq
\mu^2+ \mathsf{\Sigma}_{\rm sr}({\boldsymbol{0}})=0,
\eeq
where the sunrise mass shift  in the self-energy   reads
\beq
\label{eq:sunrise}
\mathsf{\Sigma}_{\rm sr}({\boldsymbol{0}}) = -\frac{3}{2}\frac{\lambda^2}{(aN)^2}{\sum_{j_1,j_2=1}^N} {1 \over S(j_1,j_2)},
\eeq
Here, we have substituted spatial integration by mode summation $\int \frac{{\rm d}k}{2\pi}\mapsto\frac{1}{aN}\sum_{j=1}^{N}$,  $
\omega^2(j)=
\frac{4}{a^2} \sin^2 \frac{ka}{2}+\mu^2$, and $S(j_1,j_2)=\omega(j_1)+ \omega(j_2) +\omega(j_1+j_2)$.
The contribution (\ref{eq:sunrise}) plays a key role 
as it allows to circumvent the IR divergence of the two-dimensional case. In this scenario, the procedure to obtain critical points is: 1) select a range of $\mu^2$ values. 2) For each $\mu^2$ value, obtain $\lambda$ via eq. (\ref{eq:sunrise}). 3) Use the tadpole equation (\ref{eq:tadpole}) to get $m_0^2$. We have calculated the corresponding critical line by extending the approach of~\cite{Vinas2023} to zero temperatures, obtaining the results shown in Fig.~\ref{fig_critical_line}. We are interested in the left region in which the discrete $\mathbb{Z}_2$ symmetry gets spontaneously broken, and one can look for lattice solitons interpolating between the two possible vacua.

The Heisenberg equations  for the lattice scalar field read
\begin{equation}\label{phi4_discreto_ecs}
    \begin{split}
        \frac{{\rm d}\phi_n}{{\rm d}t} & = \pi_n, \\
        \frac{{\rm d}\pi_n}{{\rm d}t} & = \frac{1}{a^2}\big(\phi_{n+1} -2 \phi_n + \phi_{n-1}\big) - m_0^2 \phi_n - \lambda \phi_n^3,
    \end{split}
\end{equation}
and, in the non-interacting limit $\lambda=0$,  lead to the following dispersion relation 
\beq 
\label{eq:dispersion_relation}
\omega(k)=\sqrt{m_0^2+\frac{4}{a^2}\sin^2\!\left(\frac{ka}{2}\right)}.
\eeq
Here, we have moved to the momentum representation of the fields $\phi_n=\sum_k\phi_k\ee^{\ii kx_n}/\sqrt{L}$ for $k=-\frac{\pi}{a}+\frac{2\pi}{L}\left(j+\frac{1}{2}\right)$. Note that, in order to account for the anti-periodic boundary conditions $\phi_{N}=-\phi_1$ that are required to accommodate an odd number of kinks/anti-kinks in the scalar field, the allowed momenta get shifted with respect to the previous values. 
A long-wavelength expansion around the BZ center  $k=0$ leads to the standard relativistic dispersion relation $\omega^2(k)=m_0^2+k^2$.  

The classical limit of the Heisenberg equations~\eqref{phi4_discreto_ecs} can be obtained by exchanging the lattice operators by $c$-numbers $\phi_n\mapsto \Phi_n$, $\pi_n\mapsto \Pi_n$, and the  commutators by Poisson brackets $\tfrac{1}{i\hbar}[\phi_n,\pi_{n'}]\mapsto \{\Phi_n,\Pi_{n'}\}_{\rm PB}=\delta_{n,n'}/a$, where  $\{F,G\}_{\rm PB}=\sum_n(\partial_{\Phi_n}F\partial_{\,\Pi_n}G-\partial_{\Phi_n}G\partial_{\,\Pi_n}F)$. The classical  solitons also appear on the lattice~\cite{KOEHLER19751515,PhysRevB.11.3535,Lizunova2021} 
and  asymptotically approach the two different SSB groundstates  with the following discretized profile
\begin{equation}\label{kink_discrete}
        \Phi_{q_{\rm t},n} = q_{\rm t}\Phi_0 \tanh{ \left( \frac{a}{\xi_0}(n-n_0)\right)}.
\end{equation}
Here, $n_0$ is the index of the soliton center, and we note that it is not forced to coincide with a lattice site. 

\begin{figure}
    \centering  
    \includegraphics[width=0.6\linewidth]{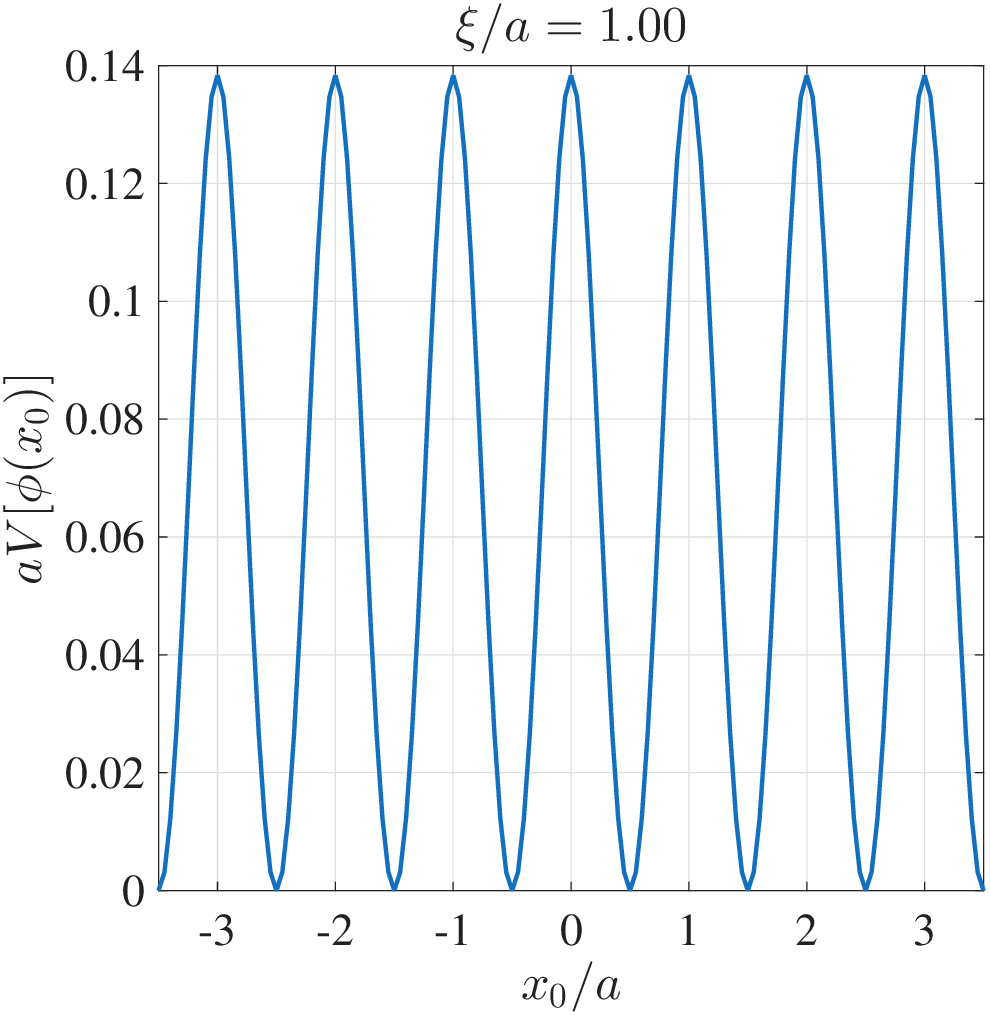}
    \caption{{\bf Peierls-Nabarro potential:} Potential energy felt by the soliton as a function of its center. Parameters used are $N=160$, $\sbgs=3$ and $\xi_0/a=1$. 
    }
    \label{fig:pn_barrier}
\end{figure} 

As already discussed for the continuum QFT, small fluctuations about the solitons present a single zero-energy mode, the Goldstone mode, describing center-of-mass translations obtained from the static profile~\eqref{kink_cont} by a Lorentz boost $\Phi_{q_{\rm t}}(x) \to \Phi_{q_{\rm t}}(\gamma_v(x-vt))$ with $\gamma_v=1/\sqrt{1-v^2}$~\cite{Kevrekidis2000}.
In the discretized model~\eqref{Hphi4}, the lowest-energy mode also describes the motion of the soliton center of mass, and can be understood as the lattice version of the Goldstone mode~\cite{Kevrekidis2000}. 
In contrast to the continuum, the displacement of this mode is now subject to a non-zero energy cost, the so-called Peierls-Nabarro (PN) barrier~\cite{RPeierls_1940,Nabarro_1947,Dmitriev2006,Peyrard1984}. The soliton mass on the lattice is thus superseded by a PN potential  
\beq
\label{eq:PN_potnetial}
V_{\rm PN}(n_0)=a\sum_n \frac{1}{2}\left(\frac{\Phi_{q_{\rm t},n+1}-\Phi_{q_{\rm t},n}}{a}\right)^2 + \frac{1}{2} m_0^2 \Phi_n^2 + \frac{\lambda}{4}\Phi_n^4,
\eeq
and now becomes a periodic function on the soliton center $V_{\rm PN}(n_0+1)=V_{\rm PN}(n_0)$.
In fact, by considering the leading harmonics in a Fourier series, one finds \cite{Woafo1993} 
\beq
\label{eq:PN}
V_{\rm PN}(n_0)\approx M_0+V_0\cos(2\pi n_0)
\eeq
Note that in the long-wavelength limit $V_0\propto\ee^{-\pi\xi_0/a}$ is the aforementioned PN barrier, which vanishes for  $\xi_0\gg a$. 
Here, one recovers the free translations of the soliton with the constant classical  mass~\eqref{eq:soliton_mass}.
Away from this limit and for $V_0>0$, we see that the rest mass energy of the lattice soliton becomes maximal when the center coincides with a lattice site $n_0$ integer, and minimal when it is centered at a link  $n_0$ half integer (see Fig.~\ref{fig:pn_barrier}). Under these conditions, only solitons centered around the lattice kinks describe stable classical solutions. In addition, one expects that in contrast to the continuous case,  an initially moving lattice soliton will dissipate energy due to resonances with the lattice phonon modes, until it gets pinned by the PN potential \cite{Kevrekidis2000}. 

\begin{figure}
    \centering    \includegraphics[width=1\linewidth]{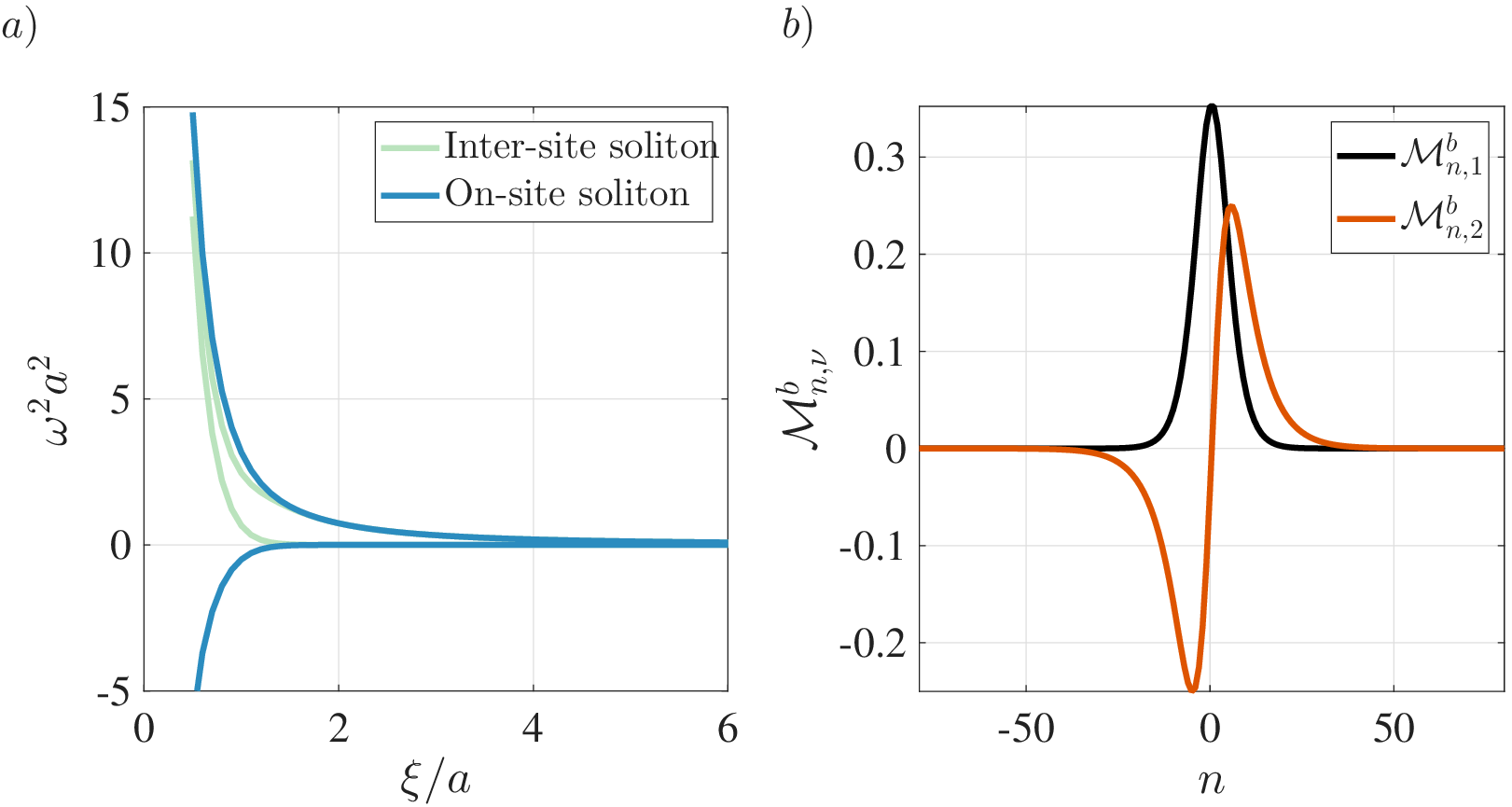}
    \caption{{\bf Low-lying excitations around the classical soliton: } a) Frequencies of the first two lowest lying modes of the equations of motion linearized around the classical soliton (eqs. \eqref{ec_dispersion_kink_lattice}) for an inter-site kink (light-blue) and an on-site kink (dark-blue). This evidences the on-site kink as linearly unstable in discrete lattices. Parameters used are: $N=160$, $\sbgs=3$. b) Amplitudes of these modes for the inter-site solution at $\xi_0/a=6$. }
    \label{lowest_lying_frequencies}
\end{figure}

As an alternative approach to identify the stable soliton configurations, we start from an initial classical configuration centered at the central lattice site $\Phi_0 \tanh(na/\xi_0)$, add an artificial damping rate $\kappa$ to the equations of motion~\eqref{phi4_discreto_ecs} in the classical limit through a term ${\rm d} \Pi_n /{\rm d}t\propto -\kappa \Pi_n$, and let the system relax to the asymptotic equilibrium state~\cite{dmitriev2018}.  We note that during this relaxation, the field only adapts its width but does not change the soliton center.  Hence, we can also obtain the solitonic solutions centered at a lattice link by relaxing from the initial state $\Phi_0 \tanh\big((n-\tfrac{1}{2})a/\xi_0\big)$. 
Next, we find the small-amplitude fluctuations by linearizing the system~\eqref{phi4_discreto_ecs}  using $\phi_n(t) = \Phi_{{\rm K},n} + \delta \phi_n(t)$~\cite{Dmitriev2006, Chelpanova2023}, where $\Phi_{{\rm K},n}$ represents the $c$-number lattice field for the steady-state solution discussed above, and $\delta\phi_n(t)$ are bosonic operators. Keeping  terms quadratic in the fluctuations, these operators fulfill  
\begin{equation}\label{ec_dispersion_kink_lattice}
    \frac{{\rm d}^2}{{\rm d}t^2}\delta \phi_n = - \left( 2 + m_0^2  + 3 \lambda \Phi_{{\rm K},n}^2 \right) \delta \phi_n + \delta \phi_{n+1} + \delta \phi_{n-1} .
\end{equation}
We obtain the corresponding spectrum looking for oscillating solutions $\delta\phi_n(t)=\ee^{\i\Omega_{\nu} t}\mathcal{M}^{\rm b}_{\nu,n}$ and,  as expected~\cite{dmitriev2018}, we find that most eigenvalues $\Omega_\nu$ follow the vacuum dispersion relation $\omega(k)$ in Eq.~\eqref{eq:dispersion_relation} with a rescaled mass $m_0^2\mapsto 2\lambda\Phi_0^2=2m_0^2$. 
However, the presence of the soliton also leads to in-gap modes $\Omega_{\nu}^2<2m_0^2$, with the lowest-lying modes being spatially localized around the soliton center. Depending on the lattice parameters, there are two or three of these modes \cite{Lizunova2021}. 
The lowest energy mode $\Omega_0$ is the center of mass or translational mode of the discrete kink, which connects with the Goldstone mode of the continuum limit, while the remaining ones $\Omega_\nu\leq\omega(k)$ represent internal vibrations of the soliton \cite{Campbell1983,dmitriev2018}.
In Fig.~\ref{lowest_lying_frequencies}, we show the frequencies of the first two modes considering both the soliton centered at the central lattice site and at a neighboring lattice link 
\cite{dmitriev2018}.
From this figure, we see that $\Omega_0^2<0$ for the site-centered soliton, giving an alternative account of the aforementioned instability for the site-centered solutions. 

\subsubsection{Fermionic sector: Dirac field}

Let us now introduce the Dirac fermions in the lattice field theory. Due to the phenomenon of fermion doubling~\cite{NIELSEN198120}, the discretization of the Dirac field can follow various prescriptions~\cite{lattice}. In this work, we opt for the staggered-fermion approach~\cite{PhysRevD.11.395} which, in the Hamiltonian formulation and for $D=1+1$ dimensions~\cite{PhysRevD.13.1043,PhysRevD.16.3031}, dispenses entirely with the effect of the doublers. Instead of using two internal components per lattice site, this approach uses a single component and 
 introduces lattice fermion fields with anti-commutation relations $\{\chi^{\phantom{\dagger}}_{n},\chi^{\dagger}_{n'}\}=\delta_{n,n'}/a\to\delta(x-y)$, effectively doubling the unit cell.  The spatial derivatives  in the JR Hamiltonian~\eqref{eq:H_JR} become finite differences that couple nearest neighbors
\begin{equation}\label{H_XX_fermions}
\begin{split}
    H_{\rm fb}= a\!\sum_{n=1}^{N}\! \!\Big(\,\ii aJ  \chi_n^\dagger \nabla_{\!\!a}^{\phantom{\dagger}}\chi^{\phantom{\dagger}}_{n} +(-1)^nm(\phi^{\phantom{\dagger}}_n)\chi_n^\dagger \chi^{\phantom{\dagger}}_{n}\Big),
    \end{split}
\end{equation}
where, in contrast to the scalar case,  we can here use symmetric differences  $\nabla_{\!a}
 \chi_n=(\chi_{n+1}-\chi_{n-1})/2a$ without introducing   next-to-nearest-neighbor couplings. Here, we have introduced a tunneling energy $J$, and incorporated the Yukawa coupling to the lattice scalar field in the mass term 
 \beq
 m(\phi_n)=m_{\rm f}+g\phi_n.
 \eeq
 For $g=0$, the mass becomes homogeneous, and the Hamiltonian becomes diagonal in momentum space $\chi_n=\sum_k\ee^{\ii kx}\chi(k)/{\sqrt{L}}$,  leading to  $H_{\rm fb}=\sum_{k\in{\rm HBZ}}\psi^\dagger(k)(\boldsymbol{d}(k)\cdot\boldsymbol{\sigma})\psi(k)$. Here,  the Dirac spinor  reads $\psi(k)=(\chi(k),\chi(k+\pi/a))^{\rm t}$ and the vector   $\boldsymbol{d}(k)=(m_{\rm f}, 0,2J\sin(ka))$ is defined on  a reduced Brillouin zone   $k=-\frac{\pi}{2a}+\frac{\pi}{L}j$ with $j=1,\cdots,\tfrac{N}{2}$ to account for the doubled unit cell. The  energy of the  free-field   modes is 
\beq
\epsilon_{\pm}(k)=\pm |\boldsymbol{d}(k)|=\pm\sqrt{m_{\rm f}^2+4J^2\sin^2(ka)}.
\eeq
 A long-wavelength expansion around $k=0$ leads to the standard relativistic dispersion of Dirac fermions with an effective speed of light $c_{\rm f}=2Ja$, such that the choice of $J=1/2a$ sets the desired natural units. This derivation also leads to the choice of gamma matrices $\gamma^0=\sigma^x,\gamma^1=-\ii\sigma^y$, and $\gamma^5=\sigma^z$. We note that the eigenstates come again in positive- and negative energy pairs $\{\ket{\epsilon_+(k)},\ket{\epsilon_-(k)}=\sigma^y\ket{\epsilon_+(k)}\}$, which can be seen as a result of $\sigma^y(\boldsymbol{d}(k)\cdot\boldsymbol{\sigma})\sigma^y=-\boldsymbol{d}(k)\cdot\boldsymbol{\sigma}$. At the level of the lattice, we see that this structure of the spectrum requires exchanging the sites of the unit cell, which will become important below to understand charge fractionalization.

 Exchanging the bare Dirac mass   $m_{\rm f}=0$ for a non-zero Yukawa coupling  $g\neq 0$ introduces an interaction vertex on the lattice model. In this case, prior to a gradient expansion  as  the one that connected  the lattice scalar field to its continuum QFT, one has to consider that the low-energy right- and left-moving Dirac fields $\psi_{\pm}(x)=\half(1\pm\gamma^5)\psi(x)$  come from  two different Fermi points $ka\in\{0,\pi\}$~\cite{Affleck:1988mua}, and 
 \beq
 \label{eq:long_wav_fermions}
 \chi_{n\pm 1}\approx \psi_{+}(x)\pm a\partial_x\psi_{+}(x)+(-1)^{n\pm 1}\big(\psi_{-}(x)\pm a\partial_x\psi_{-}(x)\big),
 \eeq
Following a similar mean-field type approximation of the Yukawa mass in the lattice soliton background   $M_{q_{\rm t},n}=g\langle\phi_n\rangle_{q_t}=q_{\rm t}g\Phi_0\tanh(a(n-n_0)/\xi_0)$, one can now look for possible fermionic zero modes localised to the discrete soliton. In fact, the lattice analogue of the zero-mode  equation~\eqref{eq:zero_mode_ode}
is a  second-order difference equation that is purely linear 
\beq
\label{eq:finite_diff_zero_mode}
{\ii J}\Psi_{q_{\rm f},n+1}+(-1)^nM_{q_{\rm t},n}
\Psi_{q_{\rm f},n}-\ii J\Psi_{q_{\rm f},n-1} = 0.
\eeq 
\noindent where the zero-mode wavefunction is in units of $\sqrt{a}$. When assuming that both the mass term and the zero-mode solution are slowly varying on the lattice scale, such that both differences give the same result and the mass term on the unit cell is the same, the $a\to 0$ limit of the above system coincides exactly with Eq.~\eqref{eq:zero_mode_ode}.

\begin{figure}
    \centering    \includegraphics[width=1\linewidth]{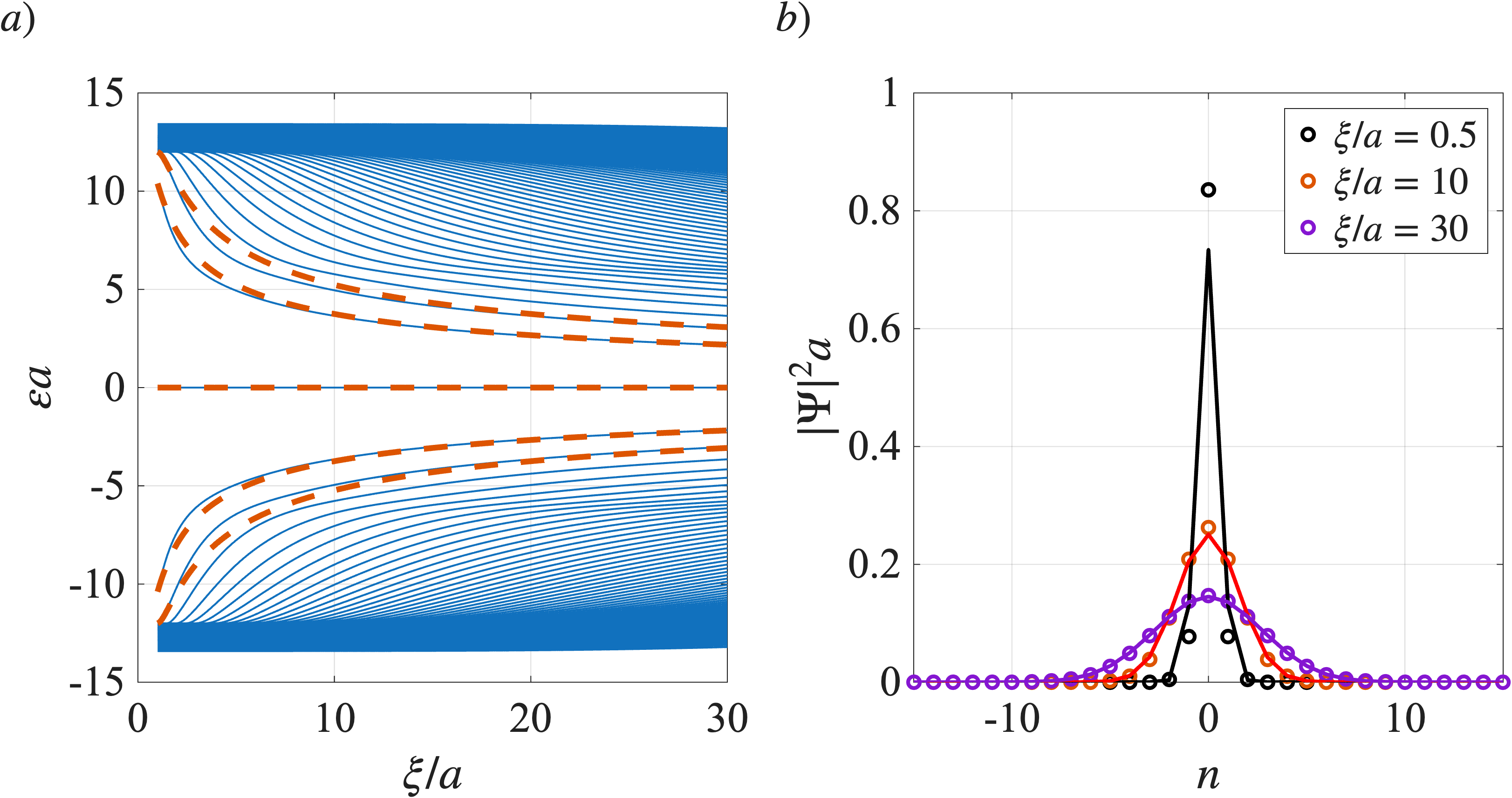}
    \caption{{\bf Fermionic spectrum and zero mode:} a) Spectrum as a function of the solution width for an on-site kink, obtained through the tight binding model of Eq.~\eqref{eq:matrix_fermions} (blue lines), and the continuum limit analytical solution for the bound states of Eq.~\eqref{cont_limit_bound_states} (red dashed lines). b) Fermionic zero-mode eigenfunctions. The circles denote numerical values obtained through Eq.~\eqref{fermionic_ground_state}, while the solid lines show the fermionic zero-mode of the continuous model in Eq.~\eqref{continuous_zero_mode}. Parameters used are $N=160$, $\sbgs=3$, $Ja=3$, $ga=2$.}
    \label{fermionic_zero_mode_frequencies}
\end{figure}

We argued in the continuum case that this zero mode can host a half-integer charge due to a spectral asymmetry, which becomes particularly transparent when considering the point-splitting  regularization~\eqref{eq:half_charge}. On the lattice, however,  the manifestation of this spectral asymmetry and its associated half charge is not straightforward, and finding the specific observable that captures it might depend on the specific discretization. The most transparent manifestation appears in the Su-Schrieffer-Hegger model~\cite{PhysRevLett.42.1698,RevModPhys.60.781,PhysRevLett.88.180401}, in which it is the nearest-neighbor tunneling of the fermions and not their local energy which is staggered or, as usually referred to in this context, dimerised. In this case, the positive and negative energy eigenstates are related by inverting the sign of the amplitudes on the even sites, and the spectral asymmetry of the unpaired mode can then be captured by simply subtracting a constant value from the lattice density, which can be readily understood as the result of normal ordering~\cite{BELL19831,PhysRevLett.91.150404}. For the lattice JR model, in contrast, this symmetry requires exchanging the eigenstate amplitudes of the even-odd lattice sites, together with complex conjugation. Let us denote the fermionic Hamiltonian
in the presence of a fixed scalar (mean) field as
\beq\label{Hfermions}
H_{\rm fb}=    a\sum_{n,l}\chi^{\dagger}_n\, [\,\,h(\boldsymbol{\Phi}_{\rm K})\,]^{\phantom{\dagger}}_{nl}\,\chi^{\phantom{\dagger}}_{l},
    \eeq
where we have introduced the tri-diagonal Hermitian matrix
\beq
\label{eq:matrix_fermions}
[\,\,h(\boldsymbol{\Phi}_{\rm K})\,]_{nl}=\ii J\delta_{l,n+1}-\ii J\delta_{l,n-1}+g(-1)^n\Phi_{{\rm K},n}\delta_{l,n}.
\eeq
 By diagonalizing it, one finds the eigenvalues in increasing order  $\{\epsilon_{\nu}:\nu\in\{1,\cdots,N\}\}$, such that the fermionic 
vacuum occupying the localised mode around the soliton  reads 
\begin{equation}\label{fermionic_ground_state}
    |{\rm g}_+\rangle = \prod_{\nu=1}^{N_{\rm f}} \gamma_{\epsilon_\nu}^\dagger \ket 0 , 
\end{equation}
\noindent where we fill up to $N_{\rm f}=(N+1)/2$ ($N_{\rm f}=N/2$) energy levels if $N$ is odd (even). Here, the fermionic operators read  $\gamma_{\epsilon_\nu}^\dagger = \sum_n \mathcal{M}^{\rm f}_{n,\epsilon_\nu} \chi^{\dagger}_n$, where $\mathcal{M}^{\rm f}$ is a unitary matrix with eigenstates arranged in the columns and fulfilling  $\sum_{n,l}(\mathcal{M}^{\rm f}_{n,\epsilon_\nu})^* [\,\,h(\boldsymbol{\Phi}_{\rm K})\,]_{nl}\mathcal{M}^{\rm f}_{l,\epsilon_\tau}=\epsilon_\nu\delta_{\nu,\tau}$. 

For  $\nu=N_{\rm f}$, one can indeed show that the corresponding eigenmode is localised around the soliton also in this discrete case, and regardless of where the soliton center lies (see Fig.~\ref{fermionic_zero_mode_frequencies}).
In this figure,  we show the change in the fermionic spectrum as one varies the soliton width (Fig.~\ref{fermionic_zero_mode_frequencies} {\bf (a)}), together with the fermionic zero mode at different widths (Fig.~\ref{fermionic_zero_mode_frequencies} {\bf (b)}). 
We observe in this figure how, as we increase $\xi$, there is a better agreement with the wavefunctions (cf. Eq. \eqref{continuous_zero_mode}) and the energies (cf. Eq.~\eqref{cont_limit_bound_states})
of the continuous model, and how several energies deviate from the bulk continuum bands $\epsilon_{\pm}(k)$ to become in-gap modes (as expected from Eq.~\eqref{cont_limit_bound_states}). Note that Eq.~\eqref {continuous_zero_mode} is derived for natural units, whereas the fermionic model can have a different Fermi velocity $c_{\rm f}=2Ja$. 
To account for this, the continuum expressions~\eqref{continuous_zero_mode} and \eqref{cont_limit_bound_states} must be rescaled as 
$g\mapsto g/ c_{\rm f}$, and $\varepsilon_n\mapsto \varepsilon_n  c_{\rm f}$,
leading to a perfect agreement with that of the lattice zero-energy solution when $\xi_0 \gg a$ ~\cite{PhysRevD.10.4130}.

For this discretization, the spectral symmetry relating positive and negative energy eigenstates is associated with the following matrix identity 
\beq
\label{eq:symmetry_JR}
\mathcal{M}^{\rm f}_{n,-\epsilon_\nu}=(\mathcal{M}^{\rm f}_{N-n+1,+\epsilon_\nu})^*.
\eeq
One can then prove that the accumulated charge can be expressed  in terms of the zero-mode   probability
\beq
\begin{aligned}
a\sum_n \langle\chi^\dagger_{n}\chi_{n}^{\phantom{\dagger}}\rangle_{{\rm g}_+}
= & \sum_n \!\!\sum_{\epsilon_\nu < 0}\!\!|\mathcal{M}^{\rm f}_{n,\epsilon_\nu}|^2+|\mathcal{M}^{\rm f}_{n,0}|^2 \\
= & \sum_n  \!\!\sum_{\epsilon_\nu < 0}\!\!\half\big(|\mathcal{M}^{\rm f}_{n,\epsilon_\nu}|^2 + |\mathcal{M}^{\rm f}_{N-n+1,-\epsilon_\nu}|^2 \big)\!+|\mathcal{M}^{\rm f}_{n,0}|^2 \\
= & \sum_n\!\! \sum_{\epsilon_\nu\lesseqgtr 0}\!\!\half|\mathcal{M}^{\rm f}_{n,\epsilon}|^2+\half|\mathcal{M}^{\rm f}_{n,0}|^2   \\ 
= & \sum_n \half\big( 1+|\mathcal{M}^{\rm f}_{n,0}|^2\big),
\end{aligned}
\eeq
where  we used Eq.~\eqref{eq:symmetry_JR} in the second line, and unitarity $\mathcal{M}^{\rm f}(\mathcal{M}^{\rm f})^{\dagger}=\mathbb{1}$ in the last one.
This calculation shows how the regularization associated with normal ordering involves the averaging of the fermion charge over two neighboring sites   
\beq
\label{eq:charge_lattice}
\rho^{\phantom{\dagger}}_{n,n+1}=\frac{a}{2}\langle\chi^{\dagger}_{n\phantom{+1}\hspace{-2ex}}\chi^{\phantom{\dagger}}_{n\phantom{+1}\hspace{-2ex}}+\chi^{\dagger}_{n+1}\chi^{\phantom{\dagger}}_{n+1}\rangle
\eeq
with  the subtraction of a constant, such that one captures the spectral asymmetry  
\beq
 q_{\rm f}=\sum_n\left(\rho_{n,n+1}-\frac{1}{2}\right)=
\frac{1}{2}\sum_n|\mathcal{M}^{\rm f}_{n,0}|^2=\frac{1}{2}.
\eeq

It is interesting to note that this unit-cell charge coincides  with a particular instance of a generic definition of the charge operator on the lattice introduced in \cite{Ranft1984}, namely 
\begin{equation}\label{charge_density}
    2Q_{n,n+1}=\sum_{l=n-n_0}^{n+n_0} \!\!\!f_{l-n}\big(\chi_l^\dagger \chi_l-{\tfrac{1}{2}}\big) + \sum_{l=n+1-n_0}^{n+1+n_0} \!\!\! f_{l-n-1}\big(\chi_l^\dagger\chi _l-\tfrac{1}{2}\big),
\end{equation}
where one averages over $n_0$ neighbors using a smooth sampling function $f_l$ of a given range. For $n_0=0$ and $f_l=1$, one can see that 
\beq
\label{eq:reg_charge}
\langle Q_{n,n+1}\rangle=\rho_{n,n+1}-\tfrac{1}{2}.
\eeq
For discrete realizations, the non-integer fermion number is not an eigenvalue of this charge operator, but just an expectation value with corrections due to finite-size \cite{Jackiw1999}. However, the variance of the fermion number should vanish in the continuum limit \cite{Jackiw1999}.

\begin{figure}
    \centering
    \includegraphics[width=1\linewidth]{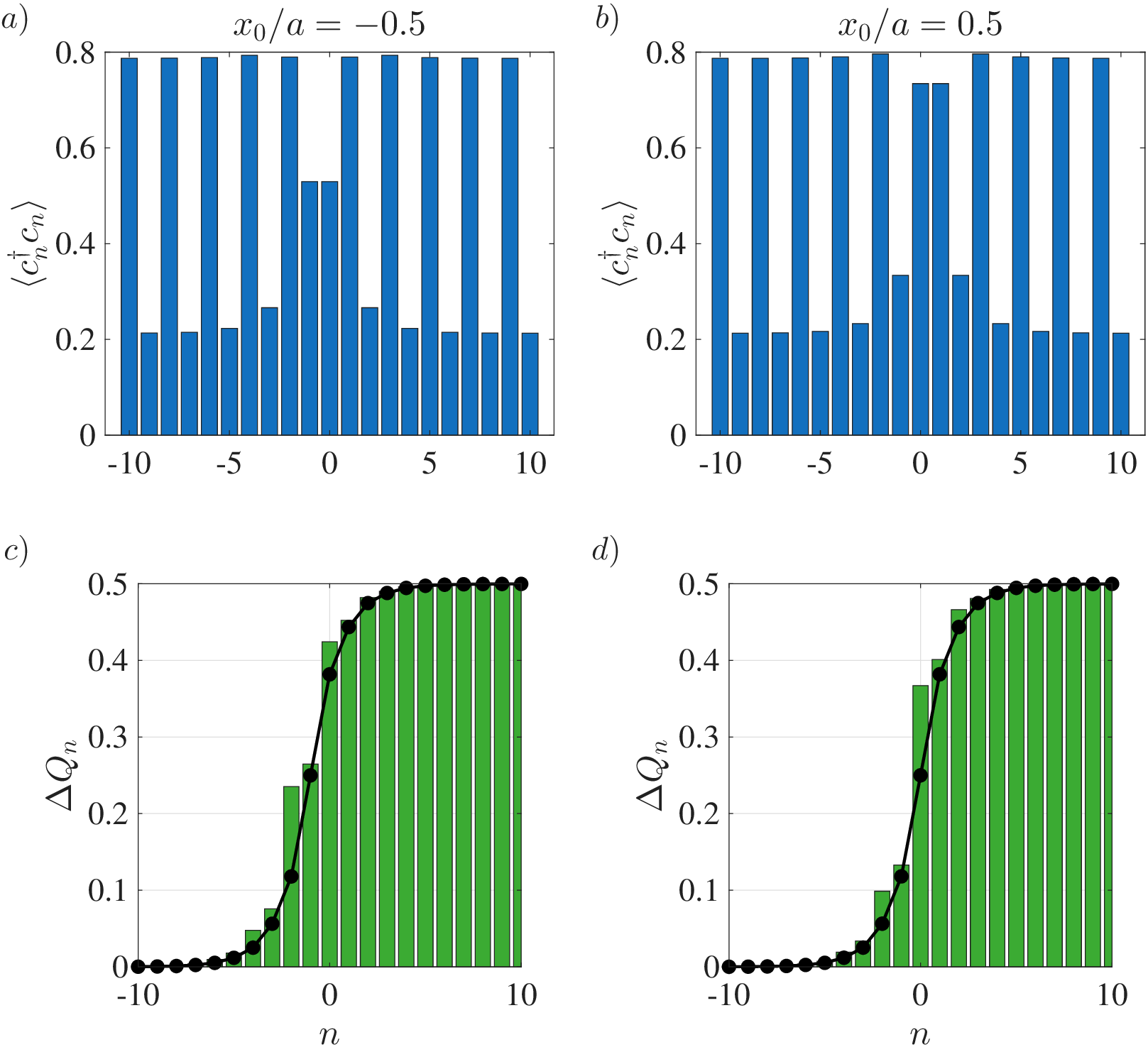}
    \caption{{\bf Charge fractionalization:} a) and b) Occupation number for two different kink positions. As explained in the text, the left configuration leads to a small PN potential well, which the kink can easily escape.  The accumulated charge density in c) and d) shows the fractional fermion number induced due to the presence of the soliton, respectively, for the kink positions in a) and b). We show for comparison the integrated density of the fermionic zero-mode $\frac{1}{2}\sum_{i=1}^n |\mathcal{M}^{\rm f}_{i,0}|^2$ in a black solid line. Parameters used are $N=160$, $\sbgs=3$, $\xi/a=1$, $Ja=3$, $ga=0.8$. 
    }\label{fig_occupation_pn_barrier}
\end{figure}

The relevant fact, however, is that this charge is localized around the zero-energy wavefunction. Due to the presence of the soliton, the perfect alternation in the fermion density stemming from the staggered external field breaks down, forming a domain wall, as we show in the top panels of Fig.~\ref{fig_occupation_pn_barrier} for two different soliton positions. Localized states with half-integer charge appear at these domain walls, hence the accumulated charge 
\beq
\label{eq:accumulated charge}
\Delta Q_n=\sum_{j=1}^n\langle Q_{j,j+1}\rangle =\sum_{j=1}^n\big(\rho_{j,j+1}-\tfrac{1}{2}\big)
\eeq
 grows to $\Delta Q_N\approx 1/2$ as one crosses the domain walls induced by the soliton background, (see the bottom panels of Fig.~\ref{fig_occupation_pn_barrier}). In these same graphs, we show how the accumulated charge density coincides with the integrated fermion density for the localized state. This is important to obtain a quantity accessible through expectation values that reflects the fermionic state density. We remark that in computing $\Delta  Q_{n}$ here and through the rest of this work, we exclude the chain boundaries to avoid effects due to the open boundary conditions.

\subsection{Trapped-ion quantum lattice simulator}\label{sec_experimental_setup}

Once several aspects of the lattice JR mode have been discussed, let us now describe a possible scheme for its experimental realization using trapped-ion Q$\ell$Ss. We consider a system of $N$  atomic ions of charge $q=+e$ and mass $m_a$, confined in a linear Paul trap with secular frequencies $\{\omega_\alpha:\,\,\alpha=x,y,z\}$~\cite{RevModPhys.75.281,doi:10.1063/1.5088164}. The ions form a Coulomb crystal with equilibrium positions resulting from the balance between the trapping forces and the Coulomb repulsion~\cite{morigi2025ioncoulombcrystalsexotic}.
Expanding around the equilibrium positions $\boldsymbol{r}_n=\boldsymbol{R}^0_n+\boldsymbol{u}_n$ with the corresponding momenta $\boldsymbol{p}_n$ 
yields an effective harmonic description for  small displacements
\begin{equation}
\label{eq:quad_ions}
H_{h}=\sum_{n,\alpha}\left(\frac{p_{n,\alpha}^2}{2m_a}+\frac{1}{2}m_a\omega_\alpha^2 u_{n,\alpha}^2+\frac{1}{4}\sum_{l\neq n}k_{nl}^{\alpha\beta}(u^{\phantom{\dagger}}_{n,\alpha}-u^{\phantom{\dagger}}_{l,\beta})^2\right).
\end{equation}
Here, we have introduced a   spring-coupling tensor
\begin{equation}
\label{vs}
k_{nl}^{\alpha\beta}=-\frac{ e^2\big(3(\boldsymbol{R}^0_{n}-\boldsymbol{R}^0_{l})_{\alpha}(\boldsymbol{R}^0_{n}-\boldsymbol{R}^0_{l})_{\beta}-\delta_{\alpha\beta}(\boldsymbol{R}^0_{n}-\boldsymbol{R}^0_{l})^2\big)}{4\pi\epsilon_0|\!|\boldsymbol{R}^0_{n}-\boldsymbol{R}^0_{l}|\!|^5},
\end{equation}
where $\epsilon_0$ is the vacuum permittivity. In the limit $\omega_y\gg\omega_z\gg\omega_x$, the ions self-assemble in a linear chain $\boldsymbol{R}^0_{n}=x_n\mathbf{e}_x$ and, as a result, the effective springs do not couple the vibrations along different  axes 
$k_{nl}^{zz}=k_{nl}^{yy}=-\tfrac{1}{2}k_{nl}^{xx}=-e^2/{4\pi\epsilon_0|x_n-x_l|^3}$. It is by diagonalizing these matrices that one finds the corresponding normal modes arranged in longitudinal and transverse phonon branches~\cite{James1998, Marquet2003}. 
We note that the negative sign of the transverse couplings $k_{nl}^{zz}$ implies that the phonon branch is inverted, and high-wavelength modes describe the low-frequency excitations. In fact, as the ratio $\kappa_x=(\omega_x/\omega_z)^2$ increases, the highest wavelength mode becomes unstable at a critical value $\kappa_{x,\mathrm c}(N)$~\cite{PhysRevLett.70.818, PhysRevLett.71.2753}, and the linear chain transitions to a {zigzag} ladder by spontaneously breaking a $\mathbb{Z}_2$ reflection symmetry $u_{n,z}\to -u_{n,z}$~\cite{PhysRevA.45.6493} as depicted in Fig.~\ref{fig:trapped_ion_scheme}.

This structural phase transition~\cite{PhysRevB.11.3535,PhysRevB.13.4877,PhysRevB.77.064111,PhysRevLett.105.075701,PhysRevX.7.041012}   belongs to the same universality class as that of the above $\lambda\phi^4$ model~\eqref{eq:JR_action} in $D=1+1$ dimensions. In fact, in light of our previous discussion of the standard lattice regularization of the $\lambda\phi^4$ model~\eqref{Hphi4}, one can draw a clear parallelism  by introducing the Fourier transform
$
u_{n,z}=\frac{1}{\sqrt{L}}\!\sum_{k\in\mathrm{BZ}}\!\ee^{\mathrm{i}kan}u_z(k)
$, $
p_{n,z}=\frac{1}{\sqrt{L}}\!\sum_{k\in\mathrm{BZ}}\!\ee^{\mathrm{i}kan}p_z(k),
$
which diagonalises the quadratic Hamiltonian~\eqref{eq:quad_ions}, yielding the phonon dispersion relation
\begin{equation}
\omega(k)=\omega_z\sqrt{1-\kappa_x\!\left(\frac{\ell}{a}\right)^{\!\!\!\!3}\,\sum_{r=1}^{N/2}\!\frac{4}{r^3}\sin^2\!\left(\frac{ka}{2}r\right)},
\label{eq:dispersion_ions_short}
\end{equation}
where we have introduced $\ell^3=e^2/4\pi\epsilon_0m_a\omega_x^2$.
This expression parallels the  dispersion of a discretised Klein-Gordon field~\eqref{eq:dispersion_relation} which, after switching back to SI units~\cite{PhysRevX.7.041012}, reads
\begin{equation}
\omega(k)=\sqrt{\frac{m_0^2c^4}{\hbar^2}+\frac{4c_{\rm  b}^2}{a^2}\sin^2\!\!\left(\frac{ka}{2}\right)}.
\label{eq:lattice_KG_short}
\end{equation}
 A long-wavelength expansion of Eq.~\eqref{eq:dispersion_ions_short} around the zigzag mode $k=\pi/a$ yields the continuum Klein--Gordon form
$\omega^2(k)\approx m_0^2c_{\rm  b}^4/{\hbar^2}+c_{\rm  b}^2 k^2$
where the effective transverse sound velocity plays the role of a relativistic light speed,
\begin{equation}
c_{\rm  b}^2=a^2\omega_x^2\!\left(\tfrac{\ell}{a}\right)^{\!3}\eta_{N}(1),\hspace{2ex} 
\eta_{N}(s)=\sum_{r=1}^{N/2}\frac{1}{r^s}(-1)^{r+1},
\label{eq:ct}
\end{equation}
and the bare mass term is given by
\begin{equation}
\label{eq:m0}
m_0^2=\frac{\hbar^2}{c_{\rm  b}^4}\!\left(\omega_z^2-\tfrac{7}{2}\omega_x^2\!\left(\tfrac{\ell}{a}\right)^{\!3}\zeta_{N}(3)\right),\hspace{2ex} 
\zeta_{N}(s)=\sum_{r=1}^{N/2}\frac{1}{r^s}.
\end{equation}
The instability point of the linear--zigzag transition is thus identified with $m_0^2=0$  as in the classical $\lambda\phi^4$ model~\eqref{sol_vacio}. This classical critical point occurs at the following ratio of the trap frequencies $\kappa_{x,{\rm c}}=2a^3/7\ell^3\zeta_{N}\!(3)$.

To construct the effective field theory, we define coarse-grained fields following a similar gradient expansion as in the case of staggered fermions, where we already had to separate a fast-oscillating part coming from the non-zero momenta from a slowly-varying envelope playing the role of the field~\eqref{eq:long_wav_fermions}. In this case, we define the scalar fields as
\begin{equation}
\phi(x)=\frac{(-1)^n}{\sqrt{m_aa^3}}u^{\phantom{\dagger}}_{n,z},\hspace{2ex}
\pi(x)=(-1)^n\sqrt{m_aa}\,p^{\phantom{\dagger}}_{n,z},
\label{eq:field_map}
\end{equation}
which satisfy $[\phi(t,x),\pi(t,y)]=\mathrm{i}\hbar\delta(x-y)$. 
Including higher orders in the expansion of the Coulomb interaction that led to Eq.~\eqref{eq:quad_ions} generates a stabilizing quartic interaction
\begin{equation}
V_{\rm int}=\frac{1}{2}\!\sum_{n\neq l}\!\frac{\beta_{nl}^{zz}}{4}\big(u^{\phantom{\dagger}}_{n,z}-u^{\phantom{\dagger}}_{l,z}\big)^{\!\!4},\hspace{2ex}
\beta_{nl}^{zz}=\frac{3e^2}{8\pi\epsilon_0}\frac{1}{|x_n-x_l|^5},
\end{equation}
which, after coarse graining~\eqref{eq:field_map}, gives the self-interaction 
\begin{equation}
\lambda=\frac{243\,\zeta_{N}(5)}{4K^4}m_a^3\omega_x^2\ell^3,
\qquad
K=\frac{m_aac_{\rm  b}}{\hbar}.
\label{eq:lambda_final}
\end{equation}

Tuning $\kappa_x$ across the above critical value $\kappa_{x,{\rm c}}$ drives a transition where the effective potential changes from a single to a double well, breaking the $\mathbb{Z}_2$ symmetry $\phi(x)\!\to\!-\phi(x)$. According to our previous discussion around Eq.~\eqref{eq:tadpole}, quantum fluctuations will shift this critical point, defining a critical line $\kappa_{z,\mathrm c}(\lambda_0)$ analogous to that displayed in Fig.~\ref{fig_critical_line}. 
If we are close to this critical line, the characteristic  length scale obtained through  the Compton wavelength  $\lambda_{\rm C}=\hbar/m_0c_{\mathrm t}$ is much larger than the ion-ion distance playing the role of the lattice spacing $a$, and the trapped-ion Q$\ell$S reproduces the long-wavelength dynamics of the $\lambda\phi^4$ QFT
\begin{equation}
H=\!\int\!{\rm d}x\!\left(\frac{c_{\rm  b}^2}{2\hbar^2}\pi^2+\frac{\hbar^2}{2}(\partial_x\phi)^2+\frac{m_0^2c_{\rm  b}^2}{2}\phi^2+\frac{\lambda}{4}\phi^4\right),
\label{eq:lambda_phi4_final}
\end{equation}
with parameters $(c_{\rm  b},m_0^2,\lambda)$ directly tunable through trap frequencies and geometry~\cite{a2,Vinas2023}. One can now set $c_{\rm b}=\hbar=1$ to connect to the previous relativistic QFT in natural units. The trapped-ion system near the zigzag transition can thus be understood as an experimental regularization of a $\lambda\phi^4$ QFT.

Before moving to the discussion about how to include Dirac fermions in the Q$\ell$S, let us mention some caveats of this derivation. First of all, Coulomb crystals are known to have an inhomogeneous ion-ion spacing $a(x)$ that increases towards the crystal boundaries~\cite{PhysRevLett.71.2753}. We note that one can still derive a coarse-grained description of the structural phase transition by allowing for space-dependent speed of sound, bare mass, and quartic coupling~\cite{PhysRevLett.105.075701, PhysRevLett.109.010501}. This causes the soft zigzag mode to destabilize first at the trap center, and then propagate outward as a moving critical front leading to an inhomogeneous onset of the spontaneous SSB order. Moreover, in the zigzag phase, Eq.~\eqref{vs} predicts a coupling of the previously independent axial and transverse phonon branches, which will also contribute to a renormalization of the effective sound velocity.
For parameters in the zigzag phase but close to the critical conditions, the $\lambda\phi^4$ model description still remains accurate, but as one goes deeper and deeper in the zigzag phase,  the single relativistic scalar field theory should be upgraded to a multi-component QFT. 

Finally, we mention the last caveat related to the periodically driven nature of the ion chain. In particular, the time-dependent radio-frequency fields of a Paul trap introduce an additional parametric oscillation of the ions known as micromotion~\cite{RevModPhys.75.281}, which modulates the above Coulomb couplings at multiples of the RF frequency.  While this modulation is already accounted for in the pseudo-potential approximation, leading to several micromotion sidebands in addition to the effective trap frequencies $\omega_\alpha$, it might introduce further effects when the ion positions are not aligned with the null of the radio-frequency fields, as is the case of the zigzag ladder.  A more complete description including this excess micromotion~\cite{10.1063/1.367318} would correspond to a Floquet version of the effective Hamiltonian~\cite{Bermudez_2017}, where the secular trap frequencies and elastic constants acquire periodic corrections. To preserve the validity of the $\lambda\phi^4$ mapping, the driving frequency must remain far off resonant from all these additional harmonics to avoid parametric resonances and heating~\cite{PhysRevA.39.4362}, ensuring that micromotion only leads to a renormalization of the effective mass and speed of sound.

After this discussion, let us now move on to the trapped-ion realization of the Dirac fields. So far, we have not made any explicit use of the atomic structure of the ions. We consider two hyperfine or Zeeman sub-levels in the groundstate manifold of the ion to encode a pair of pseudo-spin states per ion $\ket{\uparrow_n},\ket{\downarrow_n}$, and define the  tensor-product Pauli operators
\beq
\sigma_n^z=\ket{\uparrow_n}\!\bra{\uparrow_n}-\ket{\downarrow_n}\!\bra{\downarrow_n},\hspace{1ex} \sigma^{+}_n=\ket{\uparrow_n}\!\bra{\downarrow_n},\hspace{1ex} \sigma^{-}_n=\ket{\downarrow_n}\!\bra{\uparrow_n},
\eeq
such that $\sigma^{x}_n=\sigma^{+}_n+\sigma^{-}_n$ and  $\sigma^{y}_n=-\ii\sigma^{+}_n+\ii\sigma^{-}_n$. Note that the two spin states can already be identified with particle occupation numbers 
$\ket{0_n},\ket{1_n}$, but an additional Jordan-Wigner transformation~\cite{Jordan1928} must be applied in order to recover the correct fermion statistics, which reads as follows
\beq
\label{eq:JW}
\sigma_n^{+}
= \sqrt{a}\chi_n^{\dagger}\;
\ee^{\ii\pi\left(\tfrac{n}{2}+ a\sum_{\ell=1}^{n-1} \chi_l^{\dagger}\chi_l^{\phantom{\dagger}}\!\right)}=(\sigma_n^{-})^\dagger,\hspace{2ex}
\sigma_n^{z} = 2 a \chi_n^{\dagger}\chi_n^{\phantom{\dagger}} - 1.
\eeq
We note that, when working with  lattice fermion fields that have dimensions $[\chi_n]=\mathsf{L
}^{-1/2}$, the lattice spacing appears explicitly in the Jordan-Wigner transformation. In addition,  we have included a local phase that can be interpreted as a gauge transformation $\chi_n\to\ee^{\ii e\Lambda_n/\hbar}\chi_n$  with $\Lambda_n=\hbar \pi n/2e$, which becomes useful in the mapping to the Dirac field.

These spins can be coupled to the phonons of the ion crystal using additional lasers or microwaves in the near field. We start by considering such couplings in a so-called  M{\o}lmer-S{\o}rensen (MS) scheme~\cite{PhysRevLett.82.1971, PhysRevA.62.022311}, which introduces a state-dependent force that is responsible for inducing phonon-mediated spin-spin interactions in the far off-resonant limit~\cite{RevModPhys.93.025001}. These interactions will be the key to encoding the dynamics of Dirac fermions in the trapped-ion Q$\ell$S. In particular, we now discuss how the spins acquire an XY exchange coupling mediated by virtual phonons along the tightly confined transverse direction $\omega_y\gg\omega_z\gtrsim\omega_x$; then explain how a staggered ac-Stark shift implements a Yukawa coupling; and finish by discussing the coarse graining that connects this lattice model to the JR field theory. Paralleling the discussion of the scalar field, we close with some caveats regarding the continuum limit in the presence of long-range spin-spin couplings and by stating the most relevant corrections that contribute with four-Fermi interactions of the Dirac fermions.

In the following, we set $\hbar=1$. Considering the small transverse displacements along the $y$ axis (see Fig.~\ref{fig:trapped_ion_scheme}), we note that the spring  tensor~\eqref{vs} does not couple the corresponding phonon modes with the vibrational modes along the plane in which the zigzag ladder lies. By diagonalizing the dipolar Hessian matrix $k_{nl}^{yy}=-e^2/{4\pi\epsilon_0|\!|\mathbf{R}^0_n-\mathbf{R}^0_l|\!|^3}$~\cite{James1998, Marquet2003}, assuming once more a homogeneous lattice spacing as a starting point, one finds the transverse  phonon branch
\begin{equation}
\nu\!(k)=\omega_y\sqrt{1-\kappa_y\!\left(\frac{\ell}{a}\right)^{\!\!\!\!3}\,\sum_{r=1}^{N/2}\!\frac{4}{r^3}\sin^2\!\left(\frac{ka}{2}r\right)},
\label{eq:dispersion_ions_y}
\end{equation}
where the corresponding trap frequency ratio now fulfills $\kappa_y=(\omega_x/\omega_y)^2\ll 1$, such that the $y$ phonon branch $H_y=\sum_k\big(\tfrac{1}{2m_a}|p_y(k)|^2+\tfrac{1}{2}m_a\nu^2\!(k)|u_y(k)|^2\big)=\sum_k\nu\!(k)(b^\dagger_k b_k+\tfrac{1}{2})$ has a small spread $\nu\!(k)\in[\omega_y,\omega_y-\Delta\nu]$. We now consider using a pair of laser beams in a Raman configuration with opposite detunings from the carrier transition $\pm\delta_{L}$ to generate a state-dependent force that virtually excites phonons from the whole vibrational branch. In the interaction picture and after a Lamb-Dicke expansion~\cite{RevModPhys.75.281}  keeping  leading-order contributions, i.e., the red- and blue-sidebands, one obtains
\begin{equation}
V_y(t)\approx\sum_{nk} \mathcal{F}_{nk}\Delta y_0\;\sigma_n^{\varphi_s}b_k \ee^{\ii(\varphi_m-\delta(k))\,t}+{\rm H.c.},
\label{eq:HintMS}
\end{equation}
where we have introduced
$
\sigma_n^{\varphi_s}=\ii\ee^{\ii\varphi_{s}}\sigma_n^{+}+\ii\ee^{-\ii\varphi_s}\sigma_n^{-}$ in terms of the sum of the two Raman-beam phases $\varphi_s=(\varphi_r+\varphi_b)/2$, and $
\Delta y_0={1}/{\sqrt{2m_a\omega_y}}$ is a length scale associated to the width of the oscillator groundstate wavefunction. The combination of the two Raman beams exerts a force of amplitude $
\mathcal{F}_{nk}=\frac{\Omega_L}{2}\,\Delta \boldsymbol{k}_{\rm L}\cdot{\bf e}_y\sqrt{\omega_y/\nu(k)}\mathcal{M}^{\rm s}_{nk}$, where $\Omega_L$($\boldsymbol{k}_{\rm L}$) is the Rabi frequency (wavevector) of the two-photon Raman-assisted transition, and   $\mathcal{M}^{\rm s}_{nk}=\ee^{\ii kx_n}/\sqrt{N}$  is the corresponding normal mode in the homogeneous approximation. The lasers exert a linear force that displaces the normal modes in phase space in a direction that depends on the difference of the Raman-beam phases $\varphi_m=(\varphi_r-\varphi_b)/2$ and the spin state, describing circular trajectories depending on the detuning 
$\delta(k)=\nu\!(k)-\delta_{L}$.

Operating deep in the dispersive  regime,
\begin{equation}
\label{eq:dispersive}
|\mathcal{F}_{nk}\Delta y_0|,\,h_0\ll \delta(k)\ll \omega_y,
\hspace{2ex}
\max|J_{nl}|\ll 2h_0,
\end{equation}
we can time-average over the fast phonon dynamics and obtain a purely spin model with exchange couplings
\begin{equation}
J_{nl}=-\sum_{k}\mathcal{F}_{nk}\Delta y_0\,\frac{1}{\delta(k)}\,\mathcal{F}_{mk}^{*}\Delta y_0+{\rm c.c.},
\label{eq:JijMS}
\end{equation}
which have the typical expression of an interaction that is mediated by the free phonon propagator~\cite{a2,Vinas2023}. In fact,
pushing this analogy further allows for a neat understanding of the decay of the spin-spin couplings with distance. The combination of a dimensionally-reduced free-boson propagator with additional dipolar contributions that stem from the long-range nature of the spring-coupling tensor leads to 
\beq
\label{eq:spin_spin_couplings_ions}
J_{nl}=	J_{0}\!\left(\frac{\omega_{x}^4\eta_{N}(1)}{(\omega_{{y}}^2-\delta_{L}^2)^2}\frac{\ell^3}{|{ x}_n-{ x}_l|^3}-(-1)^{n-l}\frac{\tilde{\lambda}_{\rm C}a^2}{\ell^3}\ee^{-\frac{|{ x}_n-{ x}_l|}{\tilde{\lambda}_{\rm C}}}\right)\!,
\eeq
where  $
J_{0}={2 \Omega_{\rm L}^2\eta_{{x}}^2}/\omega_{{x}}\eta_{N}(1),
$ and $\eta_{{x}}=\Delta k_{{\rm L}}/\sqrt{2m_a\omega_x}$ is the Lamb-Dicke parameter. In this expression, we have introduced an effective Compton wavelength 
\beq
\label{eq:lambda}
\tilde{\lambda}_{\rm C}=\frac{1}{m_0^2c_{\rm  b}^4-\delta_{L}^{2}},
\eeq
which depends on the detuning of the Raman beams with respect to the zigzag mode frequency~\eqref{eq:m0}, and controls the faster or slower decay of the spin-spin couplings. For very large detunings of the laser with respect to the whole vibrational branch, the couplings decay with a dipolar tail. On the other hand, as one gets closer, the contribution coming from the free propagator of the free boson field can become the leading term, leading to a slower decay that can be approximately fitted to a power law with a tunable exponent.

In Eq.~\eqref{eq:dispersive}, we have 
also introduced a strong transverse field $h_0$ that can be controlled by an ac-Stark shift. When this term is much stronger than the spin-spin couplings~\cite{Jurcevic2014, Richerme2014},  it inhibits terms in the MS exchange coupling that do not conserve $S_z=\tfrac{1}{N}\sum_n\sigma_n^z$, leading to an effective XY spin model 
\begin{equation}
V_{\mathrm{eff}}=\sum_n\sum_{l>n}J_{nl}^{\phantom{+}}\,\sigma_{n\phantom{l}}^{+}\sigma_l^{-}+{\rm H.c.},
\label{eq:HeffXY}
\end{equation}
which can be obtained by means of a Schrieffer-Wolff transformation on Eq.~\eqref{eq:HintMS} followed by a rotating-wave approximation that rests on $\max|J_{nl}|\ll 2h_0$.
Using the Jordan-Wigner transformation~\eqref{eq:JW}, this conservation law leads to the global U(1) symmetry in an effective  fermionic model 
\beq
\label{eq:fermions_long_range}
V_{\rm eff}=a\sum_n\sum_{l>n}\left(\ii J_{nl}^{\phantom{\dagger}}\chi_n^\dagger \chi_l^{\phantom{\dagger}} +{\rm H.c.}\right)+V_{\rm int}(\{\chi^\dagger_r,\chi^{\phantom{\dagger}}_r\}),
\eeq
where one finds additional non-linearities due to the Jordan-Wigner string
\beq
V_{\rm int}=a\sum_{n}\sum_{r>0}J_{n,n+r}^{\phantom{\dagger}}\chi_n^\dagger \left(\Pi_{r}(1-2a\chi^\dagger_r\chi^{\phantom{\dagger}}_r)-1\right)\chi_{n+r}^{\phantom{\dagger}}+{\rm H.c.}.
\eeq

The long-range terms in Eq.~\eqref{eq:fermions_long_range}   can still be seen as a regularization of the Dirac field that is analogous to the staggered fermion formulation~\eqref{H_XX_fermions}. In fact,   the long-range tail of the tunnelings simply changes the effective speed of light derived from  a similar gradient expansion 
\beq
\label{eq:cs}
c_{\rm f}=2a J_{n,n+1}\to c_{\rm f}=2a\sum_{r>0}
(2r-1)(-1)^{r-1}J_{n,n+(2r-1)}.
\eeq
For large detunings such that the spin-spin coupling is dominated by the dipolar decay, one would arrive at an effective speed of light that can be expressed in terms of the discrete version of Riemann $\eta$ function in analogy to what happened for the scalar field~\eqref{eq:ct}. Using the scale-invariance of this fixed point associated to free Dirac fermions, one can simplify the non-linear interactions, neglecting all terms that are irrelevant in the renormalization group sense~\cite{RevModPhys.66.129}. In this particular case, one would only keep a 4-Fermi term 
$V_{\rm int}\approx -a\sum_{n}2J_{n,n+2}a\chi_n^\dagger \chi^\dagger_{n+1}\chi^{\phantom{\dagger}}_{n+1}\chi_{n+2}^{\phantom{\dagger}}+{\rm H.c.}$. 

The last step to obtain a trapped-ion Q$\ell$S that mimics the JR model is the engineering of a Yukawa coupling to the scalar field~\eqref{eq:H_JR}, which can be obtained through a staggered fermion-boson coupling~\eqref{H_XX_fermions}. To engineer this term, we consider another pair of laser beams that is now in a cross-beam ac-Stark configuration that can be exploited to induce a state-dependent dipole force in the $\sigma^z$ basis. In contrast to the MS scheme~\eqref{eq:HintMS}, this force now couples to the in-plane vibrational modes (see Fig.~\ref{fig:trapped_ion_scheme}), in particular to the coarse-grained zigzag mode that encodes the scalar $\lambda\phi^4$ field of our Q$\ell$S. When the ion crystal transitions to a zigzag configuration, breaking the $\mathbb{Z}_2$ symmetry,  the ions lie off the null of the rf trap potential leading to an additional modulation due to excess micromotion~\cite{Bermudez_2017}. We consider this effect in the derivation of the Yukawa coupling. 

We consider a pair of counter-propagating laser beams with the same frequency but different phases $\phi_1-\phi_2=\pi/2$, which are far detuned from any dipole-allowed transition. These lasers thus generate a dipole  potential that, provided one uses selection rules so that they only couple to the atomic level $\ket{\uparrow_n}$, lead to the following cross-beam term 
\begin{equation}
V_{z}\approx\sum_{n}\tfrac{\tilde{\Omega}_L}{2}\ket{\uparrow_n}\!\bra{\uparrow_n}J_0\big(|\beta_n|\big)(-1)^n\sin\big({\Delta \boldsymbol{\tilde{k}}}_L\cdot{\bf e}_z\sqrt{m_aa^3}\phi(x)\big).
\label{eq:HintLS}
\end{equation}
Here, $\tilde{\Omega}_L$ ($\Delta\boldsymbol{ \tilde{k}}_L$) is the two-photon ac-Stark shift (wavevector) of the off-resonant laser driving, and we have already introduced the effective scalar field in Eq.~\eqref{eq:field_map}. We have also included the leading effects of excess micromotion in a semi-classical parameter $\beta_n=\tfrac{q_z}{2}\boldsymbol{\Delta \tilde{k}}_L\cdot{\bf e}_z\sqrt{m_aa^3}(-1)^n\langle\phi(x)\rangle$ that quantifies how the SSB groundstate or any soliton configuration $\langle\phi(x)\rangle=\pm\Phi_n$ involves equilibrium ion positions that are not aligned with the null of the rf trap along the $x$ axis, such that additional micromotion kicks in. For a microscopic trap parameter $q_z\ll 1$, the leading order effect of this micromotion is a Floquet-type renormalization of the coupling that is controlled by the zero Bessel function of the first kind $J_0(x)$. Fortunately, the alternation of the equilibrium positions in the zigzag chain does not lead to any additional staggering through this renormalization, as $J_0(-x)=J_0(x)$, which has allowed us to write the absolute value of the modulation parameter in Eq.~\eqref{eq:HintLS}. On the contrary, the staggering that appears in this expression is due to the sinusoidal interference pattern since $\sin(-x)=-\sin(x)$. It is also important to note that, within the semi-classical approximation $\phi(x)=\Phi_n+\delta\phi(x)$, the leading order term of this light potential does not affect the staggering either since up to an irrelevant $c$-number that could be reabsorbed in the definition of $h_0$, $\sin(c\Phi_n+c\delta\phi(x))\approx\cos(c\Phi_n)\sin(c\delta\phi(x))\approx \cos(c|\Phi_n|)c\delta\phi(x)$. In this limit, we obtain a linear Yukawa coupling of the fermion mass term to the fluctuations of the scalar field, since we can write 
$V_z=g\sum_n(-1)^n\tfrac{1}{2}(1+\sigma_n^z)\delta\phi(x)$. After a Jordan-Wigner mapping~\eqref{eq:JW} and the gradient expansion yields the desired fermion-boson Yukawa vertex with
\beq
\label{eq:g}
g=\tfrac{\tilde{\Omega}_L}{2}J_0\big(\tfrac{q_z}{2}\Delta\tilde{\boldsymbol{k}}_L\cdot{\bf e}_z\,z_0\big)\cos(\Delta\tilde{\boldsymbol{k}}_L\cdot{\bf e}_z\,z_0),
\eeq
where $z_0$ is the transverse coordinate of the classical zigzag configuration.
 Just as the fermionic part~\eqref{eq:fermions_long_range} contains additional non-linearities due to the long-range couplings, here the full Yukawa term has further non-linear contributions to the vertex which, however, still respect the $\mathbb{Z}_2$
 symmetry $\phi\to-\phi,\psi\to\gamma^5\psi,\overline{\psi}\to-\overline{\psi}\gamma^5$.

Before closing this subsection,  we address new caveats of the trapped-ion Q$\ell$S that appear when including these Yukawa-coupled Dirac fermions. As discussed in this section, the microscopic trapped-ion lattice model that regularises the Yukawa-coupled scalar and Dirac fields by implementing long-range phonon-mediated spin-spin interactions   contains additional terms with respect to the standard lattice discretization with staggered fermions and coupled harmonic oscillators. We have seen that in the zigzag phase, additional scalar field components and micromotion modulations will appear in the bosonic sector. To map the spin chain to the staggered fermions, we have seen that further non-linearities in both the fermion sector and the fermion-boson coupling will also appear. 
All of these additional terms make the  Q$\ell$S   more complex but potentially also more interesting. We leave a careful study of these additional driven non-linearities to a future study, and now focus on the JR lattice model $H_{\rm b}+H_{\rm fb}$ of Eqs.~\eqref{Hphi4} and Eq.~\eqref{H_XX_fermions}, considering the parameter mappings in Eqs.~\eqref{eq:ct},~\eqref{eq:m0},~\eqref{eq:lambda}~\eqref{eq:cs} and~\eqref{eq:g}.

\section{\bf Adiabatic and phase-space methods}
\label{sec:methods}

As already noted in the introduction, all the approximations underlying the above discussion not only neglect quantum fluctuations in the scalar field, but also any possible back-reaction from the Dirac field onto the solitons. The goal of the following sections is to go beyond these crude assumptions by gradually increasing the complexity of the required methods. We note that alternative methods have address some aspects of these quantum corrections in the continuum limit, using for instance an expansion of the effective action in fermion loops~\cite{PhysRevD.31.2605}. The lowest-order term yields the vacuum polarization ~\cite{PhysRevLett.47.986}, which is the only possible correction when the scalar field is treated as a background field. Hence,  higher-loop terms lead to a back-reaction on the soliton provided that the scalar field is dynamical and not a mere fixed background. 

In the following, we introduce alternative tools that will allow us to address back-reaction in real-time dynamics.
We start by discussing an adiabatic approximation to $H_{\rm b }+ H_{\rm fb}$ in which the fermion fields instantaneously adapt to the changes of the slower scalar field. This allows us to understand some of the effects of this back-reaction through modifications of the PN potential~\eqref{eq:PN}. We then move to a more complete picture for real-time dynamics using the truncated Wigner approximation in which quantum fluctuations are effectively accounted for in a semi-classical expansion by averaging over the classical trajectories of the scalar feld with different initial conditions. For each of these trajectories, the fermion field evolves under a fully quadratic problem, and can thus be exactly integrated without making any adiabaticity assumption. 

\subsection{Born-Oppenheimer approximation and fermion-soliton back-reaction}\label{PN_barriers}

\begin{figure}
    \centering
    \includegraphics[width=0.6\linewidth]{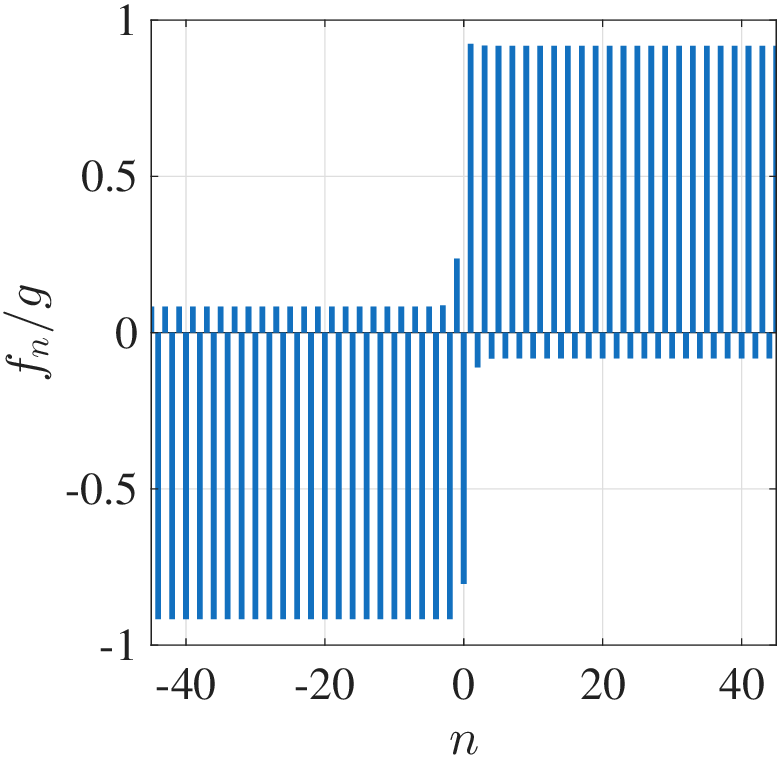}
    \caption{{\bf Fermion back-reaction:} Staggered external force per site $f_n= g  (-1)^n \langle \chi_n^\dagger \chi_n \rangle$. The presence of the kink in the scalar field inverts the direction of the force exerted onto each oscillator after the kink position. Parameters used are $N=160$, $\xi/a=1$, $\sbgs=3$, $Ja=3$, $ga=2$. }
    \label{fig_external_force}
\end{figure}

\begin{figure*}
        \centering    \includegraphics[width=0.85\linewidth]{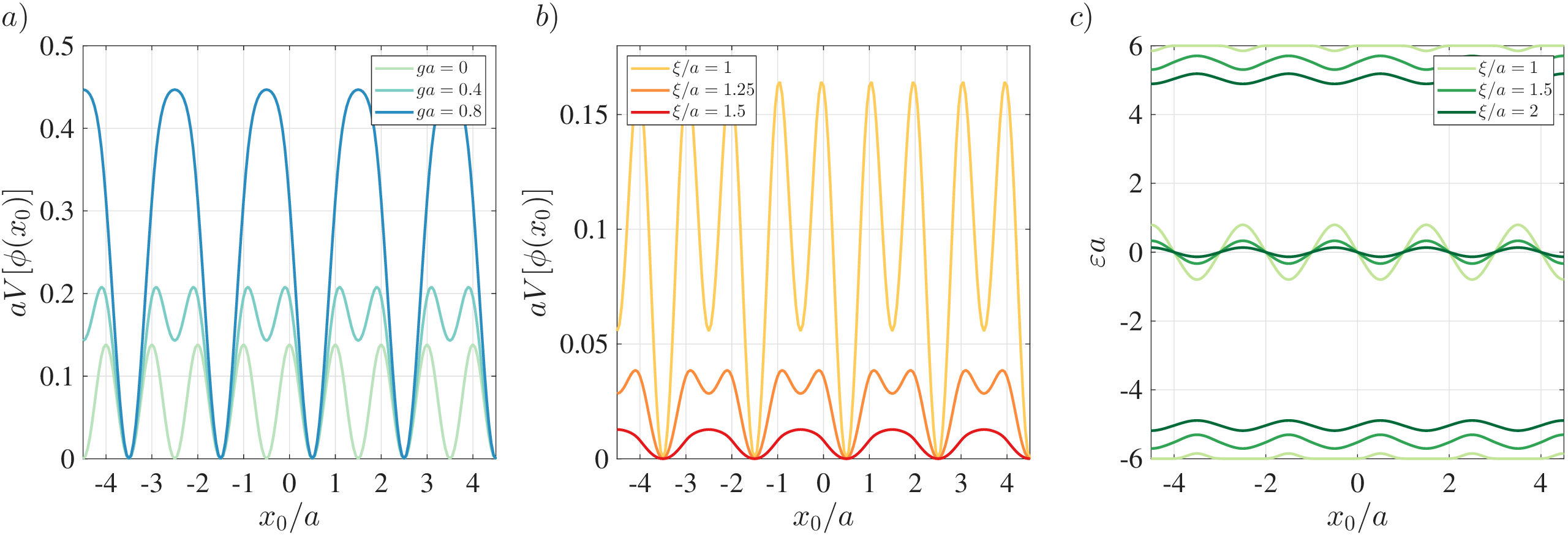}
    \caption{{\bf Peierls-Nabarro potential with back-reaction and fermionic zero-mode energy:} Potential energy as a function of kink position for a) Fixed width $\xi/a=1.00$, b) fixed interaction strength $ga=0.20$. Parameters used are $N=160$, $\sbgs=3$, $Ja=3$.  To show the potential barriers on the same scale, we are subtracting a constant energy value that depends on $g$ and coincides with the soliton mass $M_0$ in Eq.~\eqref{eq:soliton_mass} when $g=0$. c) Periodic behavior of the energy of the zero-mode as well as the consecutive ones as a function of the background scalar field position on the lattice.}
    \label{fig_pn_barriers}
\end{figure*} 

In this subsection, to start analyzing the back-reaction of the fermions on the scalar field, we follow an adiabatic approximation where the fermion adapts rapidly to the changes of the scalar field, keeping itself in the instantaneous ground state as the soliton evolves. The Heisenberg equations of motion for the scalar field [Eq.~\eqref{phi4_discreto_ecs}] must now be supplemented with a  force that depends on the staggered fermion density
\begin{equation}\label{ecs_mov_kink_forzado}
\begin{aligned}
        \frac{{\rm d}{\phi}_n}{{\rm d}t} & = \pi_n, \\
        \frac{{\rm d}{\pi}_n}{{\rm d}t}  & = \frac{\phi_{n+1} -2 \phi_n + \phi_{n-1}}{a^2} - m_0^2 \phi_n 
         - \lambda \phi_n^3 - g  (-1)^n  \chi_n^\dagger \chi^{\phantom{\dagger}}_n. 
\end{aligned}
\end{equation}
For the fermion fields, the Heisenberg equations read
\beq
\begin{split}\label{ecs_fermions}
  \frac{{\rm d}{\chi}_n}{{\rm d}t}  & = {J}\big(\chi_{n+1} - \chi_{n-1}\big) - \ii g (-1)^n \phi_n\chi_n, \\
        \frac{{\rm d}{\chi}^\dagger_n}{{\rm d}t}  & = {J}\big(\chi^\dagger_{n+1} - \chi^\dagger_{n-1}\big) + \ii g (-1)^n \phi^{\phantom{\dagger}}_n\chi^\dagger_n. 
    \end{split}
\eeq
The Born-Oppenheimer approximation considers that the scalar field is much slower than the Dirac fermions, which adapt instantaneously to the changes of the scalar field.  Combining this with the classical approximation for the soliton, $\phi_n,\pi_n\to\Phi_n,\Pi_n$, one  needs to diagonalize the fermionic problem for each instantaneous value of the classical soliton, filling all negative-energy levels together with the zero mode. The instantaneous fermion groundstate is  
\beq
|{\rm g}_+ (t)\rangle= \prod_{\epsilon_\nu\leq 0}\gamma _{\epsilon_\nu}^\dagger(t) \ket{0}=\prod_{\epsilon_\nu\leq 0}\sum_n\mathcal{M}^{\rm f}_{n,\epsilon_\nu} ({\boldsymbol{\Phi}(t)})\chi^\dagger_n\ket{0},
\eeq
 where the matrix  $\mathcal{M}^{\rm f} ({\boldsymbol{\Phi}(t)})$
contains the eigenvectors arranged in columns, which are  obtained by solving
\beq
\label{eq:instantaneous_fermionic_basis}
J(\mathcal{M}^{\rm f}_{n+1,\epsilon_\nu} -\mathcal{M}^{\rm f}_{n-1,\epsilon_\nu}) -\ii g(-1)^n\Phi_n(t)\mathcal{M}^{\rm f}_{n,\epsilon_\nu}=\epsilon_\nu\mathcal{M}^{\rm f}_{n,\epsilon_\nu}.
\eeq
The field equations for the classical soliton are then obtained from Eq.~\eqref{ecs_mov_kink_forzado} by substituting $\phi_n,\pi_n\to\Phi_n,\Pi_n$ with the corresponding Poisson brackets, and projecting the fermion-boson coupling to the above instantaneous fermionic groundstate.

In essence, this Born-Oppenheimer approximation introduces a back-reaction in the form of an inhomogeneous force on the soliton  by modifying the classical potential~\eqref{eq:classical_potential} to $V_{\rm BO}(\Phi)=V_{\rm cl}(\Phi) +\delta V_{\rm BO}(\Phi)$ with a non-linear contribution
\beq
\label{BO_approx}
\begin{split}
\delta V_{\rm BO}(\Phi)&=a\sum_ng(-1)^n\Phi_n  \langle {\rm g}_+(t)|\chi_n^\dagger \chi^{\phantom{\dagger}}_n|{\rm g}_+(t)\rangle\\
&=a\sum_n\sum_{\epsilon_{\nu}\leq 0}\sum_{\epsilon_{\mu}\leq 0}g(-1)^n\Phi_n  (\mathcal{M}^{\rm f}_{n,\epsilon_\mu}({\boldsymbol{\Phi}}))^*\mathcal{M}^{\rm f}_{n,\epsilon_\nu} ({\boldsymbol{\Phi}}).
\end{split}
\eeq
To understand how this force modifies the PN barrier, we now estimate the energy cost of rigidly moving the kink solution throughout the lattice, as in the context of polyacetylene~\cite{Su1979}, but here taking into account the external force term due to the mean-field-decoupled Yukawa coupling with the instantaneous fermionic groundstate. 
In the symmetry-broken groundstate, the local fermion occupation  $a\langle \chi_n^\dagger \chi^{\phantom{\dagger}}_n\rangle$ changes from a uniform value of $1/2$ at $g=0$, to a sequence of empty-occupied sites for the SSB groundstate with $g\gg J$, leading to a staggered force that back reacts on the scalar field. 
In contrast, when the scalar field is in a kink configuration [Eq.~\eqref{kink_discrete}], the perfect alternation of the force breaks down when there are two consecutive sites with similar occupation numbers, which occurs precisely at the kink position (see Fig.~\ref{fig_occupation_pn_barrier}).
As a result, the scalar field on either side of the kink experiences staggered forces with opposite signs, as shown in Fig.~\ref{fig_external_force}.

To understand quantitatively the effect of this staggered adiabatic force, we estimate the Peierls-Nabarro potential by numerically calculating the kink's mass [Eq.~\eqref{eq:PN_potnetial}] as a function of the kink position, which we move through the lattice with its associated localized fermionic mode. 
The kink centered at $x_0$  is obtained by first finding the steady-state classical kink $\Phi_{{\rm K},n}$ using an artificial damping rate as already discussed above. Then, we interpolate the values corresponding to this kink centered at a new position $x_0$. We do so by assuming $\Phi_{{\rm K},n}$ located in $x_n=na+x_0$, and then interpolating the values at sites $n$. We assume that for each kink position $x_0$, the external force acting on each site is given by the ground state expectation value of the fermion number operator, obtained through Eq.~\eqref{fermionic_ground_state}.
As shown in Fig.~\ref{fig_pn_barriers}a), the PN potential for $g\neq 0$ with back-reaction retains the property of the uncoupled case: the lowest energy configuration is that of a kink centered at a lattice link, which is consistent with the linear stability analysis of Fig.~\ref{lowest_lying_frequencies}. However, the back-reaction from the instantaneous fermionic groundstate with the populated zero mode leads to a qualitative change: the PN potential shows a doubled period, which can be directly traced back to the effective staggered force of Fig.~\ref{fig_external_force} that yields an enlarged two-site unit cell. 
Below a critical value for $g$, there are two distinct energy barriers every two lattice sites. One barrier increases with $g$, effectively confining the soliton, while the other decreases, and allows the kinks with center at even lattice links to easily escape. 
Above a critical value of $g$, the small energy barrier vanishes completely, turning these kink solutions into unstable ones. In this regime depicted by the blue line of Fig.~\ref{fig_pn_barriers}a),  we obtain a PN potential similar to the uncoupled case, but clearly displaying twice the lattice period since the even lattice kinks can no longer host a stable kink.

\begin{figure}
    \centering
    \includegraphics[width=1\linewidth]{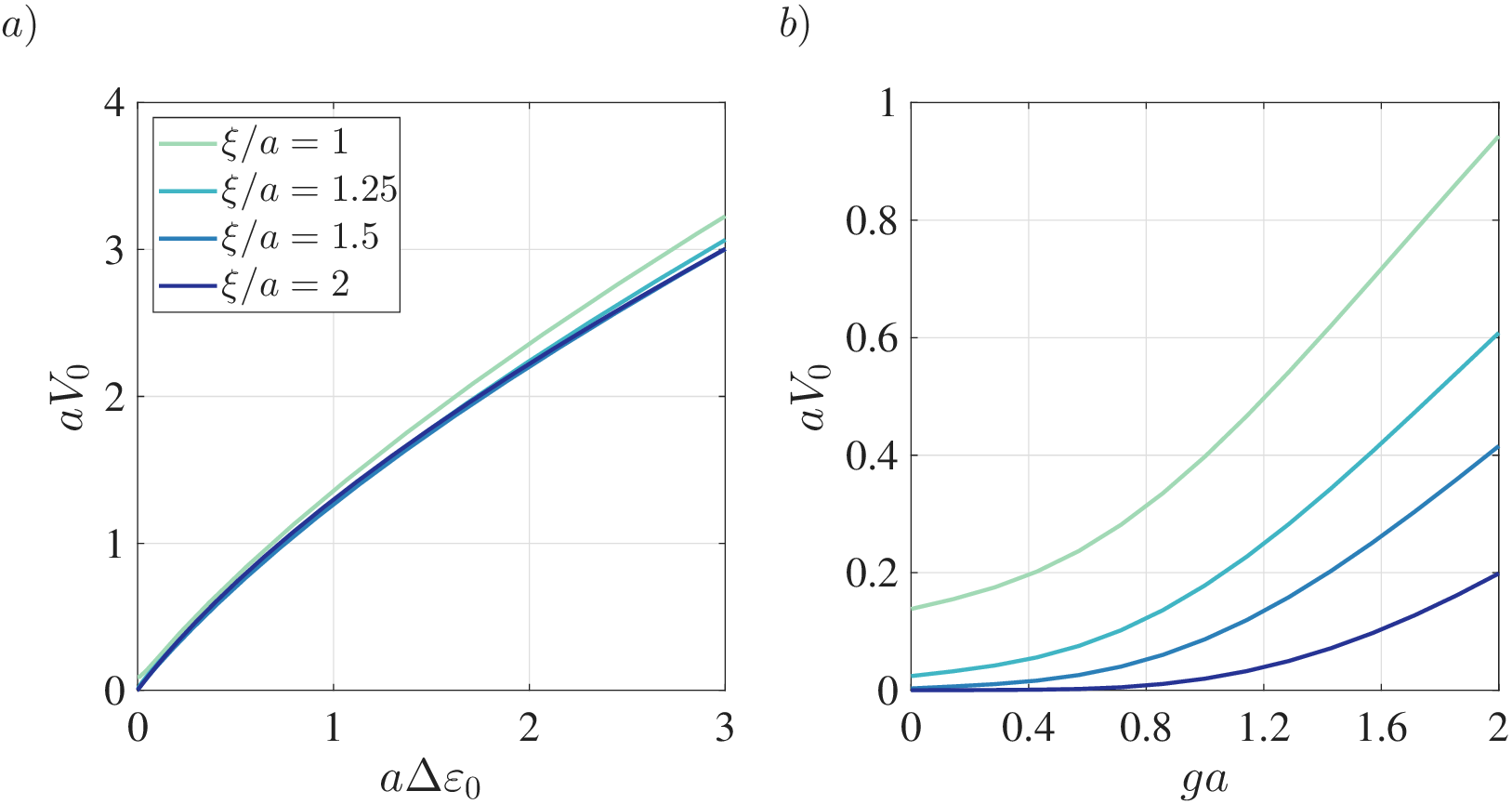}
    \caption{{\bf Peierls-Nabarro barrier behaviour:}  Energy of the PN barrier as a function of a) the energy of the fermionic zero-mode, and b) interaction strength $g$.
    Parameters used are $N=160$, $\sbgs=3$, $Ja=3$, $ga=2$. }
\label{fig_EPN_vs_energy_zero_mode}
\end{figure}

To understand the origin of the changes to the PN potential, we examine the energy associated with the occupation of the zero-mode, where we recall that different charge distributions exist depending on where the center of the soliton resides. We find that the PN barrier is reduced (increased) when the two sites at the kink position are half (fully) occupied, a situation corresponding to the left (right) panel of Fig.~\ref{fig_occupation_pn_barrier}. 
The energy of the fermionic zero-mode is calculated by diagonalizing the  Hamiltonian in Eq.~\eqref{eq:matrix_fermions} for each kink position $x_0$.
We find that this energy is precisely zero for the unstable kink configurations centered at a lattice site. In contrast, we find that the kinks centered at the links to the left and right of a given lattice site are not degenerate, and the energy of the zero-mode oscillates with the kink position consistently with the PN potential barrier, as we show in Fig. \ref{fig_pn_barriers}-c). 
In fact, the largest contribution to the back-reaction on the PN potential stems from the occupation of this zero mode. As we show in Fig.~\ref{fig_EPN_vs_energy_zero_mode}-a), the height of the PN barriers $V_0$, estimated as the energy difference between the highest- and lowest kink energies,  and the energy of the fermionic zero-mode are correlated and display the same order of magnitude for $ga=1$, while scaling almost independently of $\xi_0$.

In Fig.~\ref{fig_pn_barriers}-b), we observe that increasing the soliton width $\xi_0$ reduces the overall height of the PN barriers. This is expected since increasing $\xi_0/a$ is one of the conditions to approach the continuum limit, where we should recover translational invariance and no PN barriers. 
This is seen more clearly in Fig. \ref{fig_EPN_vs_energy_zero_mode}-b), where we show the  PN barrier calculated as the larger energy difference between the unstable on-site kink and the inter-site kink (centered at $x_0=0.5a$ for this specific set of parameters).
For fixed width, we observe in Fig.~\ref{fig_EPN_vs_energy_zero_mode}-b) that this PN barrier increases with $g$.
As expected, for vanishing interaction strength, the PN barrier goes to zero for large $\xi_0$, where the lowest-lying eigenmode becomes frictionless.
{We remark that, although  the energy of the center of mass mode approaches zero in the limit $\xi_0\gg a$ (as in Fig.~\ref{lowest_lying_frequencies}-a)) indicating the onset of translational invariance, 
the  PN barrier obtained for strong interaction strengths $g\sim J$ is still non-zero for large values of $\xi_0$, such that the soliton will not move as a free particle in this regime.}

\subsection{Truncated Wigner approximation and  fermionic Gaussian state ensembles}

The previous Born-Oppenheimer results require the fermions to evolve much faster than the bosons, and the latter to be accurately described at the classical level. In this section, we go beyond these two limitations, allowing for both non-adiabatic effects and, importantly, also including quantum fluctuations of the scalar field using a semiclassical method based on the truncated Wigner approximation (TWA), introducing back $\hbar$ in our equations. The TWA allows to calculate the dynamics of various expectation values through averages
over random classical trajectories in phase space \cite{Berges2007, Zhu2019}. We note that the TWA has been recently used to study real-time dynamics of purely bosonic field theories on the lattice, either of the $\lambda\phi^4$ model \cite{Szasz2022} or the Sine-Gordon model, whose discrete version is sometimes referred to as the Frenkel-Kontorova model~\cite{Chelpanova2023, Chelpanova2024, Calliari2025}. 
Here, one common approach consists of considering the set of discrete DoF as a collective spin described in a continuous variable phase space, approaching a semiclassical limit for large $N$ \cite{Altland2009, Huber2021}. 
There is, however, a different approach in which each single discrete level system is considered under a semiclassical approximation. 
The latter has been used in quantum optics and condensed matter systems \cite{Schachenmayer2015, Davidson2017}.
The TWA has also been used to explore problems of continuous variables coupled to discrete variable degrees of freedom (DoF), as in electron-phonon systems~\cite{Paprotzki2024} (with a continuous variable phase space) or spin-boson models~\cite{Bond2024} (using a discrete variable phase space). 

We now move to the discussion of our method, which integrates the TWA with fermionic Gaussian states \cite{Surace2022}, and draw some further connections to these approaches.
The focus of the TWA is on introducing a controlled semiclassical expansion of the dynamics using phase-space representations~\cite{Carmichael1999, Polkovnikov2003}.
In this framework, the phase-space coordinates $\boldsymbol{\Phi}=(\Phi_1,\cdots,\Phi_N),\boldsymbol{\Pi}=(\Pi_1,\cdots,\Pi_N)$ are used to define the field and momentum eigenbasis $\ket{\boldsymbol{\Phi}}=\otimes_n\ket{\Phi_n},\ket{\boldsymbol{\Pi}}=\otimes_n\ket{\Pi_n}$,   such that $\braket{\boldsymbol{\Phi}|{\boldsymbol{\Pi}}}=\ee^{\ii\boldsymbol{\Phi}\cdot\boldsymbol{\Pi}/\hbar}/(2\pi)^{N/2}$. The many-body state $\rho$ is described in phase space by its Wigner distribution 
\beq
\label{eq:wigner}
W(\boldsymbol{\Phi},\boldsymbol{\Pi})=\frac{1}{(2\pi)^N}\int\! {\rm d}^N{\zeta} \langle \boldsymbol{\Phi}-\boldsymbol{\zeta}/2 | \rho |\boldsymbol{\Phi}+\boldsymbol{\zeta}/2 \rangle \ee^{\ii \tfrac{\boldsymbol{\Pi}\cdot \boldsymbol{\zeta}}{\hbar}},
\eeq
which is a quasi-probability distribution normalised to unity, and can attain negative values that signal the quantum-mechanical nature of the system~\cite{Kenfack_2004}.

 Expectation values of observables expressed in terms of the lattice fields $O(\{\phi_m,\pi_n\})$ can be obtained through their Weyl symbols $O_W$~\cite{Polkovnikov2010}, which result from a mapping  to  ordinary functions defined in 
phase space. For operators $O(\{\phi_m,\pi_n\})=O_1(\{\phi_m\})+O_2(\{\pi_n\})$, this phase space representation of operators is obtained by simply replacing $\phi_n,\pi_n\mapsto\Phi_n,\Pi_n$. The  evolution of the observable follows
\begin{equation}
\label{eq:exp_values_W}
    \langle O(t) \rangle= \int\!\!{\rm d}^N{\Phi}\!\!\int\!\! {\rm d}^N{\Pi}\,\,\,O_W(\boldsymbol{\Phi},\boldsymbol{\Pi})\,\,W(\boldsymbol{\Phi},\boldsymbol{\Pi},t),
\end{equation}
where the Wigner-Moyal equation~\cite{Hillery1984} dictates the evolution of the quasi-probability distribution in phase space 
\beq
\partial_t W(\boldsymbol{\Phi},\boldsymbol{\Pi},t) =\{H(\boldsymbol{\Phi},\boldsymbol{\Pi}),W(\boldsymbol{\Phi},\boldsymbol{\Pi},t)\}_{\rm MB}
\eeq
with the so-called  Moyal bracket
\begin{equation}\label{moyal_bracket}
    \{H,W\}_{\rm MB}= \frac{2}{\hbar}H_W(\boldsymbol{\Phi},\boldsymbol{\Pi}) \sin\left(\frac{\hbar}{2}\Lambda\right) W(\boldsymbol{\Phi},\boldsymbol{\Pi}).
\end{equation}
Here,  we have introduced  the differential Poisson operator 
\beq
\Lambda=\overleftarrow{\boldsymbol{\nabla}}_{\Phi}\cdot\overrightarrow{\boldsymbol{\nabla}}_{\Pi}-\overleftarrow{\boldsymbol{\nabla}}_{\Pi}\cdot\overrightarrow{\boldsymbol{\nabla}}_{\Phi},
\eeq
where ${\boldsymbol{\nabla}}_{\boldsymbol{v}}=(\partial_{v_1},\partial_{v_2},\cdots,\partial_{v_N})$ is taken with respect to the field and momentum variables in phase space, and the arrows indicate if they act on phase-space functions to the left (i.e. Hamiltonian) or to the right (i.e. Wigner distribution).
The equation of motion~\eqref{moyal_bracket} is usually called Moyal equation in analogy with classical probability theory \cite{Risken1996}.

The TWA is obtained by expanding the right-hand side of Eq.~\eqref{moyal_bracket} in powers of  $\hbar$.
At lowest order, $\sin(\hbar\Lambda/2)\approx \hbar\Lambda/2$ and the Moyal bracket becomes a Poisson bracket, mapping the Moyal-Wigner equation to a classical Liouville equation  $\partial_tW-\{H,W\}_{\rm PB}=0$. 
 Liouville's theorem implies that the phase-space volume element is conserved under a Hamiltonian incompressible flow $W(\boldsymbol{\Phi},\boldsymbol{\Pi},t){\rm d}^N{\Phi}{\rm d}^N{\Pi}=W(\boldsymbol{\Phi}_0,\boldsymbol{\Pi}_0){\rm d}^N{\Phi}_0{\rm d}^N{\Pi}_0$, where the time-reversed phase-space trajectories connect to the initial conditions $\boldsymbol{\Phi}(-t)=\boldsymbol{\Phi}_0,\boldsymbol{\Pi}(-t)=\boldsymbol{\Pi}_0$ \cite{Arnold1989}. Therefore, the observable dynamics within the TWA follows 
\begin{equation}
\label{eq:TWA_observables}
    \langle  O(t) \rangle= \int\!\!{\rm d}^N{\Phi}_0\!\!\int\!\! {\rm d}^N{\Pi}_0\,\,\, W(\boldsymbol{\Phi}_0,\boldsymbol{\Pi}_0) O_W(\boldsymbol{\Phi}(t),\boldsymbol{\Pi}(t)).
\end{equation}

\noindent Thus, the effect of quantum fluctuations to leading order in $\hbar$ amounts to a stochastic sampling of initial conditions $\boldsymbol{\Phi}_0,\boldsymbol{\Pi}_0$ from the initial state Wigner distribution $W(\boldsymbol{\Phi}_0,\boldsymbol{\Pi}_0)$, followed by the subsequent deterministic propagation of trajectories $\boldsymbol{\Phi}(t),\boldsymbol{\Pi}(t)$ following the classical equations of motion.

We remark that TWA is exact for quadratic systems since, in this case, there would be no terms beyond the second-order derivatives involved in the Moyal bracket, such that phase-space trajectories evolve independently without interference. The TWA is also very accurate for generic systems at short evolution times \cite{Polkovnikov2003}. Let us note that, for the trapped-ion implementation, we have introduced an effective rigidity parameter $K$~\eqref{eq:lambda_final} that, at first sight, might appear as irrelevant as it can be rescaled through a canonical transformation~\cite{a2}. However, the inverse square of this parameter can also be seen as an effective Planck constant $\hbar_{\rm eff}$, such that a loop expansion of the partition function can be organised in inverse powers of $K$, and analysed from the perspective of the renormalization group~\cite{PhysRevB.89.214408}. This leads to a renormalization-group flow from a classical mean-field scaling towards a quantum-mechanical Ising one~\cite{PhysRevB.89.094103,PhysRevLett.116.225701}. From the perspective of the TWA,  one thus expects the leading-order semiclassical approximation~\eqref{eq:TWA_observables} to be particularly accurate when $K\gg1$ and $\hbar_{\rm eff}\ll 1$, even if we are in a regime of SSB and there are solitonic excitations in the field.
Before discussing how the Yukawa-coupled Dirac field can be incorporated in this TWA framework, let us describe how to define the initial Wigner distribution, which encodes the quantum fluctuations that go beyond the classical approximation of the previous sections.

\subsubsection{Initial Wigner distribution  of the  $\lambda\phi^4$ field}

We consider that the initial state of the scalar field is the vacuum of the Hamiltonian field theory resulting from a second-order expansion around the classical kink solution, leading to Eq.~\eqref{ec_dispersion_kink_lattice}. 
Explicitly, initial field configurations are described by 
\begin{equation}
\begin{aligned}
    \boldsymbol{\Phi}  = \boldsymbol{\Phi_{K}} +\boldsymbol{\Phi}_0, \hspace{2ex}
    \boldsymbol{\Pi} & =\boldsymbol{\Pi}_0
\end{aligned}
\end{equation}
\noindent where again $\boldsymbol{\Phi}_0$ and $\boldsymbol{\Pi}_0$ describe the phase-space representation of the fluctuation conjugate operators $\delta\boldsymbol{\phi}$, $\boldsymbol{\pi}$ around the steady-state classical values found for the initial soliton state. To account for initial state fluctuations, we fix the initial Wigner distribution by writing the Hamiltonian after a second-order expansion 
\begin{equation}
\label{eq:quadratic_expansion_soliton}
H_{\rm b}\approx \frac{a}{2} \boldsymbol{\pi}^2   + \frac{1}{2a}  {\delta \boldsymbol\phi}^{\rm t} \, K(\boldsymbol{\Phi}_{{\rm K}}) {\delta \boldsymbol{\phi}}
\end{equation}
where $\boldsymbol{\pi}=(\pi_1,\cdots,\pi_N)^{\rm t}, \delta\boldsymbol{\phi}=(\delta \phi_1,\cdots,\delta\phi_N)^{\rm t}$, and
\beq
\label{eq:el_matrix}
K (\boldsymbol{\Phi}_{{\rm K}})=\left( 2 + m_0^2 a^2  + 3 a^2 \lambda \Phi_{{\rm K},n}^2 \right)  \delta_{n,n} + \delta_{n,n+1}  + \delta_{n,n-1} 
\eeq
is the resulting elasticity matrix encoding the potential energy around the kink solution.
As a result of the kink profile, this linearization explicitly breaks translational invariance, such that quasi-momentum is no longer a good quantum number. One can, nonetheless, find the ground state by an orthogonal change of the basis to that of the normal modes $ \sum_{nl}\mathcal{M}^{\rm b}_{n\nu}[K (\boldsymbol{\Phi}_{\rm K})]_{nl}\mathcal{M}^{\rm b}_{l\tau}=\Omega^2_\nu\delta_{\nu,\tau}$, 
such the quadratic Hamiltonian decouples into independent harmonic oscillators 
$H_{\rm b}=\sum_\nu \frac{\Omega_\nu}{2} ({P}_\nu^2 + {Q}_\nu^2)$. 
The Wigner function of the initial  state is thus a  product of Gaussian distributions in the normal mode coordinates ${Q}_\nu=\sum_n\mathcal{M}^{\rm b}_{n\nu}{\Phi}_{0,n}, \;  {P}_{\nu}=a\sum_n\mathcal{M}^{\rm b}_{n\nu}{\Pi}_{0,n}  $, namely
\begin{equation}\label{harmonic_chain_ground_state}
W(\boldsymbol{\Phi}_0,\boldsymbol{\Pi}_0) \propto \prod_{\nu=0}^{N-1} \exp\left( -\frac{Q_\nu^2(\boldsymbol{\Phi}_0)}{2\sigma_{Q_\nu}^2}  - \frac{P_{\nu}^2(\boldsymbol{\Pi}_0)}{2\sigma_{P_\nu}^2} \right),
\end{equation}
where $\sigma_{Q_\nu}^2=1/2\Omega_\nu$ and $\sigma_{P_\nu}^2=\Omega_\nu/2$. Accordingly, one can compute real-space field configurations by sampling the normal mode variables from this initial Wigner distribution, 
obtaining an initial state with spatial correlations. 
An example histogram of $N_0=4000$ samples from this distribution leads to  the results displayed in Fig.~\ref{fig_TWA_initial}, which shows how the quantum fluctuations lead to a certain spread around the classical solitonic profile.

Once initial configurations $\{\boldsymbol{\Phi}_0,\boldsymbol{\Pi}_0\}$ compatible with Eq.~\eqref{harmonic_chain_ground_state} are obtained, they must be evolved deterministically using the classical equations of motion,  generating in this way the aforementioned incompressible flows $\{\boldsymbol{\Phi}(t),\boldsymbol{\Pi}(t)\}$.
We remark that the discrete realization of the Goldstone mode ($\nu=0$) has a finite energy that tends to zero as the system approaches the continuum limit (e.g., for large $\xi_0$). This restricts the range of values that can be considered, as the Goldstone mode becomes largely populated for large $\xi_0$ (e.g., we need to take the SSB value $\Phi_0$ large when $\xi_0$ is large, to keep fluctuations of the scalar field comparatively smaller than the SSB ground state $\langle\delta\phi_n^2\rangle < \Phi_0^2$.

\subsubsection{Trajectory-wise fermionic Gaussian states}

After having discussed the  TWA for the scalar field, we switch the Yukawa coupling back on, $g\neq 0$, and consider the entire fermion-boson system. At a formal level, for each of the classical trajectories of the scalar field $\{\boldsymbol{\Phi}_0,\boldsymbol{\Phi}(t)\}$   sampled from the initial Wigner distribution~\eqref{harmonic_chain_ground_state}, the time evolution of the fermion field falls within the family $\{\ket{\Psi(\Gamma)}\}$ of fermionic Gaussian states (fGSs)~\cite{10.5555/2011637.2011640,Kraus_2010,Surace2022}. By definition, these states are fully determined by the two-point functions, as higher-order correlations can be obtained by using Wick's theorem. For the present case, it suffices to  consider   the following correlation matrix
\beq
\label{eq:corr_matrix_equations}
\Gamma^{\phantom{\dagger}}_{nl}=\langle\chi^{\dagger}_{n\phantom{l}\!\!}\chi^{\phantom{\dagger}}_l\rangle,\hspace{2ex}\frac{{\rm d}\Gamma}{{\rm d}t}=\Big\{\Gamma,h\big(\boldsymbol{\Phi}(t)\big)\Big\}.
\eeq
We note that the equations of motion for the correlation matrix are written in terms of the Hamiltonian matrix in Eq.~\eqref{eq:matrix_fermions}, but exchanging the static kink with the time-evolving TWA trajectory $\boldsymbol{\Phi}_{\rm K}\mapsto\boldsymbol{\Phi}(t)$ such that $h(\boldsymbol{\Phi}_{\rm K})\mapsto h\big(\boldsymbol{\Phi}(t)\big)$. One sees that the time evolution induced by a fermionic quadratic Hamiltonian preserves the Gaussian character of states \cite{Surace2022}. We also remark that at this leading TWA order, there is no further approximation in the dynamics,  as the fermionic system is purely quadratic and feels each of the TWA trajectories as a background field. The non-trivial effects come from the averaging over trajectories, on the one hand, and the back-reaction of the fermion fields on the scalar field, which occurs since the TWA trajectories are now subject to  
a staggered external force that depends on the fermion occupation number. The differential equations for these trajectories are obtained from Eq.~\eqref{ecs_mov_kink_forzado}, projecting on phase space and considering the fermionic correlation matrix
\begin{equation}\label{ecs_mov_kink_forzado_twa}
\begin{aligned}
        \frac{{\rm d}{\Phi}_n}{{\rm d}t} & = {\Pi}_n(t), \\
        \frac{{\rm d}{\Pi}_n}{{\rm d}t}  & = \frac{\Phi_{n+1} -2 \Phi_n + \Phi_{n-1}}{a^2} - m_0^2 \Phi_n 
         - \lambda \Phi_n^3 - g  (-1)^n  \Gamma_{nn}(t). 
\end{aligned}
\end{equation}
The set of differential equations~\eqref{eq:corr_matrix_equations}-\eqref{ecs_mov_kink_forzado_twa}, together with the stochastic sampling over the initial Wigner distribution~\eqref{harmonic_chain_ground_state} according to Eq.~\eqref{eq:TWA_observables} allows us to calculate the real-time dynamics of both scalar and fermion fields going beyond all the approximations of the previous sections. The whole dynamics is thus encoded in a phase-space ensemble that combines the initial Wigner distribution, the trajectories of the scalar field, and the trajectory-dependent fermionic Gaussian states $\{W(\boldsymbol{\Phi}_0,\boldsymbol{\Pi}_0),\boldsymbol{\Phi}(\Gamma(t)),\boldsymbol{\Pi}(\Gamma(t)),\ket{\psi(\Gamma(\boldsymbol{\Phi}(t),\boldsymbol{\Pi}(t))}\}$.  

This TWA-fGS approach allows us to incorporate the leading effect of quantum fluctuations, and go beyond the assumption underlying the adiabatic Born-Oppenheimer approximation and the effects of back-reaction~\eqref{BO_approx} on the PN potential. In the TWA-fGS framework, there is no need to assume that the Dirac field is much faster than the scalar field, and thus adapts instantaneously to its changes. On the other hand, a similar type mean-field decoupling between the two species is in-built in this leading-order phase-space approach, as one sees that the TWA trajectories of the scalar field are only affected by the expectation value of the fermionic correlation matrix~\eqref{ecs_mov_kink_forzado_twa}, but there is no build-up of correlations between the two species. 


In practice, we break the time evolution is small time steps $\Delta t$. After solving the scalar-field evolution during a first time step, we exactly integrate the time evolution of the fGSs to update the correlation matrix. Making use of the eigenstates of the instantaneous quadratic Hamiltonian~\eqref{eq:instantaneous_fermionic_basis}, we update the fermionic correlation matrix according to
\begin{equation}\label{correlation_matrix}
    \Gamma_{nl}(t+\Delta t) = \sum_{\nu \le N} \big(\mathcal{M}^{\rm f}_{n\epsilon_\nu}(\boldsymbol{\Phi}(t+\Delta t))\big)^* \mathcal{M}^{\rm f}_{l\epsilon_\nu}(\boldsymbol{\Phi}(t+\Delta t)).
\end{equation}
After such an update, we integrate the non-linear differential equations for the scalar field for the next time step, and proceed in an iterative fashion.
Afterwards, trajectories for both the scalar field and elements of the correlation matrix are averaged to obtain their expectation values \cite{Kasper2014, Davidson2017, Paprotzki2024}.

We note that using the Poisson bracket for the semiclassical description of fermionic modes,  as advocated in \cite{Davidson2017} for purely fermionic models,  would yield the same equations of motion as Eq.~\eqref{eq:corr_matrix_equations}, since we are considering a quadratic Hamiltonian that preserves fermion number.
A difference with the approach in \cite{Davidson2017, Iwanek2022} is that, in our case, there are no stochastic fluctuations of the initial fermionic state, as these ground state fluctuations are entirely described within the correlation matrix for a fermionic Gaussian state. 
We note, however, that the fermions inherit a stochastic behavior from their interaction with the scalar field fluctuations at $t=0$. 
Moreover, it  has been shown that neglecting fluctuations in the initial state leads to better long-time predictions within the fermionic TWA when studying disordered interacting systems \cite{Iwanek2024}.


\begin{figure*}
    \centering
    \includegraphics[width=1\linewidth]{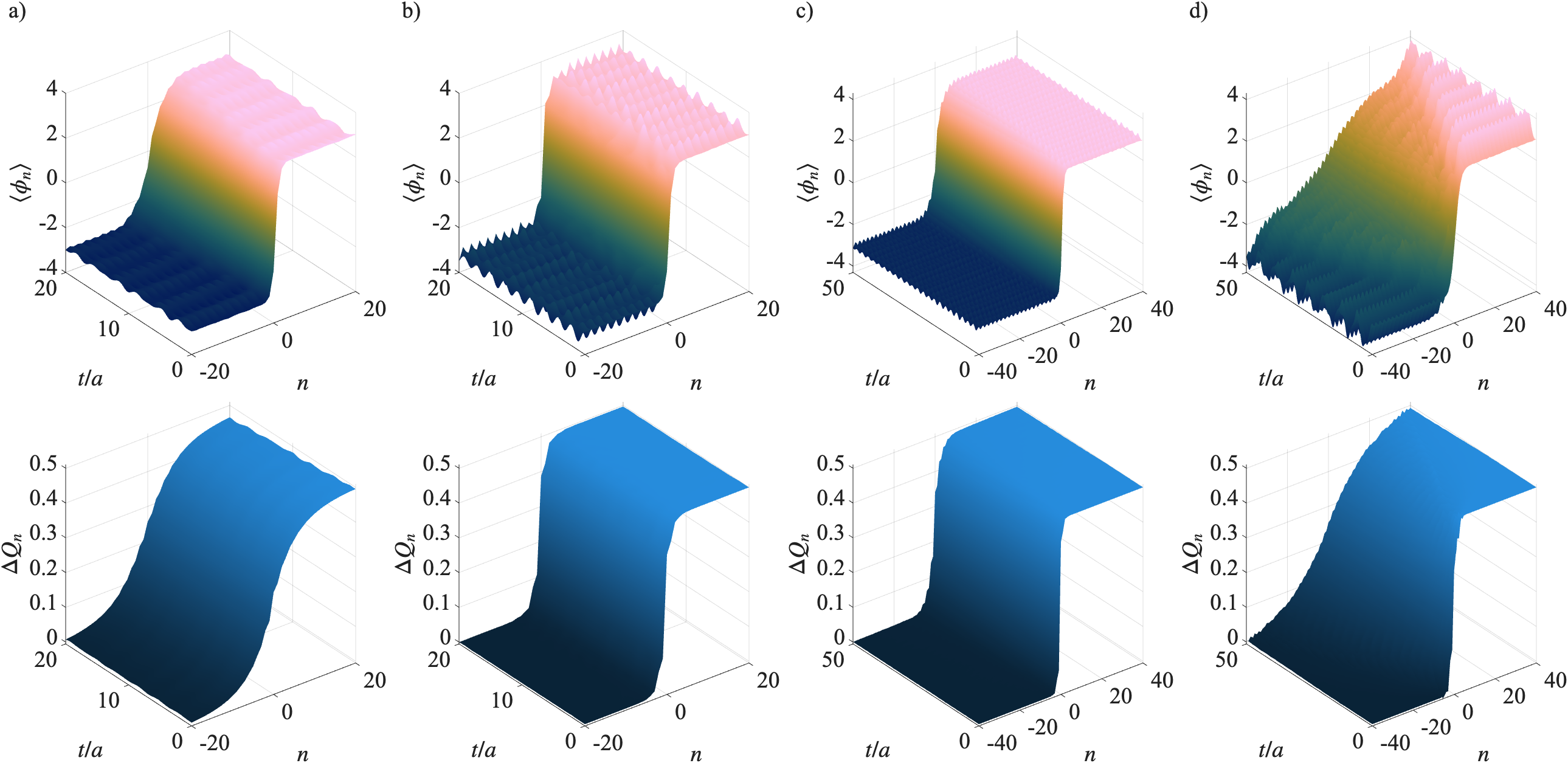}
    \caption{{\bf Single soliton  and fractional charge dynamics:} Upper panels: Expectation value of the scalar field. Bottom panels: Expectation value of integrated charge.
    Parameters used are $N=160$, $\sbgs=3$, $dt/a=0.01$, and $N_t=2000$ trajectories.  
    a) $\xi/a=1$, $Ja=10$, $ga=2/3$, 
    b) $\xi/a=1$, $Ja=10$, $ga=10/3$. 
    c) $\xi/a=1.0$, $Ja=3$, $ga=1.5$, 
    d) $\xi/a=2.5$, $Ja=3$, $ga=1.5$.}
    \label{fig_TWA_comparison}
\end{figure*} 

\section{\bf Real-time dynamics of a fractionally-charged fermion bound to a topological soliton}
\label{sec:single_soliton}

In this section, we consider an initial state expressed as the tensor product of both ground states of the linearized $\lambda\phi^4$ model around a single kink profile, and of the fermion Hamiltonian with the corresponding Yukawa coupling acting as an external background field $| \Psi_0 \rangle= |\rm g_{\rm b}\rangle \otimes |{\rm g}_{\rm f}\rangle$. On one hand, $|\rm g_{\rm b}\rangle$ is described by its Wigner distribution~\eqref{eq:wigner}, while $|\rm g_{\rm f}\rangle$ is described by its correlation matrix~\eqref{eq:corr_matrix_equations}. The initial Wigner function is the multivariate Gaussian distribution~\eqref{harmonic_chain_ground_state}, while the initial correlation matrix can be obtained by diagonalizing the quadratic fermionic Hamiltonian. In this way, we obtain the ground state at half filling for the fermions as $|{\rm g_{\rm f}}\rangle =\prod_{\epsilon_\nu\leq 0}\sum_n\mathcal{M}^{\rm f}_{n,\epsilon_\nu} ({\boldsymbol{\Phi}_0})\chi^\dagger_n\ket{0}=\ket{{\rm g}_+(t_0)}$, where we sum over the negative-energy eigenstates, including the zero mode. 
For  the initial correlation matrix, we have
\begin{equation}\label{correlation_matrix}
    \Gamma_{nl}(t_0)=\sum_{\nu \le N/2}\! \big(\mathcal{M}^{\rm f}_{n,\epsilon_\nu}({\boldsymbol{\Phi}_0})\big)^{\!*} \mathcal{M}^{\rm f}_{l,\epsilon_\nu}({\boldsymbol{\Phi}_0}).
\end{equation}
Considering this initial state, we now calculate the dynamics using our TWA-fGS framework. We consider two different situations.

\subsection{Fractional charge diffusion and localisation}

Generally, we find that for every individual TWA  trajectory of the scalar field, the kink randomly moves towards left or right until it gets trapped at a particular lattice link,  corresponding to one of the minima of the PN potential (c.f. Fig.\ref{fig_pn_barriers}). In this case, the statistical averaging of many trajectories will produce expectation values corresponding to a kink that gets broadened as time evolves, as we show in Fig.~\ref{fig_TWA_comparison}. Thus, the semiclassical approximation shows an increase in the soliton's width as a result of the leading quantum fluctuations, reminiscent of a Sine-Gordon soliton under the action of thermal noise \cite{Quintero1999, Quintero2000}.
For a finite chain, this broadened soliton profile persists as long as a trajectory does not reach the end of the chain, thus leading to bounced solutions that change the overall profile. 
We note that in Fig.~\ref{fig_TWA_comparison}-a), the kink evolves without any coupling to the fermions $g=0$, and the broadening is just the result of the fluctuations of the initial state, in particular, the population of the various normal modes around the soliton profile.
 Fig. \ref{fig_TWA_comparison}-b) demonstrates that upon switching on the Yukawa term $g\neq 0$, which effectively leads to an increase of the staggering force,  one obtains a pronounced narrowing of the kink's width broadening. 
Considering Eq. \eqref{ecs_mov_kink_forzado}, this can be understood as a back-reaction effect arising from a Peierls-Nabarro barrier, which we showed to grow with the state-dependent force $g$ (see Fig. \ref{fig_EPN_vs_energy_zero_mode}-b)). 
The spatial oscillations in the scalar field far from the kink are a consequence of the staggered external force mentioned above.

\begin{figure*}
    \centering
    \includegraphics[width=0.85\linewidth]{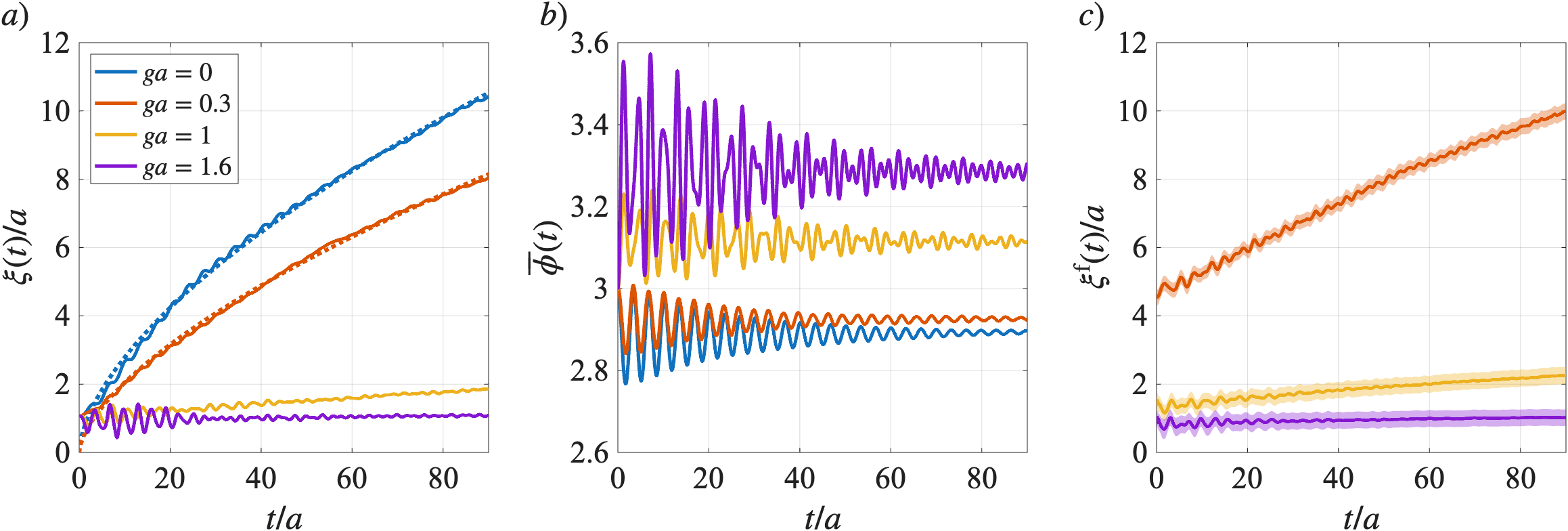}    
    \caption{{\bf Dynamical soliton and bound-fermion fits:} Fitting parameters of the TWA expectation values to the kink solution $\Phi_{{\rm K},n}= \Phi_0\tanh(a n/\xi)$, but allowing for a time-evolving width $\xi(t)$ (left panel), and amplitude $ \Phi_0(t)$ (right panel). 
    Dotted lines in the left panel show the fit with the function $\xi(t)=c t^\alpha$, with fitted exponents $\alpha\in\{0.61,\,    0.63\,\}$ for $g\in\{0,\, 0.3\,\}$, respectively. 
    Parameters are the same as in Fig. \ref{fig_TWA_comparison}-a). c) Fitting parameter $\xi_{\textrm{f}}(t)$ of the TWA expectation values obtained for the accumulated charge density $\Delta Q_n$ in Eq. \eqref{accumulated_charge_fit}. The shaded area shows the $95\%$ confidence interval for the fitted parameter.}
    \label{kink_fit} 
\end{figure*}

Let us now analyze in closer detail the time dependence of the broadening of the kink's width, and how it depends on the Yukawa coupling.
We plot in Fig.~\ref{kink_fit} the results of fitting the results of our TWA-fGS simulations as a function of time to an effective kink profile inspired by Eq.~\eqref{kink_discrete},  but allowing for a dynamical width $\xi(t)$ and amplitude $ \Phi_0(t)$, namely
\beq
\Phi_n(t)= \Phi_0(t)\tanh\left(\frac{a}{\xi(t)}(n-n_0)\right)
\eeq
 In Fig.~\ref{kink_fit}-a), we show in dotted lines the results of fitting the time-dependent width to a power law $\xi(t)=ct^\alpha$, showing that the broadening of the soliton is close to a diffusive dynamics $\xi(t)\propto\sqrt{t}$  for Yukawa couplings $g/J<1$. 
On the other hand, when the Yukawa coupling is stronger, we see that the width of the soliton does not spread diffusively anymore, but gets instead localized and oscillates around an initial value. This is a result of the large back-reaction of the fermions onto the scalar field, and coincides with the regime in which the PN barrier of Fig.~\ref{fig_pn_barriers} becomes so large that even individual TWA trajectories cannot escape from it.
In Fig.~\ref{kink_fit}-b), we find that the kink's amplitude shows oscillations while decaying towards a steady-state value. This figure also shows a crossover behaviour as a function of $g$, with a threshold above (below) which the initial amplitude increases (decreases). Altogether, these results show the interplay of quantum fluctuations and back-reaction in the problem. While the former aim at inducing a diffusive broadening of the kink, the latter introduces a pinning force that eventually precludes the kink's dynamics.  

We note that this pinning mechanism could be of particular interest in the study of kink dynamics in ion chains, as it could prevent the defect losses due to the kinks reaching the boundaries of the chain or zigzag region~\cite{Partner2013, Ulm2013}. 
Translated to the nomenclature used in the studies of trapped-ion kinks in zigzag chains \cite{Partner2013}, on-site and inter-site kinks correspond to odd and extended kinks, respectively. 
In these terms, the effect described above, where kinks are stabilized at inter-site locations due to the pinning mechanism of Eq.~\eqref{ecs_mov_kink_forzado_twa}, means that the external force exerted by the interaction with the fermion number density changes an odd kink configuration into an extended kink configuration.

Let us now address whether this kink dynamics drags the fractional fermion charge with it, also leading to a diffusive versus localised dynamics of fractionalised fermion charge. We use the definition of the regularised charge discussed around Eq.~\eqref{charge_density}, and its corresponding accumulation $\Delta Q_n$ in Eq.~\eqref{eq:accumulated charge}.
The bottom panels of Fig.~\ref{fig_TWA_comparison} show the time evolution of the accumulated charge density $\Delta Q_n(t)$, verifying how a fractional value of 1/2 is reached when integrating along a region of length that is correlated with the soliton's width, which is now spreading with time. This indicates that the fermion charge is bound and propagates with the soliton. Naturally, the binding of the fermionic zero-mode is not perfect regardless of the coupling strength.
In the strongly interacting regime, this charge distribution follows the soliton propagation adiabatically, while for weak interactions, it only exhibits the same tendency.
To quantify this relationship, we fit the accumulated charge density to the function 
\beq\label{accumulated_charge_fit}
\Delta Q_n(t)=A(t)\tanh\left(\frac{a}{\xi_\textrm{f}(t)}(n-n_0)\right)+B,
\eeq
and extract the width of the fractional charge distribution as a function of real time. 
We show $\xi_\textrm{f}(t)$ in Fig.~\ref{kink_fit}-c), for the same parameters as in Fig.~\ref{kink_fit}a), excluding   $g=0$ as there is no bound fermion mode for a vanishing Yukawa coupling. This allows for a quantitative comparison of the spatial extent of the soliton excitation and the bound fermion charge.
For weak coupling, the scalar field dynamics does not transport the fractional fermion charge tightly. 
In this case, at the initial state, the charge density (representing the fermionic zero-mode wavefunction) is not exactly prescribed by the soliton width, and they remain distinct at later times. However, in both cases,  the soliton accumulated-charge widths spread, exhibiting a similar diffusive behavior. This connects to our previous statement that, even if the fGS does not require incorporating phase-space fluctuations in the initial state, after the propagation with our TWA-fGS formalism, it inherits its stochastic features from the scalar field, in this case leading to diffusion.

As the Yukawa coupling increases, we have seen that the back-reaction modifies the scalar field evolution, eventually pinning it. In Fig.~\ref{kink_fit}-c), we see how this pinning is also reflected in the accumulated charge, which also gets localised as one increases the Yukawa coupling. In this limit, one can say that there is a composite excitation in which the half fermion is bound to the soliton field, and both remain pinned due to the back reaction.

As the soliton width $\xi(t)$ increases, the ground state fluctuations of the initial state have a larger projection onto the Goldstone mode, which causes an increased diffusion of the domain wall width. 
As commented above, for strong interaction, the fractional charge is bound to the soliton. We observe this effect in Fig.~\ref{fig_TWA_comparison}-d), where the fractional charge propagates with the soliton. 
We remark how in this regime there is a significant deviation of the scalar field expectation values from the characteristic kink profile during time evolution, as we show in the upper panel of Fig.~\ref{fig_TWA_comparison}-d).

In summary, Fig.~\ref{fig_TWA_comparison} highlights the interplay between the propagation induced by zero-point excitations of the Goldstone mode for large $\xi$ and the pinning effect of the interaction. 
For a fixed $\xi$, increasing the interaction $g$ has the effect of pinning the soliton to the PN potential wells (or decreasing the diffusion rate) and inducing 
a charge density distribution that propagates at the same rate as the soliton.

\subsection{Fractional charge    drag by a moving soliton }

Finally, before turning to the discussion of collisions of solitons carrying fractional fermion numbers, we first illustrate the dynamics of a  soliton with a non-zero initial momentum. A possible initial state in phase space that achieves this goal singles out the Goldstone mode $\nu=0$, shifting its momentum distribution by the amount $\overline{P}_0$ as follows
\begin{equation}\label{init_single_mov_soliton}
\begin{aligned}
W(\boldsymbol{\phi}_0,\boldsymbol{\Pi}_0) \propto \; 
\ee^{ -\left(\frac{Q_0^2}{2\sigma_{Q_0}^2}+\frac{(P_0-\overline{P}_0)^2}{2\sigma_{P_0}^2} \right)}
\prod_{\nu=1}^{N-1} \ee^{ - \left( \frac{Q_\nu^2}{2\sigma_{Q_\nu}^2}  +\frac{P_{\nu}^2}{2\sigma_{P_\nu}^2} \right)}.
\end{aligned}
\end{equation}
On the other hand, as already mentioned in the discussion below Eq.~\eqref{ec_dispersion_kink_lattice}, the lowest-energy mode $\nu=0$ is the one responsible for translations of the soliton center coordinate $x_0\mapsto x(t)=x_0+vt$, which can be obtained in the continuum limit by a Lorentz boost. From this perspective, this mode is treated entirely classically without allowing for any further fluctuations in phase space. This limit can be achieved by excluding the Goldstone mode from the Wigner distribution, absorbing its effect in a time-dependent classical solitonic background $\Phi_{{\rm K},n}(t)$. Equivalently, we can simply set vanishing widths $\sigma^2_{Q_0},\sigma^2_{P_0}$ in the above expression, 
describing an initial kink that is   classically moving  towards  $x_n>0$ for $\overline{P}_0<0$.

\begin{figure}
    \centering
\includegraphics[width=1\linewidth]{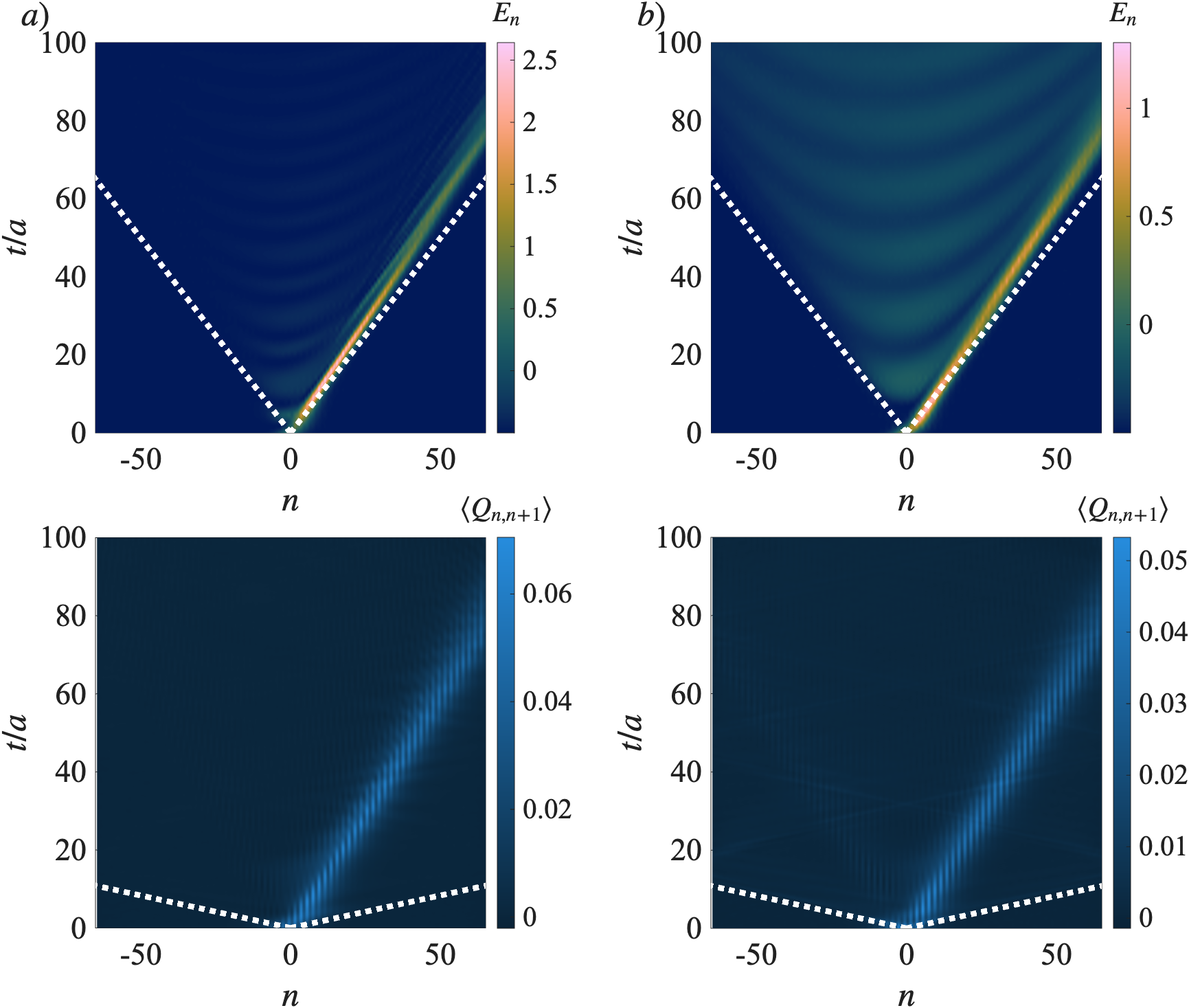}    
\caption{\textbf{Single moving soliton and associated charge density.} Top panels: Dynamics of a single soliton with an initial momentum distribution obtained through TWA. The dashed white lines represent the light cones $x=\pm c_{\rm b} t$ for excitations in the bosonic Hamiltonian, which has $c_{\rm b}=1$. 
Bottom Panels: The corresponding charge density associated with the fermionic zero-mode follows the kink dynamics. The dashed white lines represent the light cones $x=\pm c_\textrm{f}\,t$ for excitations in the fermionic Hamiltonian, which has $c_{\rm f}=2J$.
We consider the ground state excitations of the Goldstone mode, a) excluded, b) included (corresponding to the state in Eq.~\eqref{init_single_mov_soliton}). Parameters used are $\overline{P}_0=-2$, $N=160$, $\sbgs=5$, $dt/a=0.01$, $N_t=2000$, $\xi/a=5$, $Ja=3$, $ga=0.16$. \label{single_moving_soliton}}
\end{figure}

In the TWA, the leading quantum fluctuations will cause the soliton to diffuse as it propagates, broadening the classical trajectory. We want to explore if this movement drags the fractional fermion charge along with it, and explore this dragging effect quantitatively.  Moreover, we would like to understand the effect of adding quantum fluctuations on the Goldstone mode, following Eq.~\eqref{init_single_mov_soliton} with finite-width Gaussians in phase space.
As we show in the upper panel of Fig. \ref{single_moving_soliton} a), the energy density $E_n=\langle \mathcal{H}_{n}\rangle$ of the scalar sector~\eqref{Hphi4}  is initially concentrated around the soliton center.
When the initial state in Eq. \eqref{init_single_mov_soliton} neglects the effect of zero-point fluctuations in the Goldstone mode, we only find trajectories that propagate to the right, the average of which yields the clear propagation of the soliton as a quasi-particle with a net momentum encoded in the initial state. Upon a closer inspection, one can see that there are recurrent ripples in the energy distribution showing a clear cone-like distribution, which are the consequence of considering quantum fluctuations,  back-reaction, and the discretized nature of the model leading to PN barriers.  Even if the classical soliton is predicted to move entirely ballistically conserving its localised energy, the quantum version leaves a track of small ripples due to its interaction with the PN barriers, and, moreover, its width around the classical trajectory spreads due to the ample average over several trajectories. Interestingly, the lower panel of Fig.~\ref{single_moving_soliton} a) shows how the regularised fermionic charge~\eqref{eq:reg_charge}, which is concentrated on the zero mode, follows the same broadened trajectory as the soliton. We can thus conclude that the moving soliton drags the fermion charge with it.   

 Numerically, we have not confirmed the presence of resonances where, after the first bounce, the solitons can escape their mutually attracting potential. Since internal shape modes are present, this suggests that the addition of the external force breaks the resonance condition, and energy cannot be transferred back to the translational modes.

Let us now also consider the effect of quantum fluctuations on the Goldstone mode, which will lead to more pronounced deviations from the classical dynamics of a moving soliton. We now set a non-zero Gaussian width to the $\nu=0$ Gaussians of Eq.~\eqref{init_single_mov_soliton}. Interestingly, even when the fluctuations in momentum quadrature are small in comparison to the momentum shift $\overline{P}_0$, the fluctuations in the spatial one lead to phase-space trajectories that can propagate in the opposite direction. In the upper and lower panels of Fig.~\ref{single_moving_soliton} b), we show the corresponding dynamics in the scalar and fermionic sectors for the propagating soliton-fermion composite excitation. In the scalar sector, one clearly observes the effect of the additional fluctuations on the Goldstone mode: the trajectories that propagate in the opposite direction lead to a causal structure that also extends to the left of the initial soliton. This is manifested in a symmetric cone-like structure filled by ripples of the energy density, always inside the causal region defined by the propagating soliton. On the other hand,  the largest contributions to the scalar-field energy and to the fermion charge are still centered around the classical trajectory that moves to the right.  In the lower panel of the figure, we see how the regularised charge is instead not much affected by the extra fluctuations on the Goldstone mode, showing that the zero mode mostly travels with the soliton wavefront.

\section{\bf Real-time collisions of half-charged fermions on soliton-antisoliton configurations}
\label{sec:soliton_collisions}

While a single stationary soliton with its bound zero-mode demonstrates the phenomenon of charge fractionalization, and its robustness rests on the topological nature of the excitation, we have seen how quantum fluctuations, back-reaction, and lattice discreteness can lead to spreading, localization, and cone-like ripples once the soliton is quantised and both the fermion and scalar fields have quantum fluctuations. Having characterized the dynamics of initial states with a single soliton and a single half-charged fermion using our TWA-fGS approach, we now begin to study the real-time dynamics of collisions. We note that a collision between solitons leads to dynamical phenomena in which the topological nature of the scalar field configuration can be temporarily lost, and the topological protection is not guaranteed. Hence, it is not a priori clear what will happen to the bound half-charges that are dragged by the moving solitons-antisolitons. Moreover, the $\lambda\phi^4$ QFT is not an integrable model, and the collisions are not expected to be elastic. Recently, there has been an increasing interest in studying the real-time dynamics of scattering particles in (1+1) models, using the advantages of tensor network techniques \cite{Jha2025,barata2025, Papaefstathiou2025, Rigobello2021}, or proposals for digital simulation on a quantum computing platform \cite{davoudi2025}, as well as analog simulations using qudits \cite{Calliari2025}. It thus seems interesting to explore these collisions in more detail by generalizing our TWA-fGS studies to the soliton-antisoliton sector.

In this section, we show how the fermionic back-reaction significantly alters collision dynamics in several ways: it modifies the effective kink-antikink potential, and we observe the formation of stable, oscillatory, bound states of the composite kink-antikink system mediated by the Yukawa coupling to the fermions. 
Moreover, we probe the stability of the fractional charge distribution during the collision.
We first describe a classical approximation of the scalar field, and later add the effect of scalar field ground-state fluctuations. 
Beyond classical simulations, the TWA approach reveals how quantum fluctuations in the scalar field smear out the energy distribution in the system, allowing individual trajectories to escape the mutual attracting potential of the solitons.

\subsection{Classical  collisions in the continuum $\lambda\phi^4$ model}

Let us start revisiting classical soliton collisions in the continuum $\lambda \phi^4$ theory, where one can find different behaviors depending on the relative initial velocities of the colliding pair of kink and antikink solitons, which are not transparent to each other but interact during the collision. In principle, solitons can either reflect from one another or form a bound state \cite{Belova1997}.
This clearly contrasts with solitons in integrable models, such as the sine-Gordon model, which do not interact with one another and can thus pass through each other during the collision with their final velocities after interaction remaining unchanged \cite{Peyrard1983}.
Kinks and antikinks of the $\lambda \phi^4$ do not pass through each other but interact and, in doing so, there appears a finite region where the scalar field differs from its vacuum value, growing with the separation between kinks once they exchange their position. 

In general terms, above a critical relative velocity $v_c$ with which the solitons are initially approaching each other, the kinks will always reflect after the collision. Due to the non-integrable nature of the model, the initial and final velocities after a reflection will differ, with some energy being released/dispersed into other modes as phonons of the continuum portion of the spectrum \cite{Campbell1983}.
If the kinetic energy is instead below some threshold value, then a long-lived oscillatory mode appears that is spatially localized \cite{Wingate1983}, sometimes referred to as a ``bion'' \cite{Geicke1983, Belova1997}. In a simplified model of kink-antikink collisions, this state can be understood as a result of their mutual attraction. 
These oscillatory modes are long-lived excitations capable of decaying towards the non-topological vacuum state by emitting radiation \cite{Lizunova2021}.

In the intermediate region of initial kink-antikink velocities, there is an alternating sequence of these reflective and bound-state behaviors \cite{Campbell1986}. 
For velocities above $v_c$, the kinks reflect from one another after collision.
In contrast, within the intermediate region, the kinks collide and then separate a finite distance, being trapped due to their attractive potential. In this process, some energy of the Goldstone mode is transferred to the internal modes that only change the shape of the solitons, causing oscillations of the kinks. After a second collision, if enough energy is transferred back to their Goldstone modes, the kinks can escape their mutually induced potential; this is usually called a 2-bounce escape window \cite{Campbell1983}.

\subsection{Classical kink-antikink collisions  on the lattice}\label{SAS_collisions_classical}

In this section, we analyze the effect the Yukawa coupling has on the classical dynamics of the collisions on the lattice. In this context, classical means that we solve the set of ordinary differential equations~\eqref{eq:corr_matrix_equations}-\eqref{ecs_mov_kink_forzado_twa} without sampling over any phase-space distribution. Strictly speaking, the fermionic state corresponding to a filled Fermi sea is non-classical due to both quantum statistics and its entanglement content. Hence, the classicality here refers to the scalar-field sector.    To describe binary collisions, it is always possible to choose a frame of reference in which both kink and antikink move towards each other with opposite speeds, such that the $c$-number conjugate fields on the lattice at the initial instant of time read
\begin{equation}\label{initial_state_collision}
\begin{aligned}
        \Phi_{0,n} & =   \Phi_{{{\rm K}\overline{\rm K}},n}, 
        \hspace{2ex} \Pi_{0,n}  & = -\overline \Pi_0 \mathcal{M}^{\rm b, (e)}_{n,0},
\end{aligned}
\end{equation}
\noindent 
where we have introduced 
\beq
 \Phi_{{{\rm K}\overline{\rm K}},n}= - \Phi_0 \tanh\left(\frac{an-d/2}{\xi_0}\right)
        + \Phi_0 \tanh\left(\frac{an+d/2}{\xi_0}\right) - \Phi_0.
\eeq
In the two-soliton system described by Eq. \eqref{initial_state_collision}, the zero-energy eigenmode after linearization [cf. Eq.~\eqref{ec_dispersion_kink_lattice}] is doubly degenerate, representing the Goldstone modes of each soliton. When the solitons are separated by a large distance, these normal modes are independent and spatially localized at each soliton. However, for short separations, the doubly degenerate pair corresponds to the even and odd $\mathcal{M}^{\rm b,(e)}_{n,0}, \mathcal{M}^{\rm b,(o)}_{n,1}$ combinations of the Goldstone modes localized at each soliton. To obtain propagating solutions, we consider an initial state with a momentum that is proportional to the even Goldstone mode $\mathcal{M}^{\rm b,(e)}_{n,0}$.
For a kink-antikink configuration, this initial momentum distribution will cause the soliton and anti-soliton to move towards each other and eventually collide.

\begin{figure}
    \centering
    \includegraphics[width=1\linewidth]{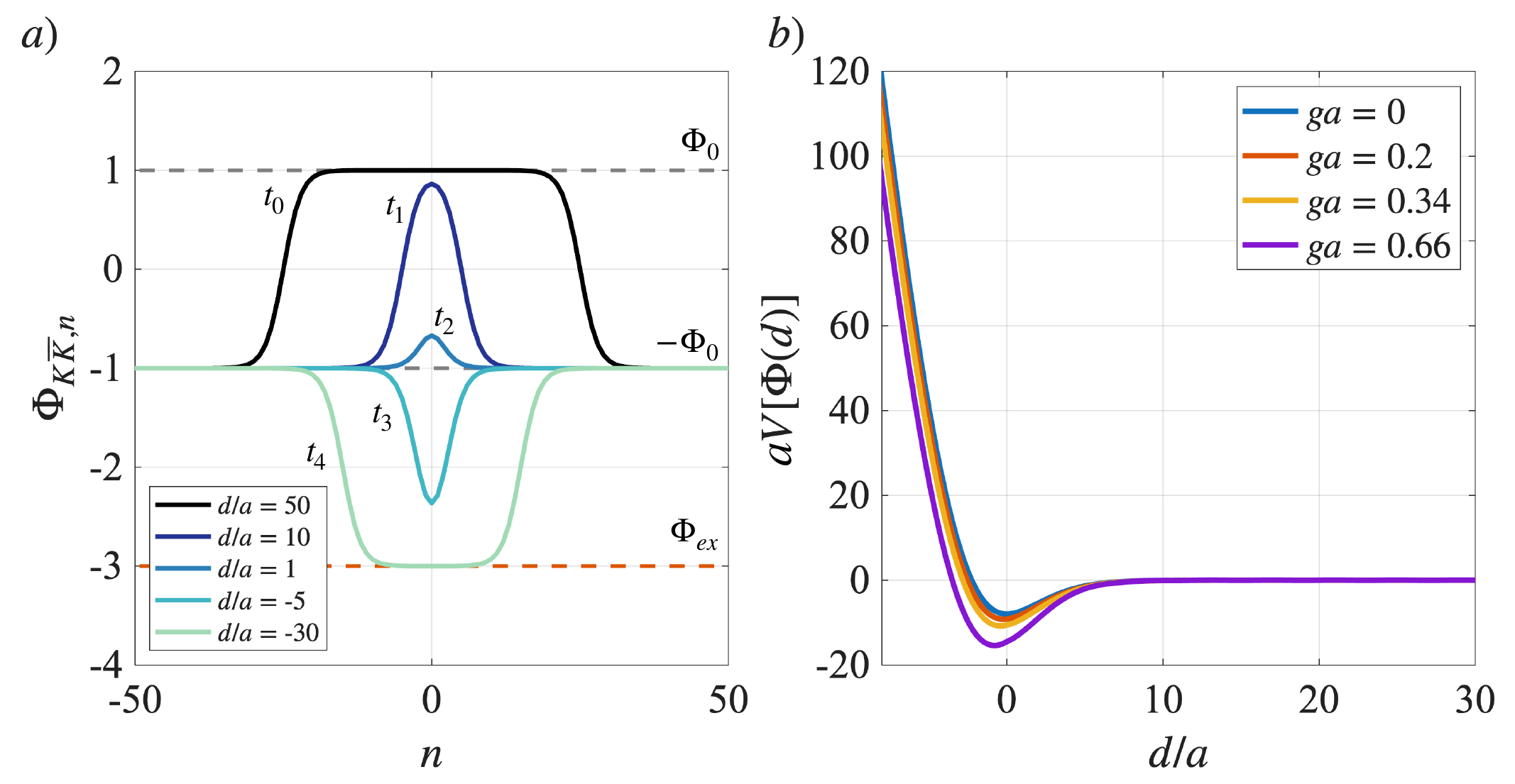}
    \caption{{\bf Kink-antikink Born-Oppenheimer potential:} a) Different field configurations for varying distances in the kink antikink system, labeled by evolving instants of time $t_0\to t_4$. The dashed gray lines indicate the two SSB vacuum expectation values connected through the kinks and anti-kinks. Note how for negative distance $d(t)$, there is an extensive region (length growing with $|d(t)|$)  with an energy $\Phi_{ex}$ different from that of the vacuum, causing the energy of this configuration to grow linearly with the distance. Parameters used are: $N=160$, $\sbgs=1$, $\xi/a=3$, $Ja=1$. b) Potential energy of the scalar field as a function of the distance between the two solitons. Here, the interaction $g$ increases the depth of the potential well acting at short distances. For each $g$, we subtract the energy of the non-interacting solitons (infinite distance). Parameters used are $N=160$, $\sbgs=3$, $\xi/a=3$, $Ja=1$.}
\label{potential_two_solitons}
\end{figure}

Following the approach used for the single-soliton case and the PN potential, we start by calculating an effective kink-antikink interaction potential $V_{{\rm K}\overline{\rm K}}(d)$ using the Born-Oppenheimer adiabatic approximation. Starting from a steady-state two-soliton configuration, varying the kink and antikink centers and interpolating, we can estimate this interaction by integrating the classical energy density, as depicted in Fig.~\ref{potential_two_solitons} for various Yukawa couplings. In the non-interacting limit $g=0$, we recover the usual attractive potential acting at short distances \cite{Kudryavtsev1975}. 
The effect of a non-vanishing Yukawa coupling is to increase the depth of this potential well as a consequence of the back-reaction, paralleling the situation discussed for the single soliton in and the PN barriers in Fig.~\ref{fig_pn_barriers}{\bf (a)}. 
This means that as the distance between solitons decreases and energy is dispersed as phonons of the continuum modes, escaping the bound state will require more energy to be transferred back to the center of mass mode as $g$ increases.
Note that, for the large values of $\xi_0/a$ used in this case, we are effectively close to the continuum limit, and this potential does not display any additional PN barrier due to the discreteness of the lattice.

\begin{figure*}
    \centering
    \includegraphics[width=1\linewidth]{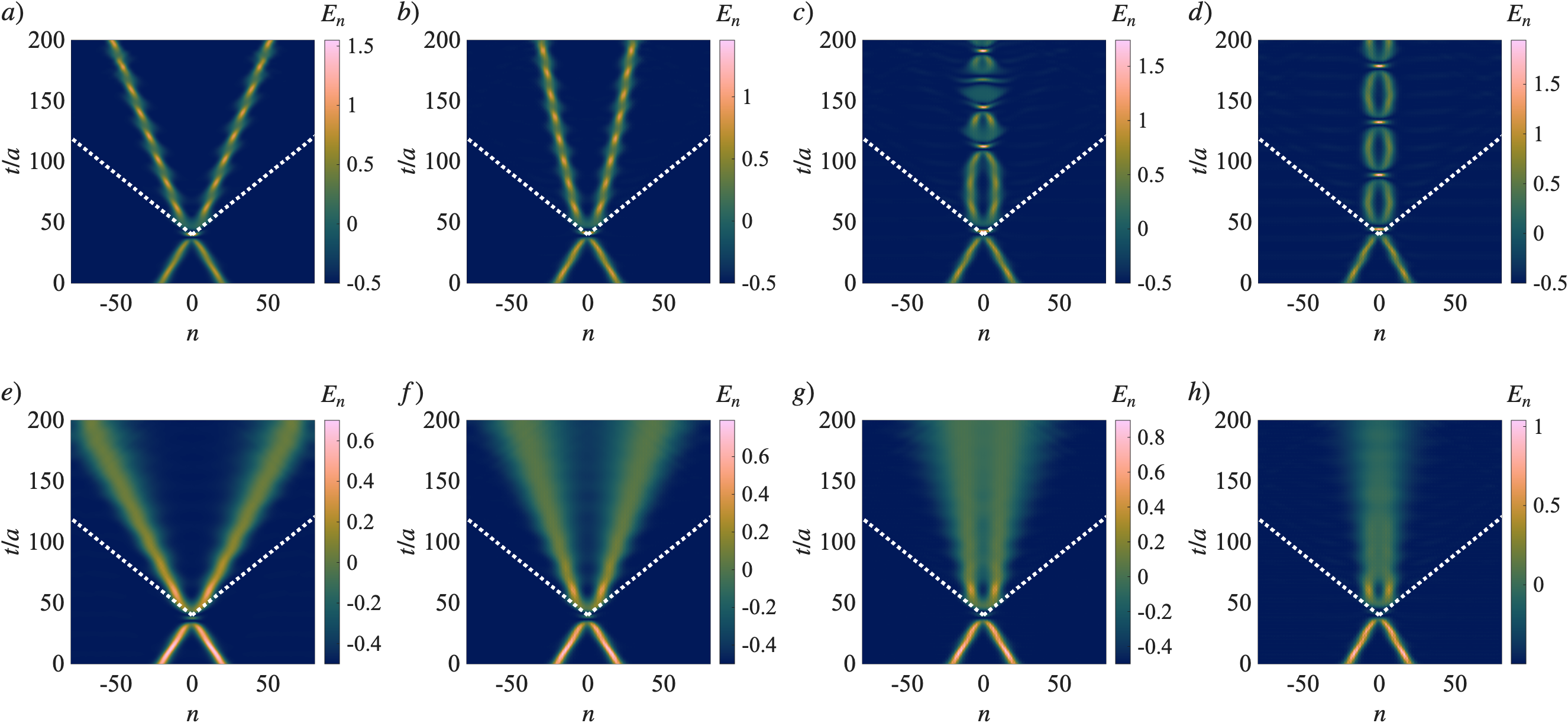}
    \caption{\textbf{ Classical and quantum-mechanical soliton collisions:} Top panels: Classical motion of kink-antikink system. Parameters used are $\sbgs=5$, $\xi/a=5$, $Ja=3$, $dt/a=0.1$, and $ga\in\{0.02, \, 0.12, \,  0.18, \, 0.24\}$ for figures a)-d) respectively. 
    Bottom panels: First quantum corrections to the classical description of kink-antikink collisions, computed using TWA-fGS with $N_t=10^4$ trajectories, for the same parameters as in Figs. a)-d).
    The white dashed lines show the effective light cones for the bosonic system, separating the causally disconnected and connected regions with respect to the collision.
    }
    \label{fig_collision_combined}
\end{figure*} 

Let us analyze the form of $V_{{\rm K}\overline{\rm K}}(d)$ in more detail. As expected, $V_{{\rm K}\overline{\rm K}}(d)\to 0$ for $d\gg a$, showing that distant solitons do not feel each other, and one can define asymptotic in-out scattering states using the knowledge gathered for the single-soliton solutions of the previous section. As the kink-antikink pair gets closer, they start to feel the deformation of the vacuum arising from each other, and effectively feel an attractive potential for $\partial_dV_{{\rm K}\overline{\rm K}}(d)<0$ for $d\sim a$. On the other hand, when the kink and anti-kink swap places $d<0$, they feel a linearly rising repulsive potential $V_{{\rm K}\overline{\rm K}}(d)=\sigma d$ for $d\ll 1$, which is due to the energy penalty of leaving an excited region of length $d$ in between them (see Fig.~\ref{potential_two_solitons}-a)). This part is similar to an effective string tension in the context of gauge theories, but only extends to the $d<0$ half plane, and in general leads to a repulsive barrier such that the kink and antikink will bounce from each other. There is, however, also the possibility that the collision hits a resonant bound state that would remain trapped inside the potential well at intermediate distances $d\sim a$.  This intuition will turn out to be useful when analyzing the dynamical results.

\subsubsection{Soliton collisional dynamics}

In Fig. \ref{fig_collision_combined}-a)-d) we show the classical time evolution of the scalar field energy density.
For vanishing Yukawa coupling $g=0$, we recover the usual dynamics for soliton-antisoliton collisions, where kinks bounce off each other at large initial speeds. We see that the energy density of the kink and antikink travels without distortion towards the colliding coordinate, here set to $n=0$. At the collision, the linearly-rising potential for $d<0$ leads to a local energy burst at the center, which is followed by the bouncing of the solitons that are expelled into outgoing directions. During the collision, an internal shape mode of the kink and antikink gets excited, which is manifested by the energy oscillations shown in the outward cone, and by the slight tilting of this cone with respect to the incoming one, showing that the propagation velocity decreases after the collision.

\begin{figure*}
    \centering
    \includegraphics[width=1\linewidth]{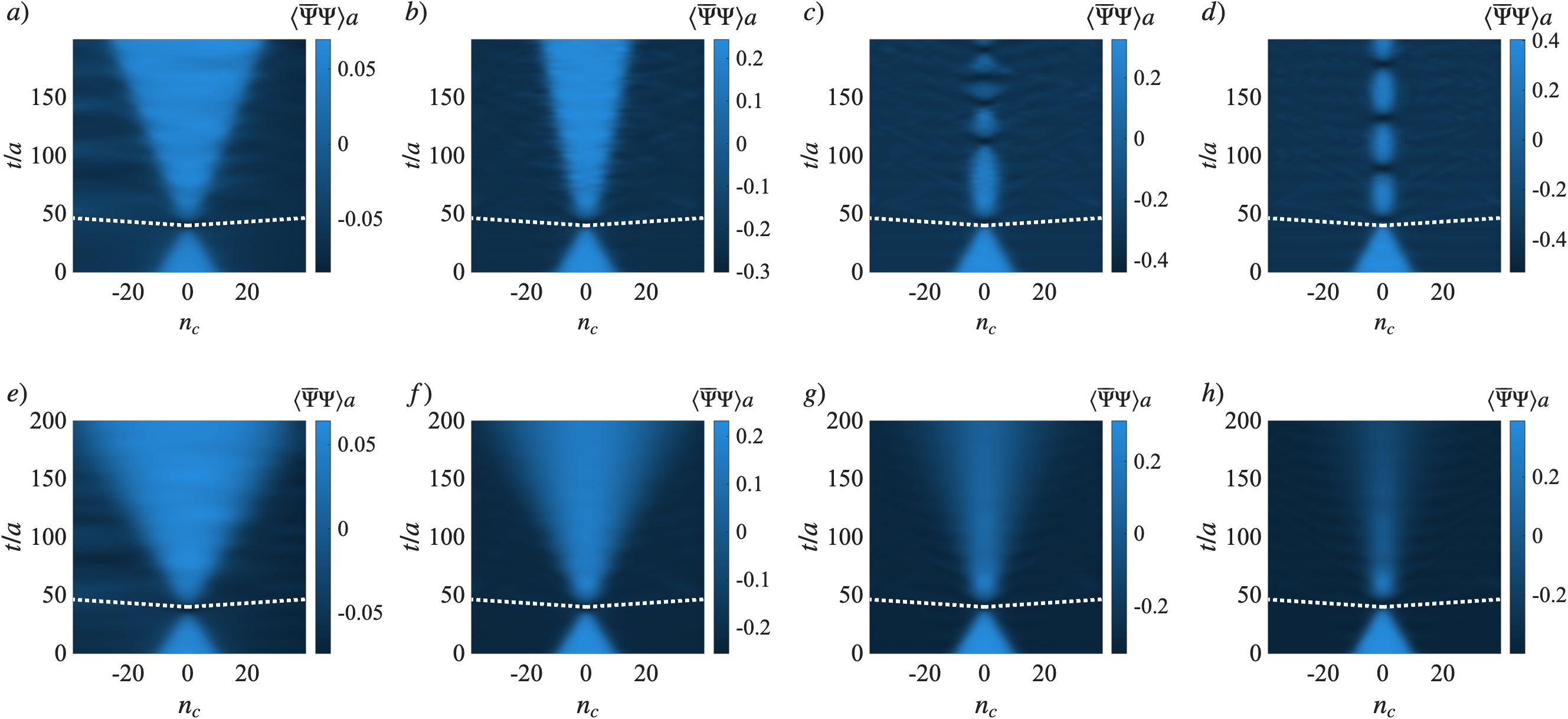}
    \caption{\textbf{Domain walls in the scalar fermion condensate during soliton collisions.} Top panels: Domain walls in the scalar fermion condensate during the kink-antikink collisions, computed using the classical scalar field dynamics and a single trajectory for the fGS. Parameters used are $\sbgs=5$, $\xi/a=5$, $Ja=3$, $dt/a=0.1$, and $ga\in\{0.02, \, 0.12, \,  0.18, \, 0.24\}$ for figures a)-d) respectively. Bottom panels: First quantum corrections to the classical dynamics of the top panels, computed with the TWA with $N_t=10^4$ trajectories. The dashed white lines show the effective light cones for the fermionic system centered at the collision.}
    \label{fig_collision_TWA_domain_walls}
\end{figure*} 

When switching the Yukawa coupling $g\neq 0$, each of these solitons carries a fractionally-charged fermion as described in the previous sections, provided that both zero-energy fermion states are occupied in the initial state.
By increasing the Yukawa coupling strength as in Fig.~\ref{fig_collision_combined}-b), we observe a similar behaviour as before, but there is a clear decrease in the kinetic energy after the collision. This can be attributed to the effect of back-reaction and an increased inertia of the kink-antikink pair due to excited fermions during the collision. 
By further increasing $g$, as in Fig. \ref{fig_collision_combined}-c), we find that after a collision and an initial bouncing, the kink and antikink can reflect back by reversing their speeds due to their mutual attraction within the potential well (see Fig.~\ref{potential_two_solitons}). 
This leads to a second collision, but,  due to the pinning mechanism induced by the Yukawa coupling, the composite state never reaches enough energy to escape the potential well after the second collision, and remains in a bound state showing persistent oscillations. This is the bound state referred to as a ``bion''~\cite{Belova1997}.
Finally, even for large values of $g$, we can clearly identify the bion, as the system develops long-lived periodic oscillations of the bound kink-antikink,  as we show in Fig. \ref{fig_collision_combined}-d). In the following subsection, we will analyze this confinement mechanism from the perspective of the fractional charge.

Within this classical approximation of the coupled scalar field and fermion system, increasing $g$ has approximately the same effect as reducing the kinetic energy of the initial state in the purely $\lambda\phi^4$ theory. This, in turn, is consistent with the adiabatic interpretation of the coupling acting as an effective Peierls-Nabarro barrier for the dynamics of each soliton. 
There is, however, a difference between the non-interacting solitons and this system, since here we have not observed an energy transfer to the Goldstone mode after the second bounce when the initial speed is below the critical velocity.  
Let us close this part by noting that, upon a closer inspection of the contour plots of the energy densities,  one can notice the appearance of ripples. These are present even in the continuum $\lambda \phi^4$, as a consequence of the kink-antikink configuration not being a steady state of the equations of motion, and demonstrate how energy is dissipated into different motional modes, resulting in the decrease in speed after the collision.

\subsubsection{Fractional-charge collisional dynamics}

To study the dynamics of the fermionic zero-mode, we start by focusing on the domain walls 
in the scalar fermion condensate 
\beq
\langle \overline{\psi}(x_n)\psi(x_n)\rangle=\langle\chi^{\dagger}_{2n-1}\chi_{2n-1}\rangle-\langle\chi^{\dagger}_{2n}\chi_{2n}\rangle,
\eeq
which are defined within the  two-site unit cells $n\in\{1,\cdots N/2\}$, in Fig. \ref{fig_collision_TWA_domain_walls}-a)-d). 
By studying the scalar fermion condensate, we observe how the domain walls approximately follow the scalar field. During the collision, the domain walls momentarily disappear when the scalar field passes through the symmetry-broken ground state and reappear after the solitons bounce back from each other, demonstrating the robustness of the fermion states. Note, however, that the magnitude of these domain walls increases with $g$. 

When looking at the integrated charge instead, after the collision, there are small domain walls propagating at approximately the maximum group velocity of the fermionic Hamiltonian in Eq.~\eqref{Hfermions} for the symmetry-broken ground state. 
This can be understood as the system being in an excited state for a short time during the collision: If we assume a conserved fermion number, then when the scalar field goes through the symmetry-broken ground state, two excited fermion states get occupied as a consequence of the scattering.

Suppose now that the maximum group velocity is larger than the soliton's final speed (upper-bounded by its initial speed). In that case, these domain walls propagate faster than the soliton, rapidly reaching the boundaries and bringing the onset of finite-size errors. 
Therefore, to observe the fractional charge using Eq.~\eqref{charge_density}, it is enough to consider solitons moving with initial speeds $v_i$ larger than the maximum group velocity, i.e., $|v_i|>|v_f|>\max(|v_g|)$. 

\begin{figure}
    \centering
    \includegraphics[width=1\linewidth]{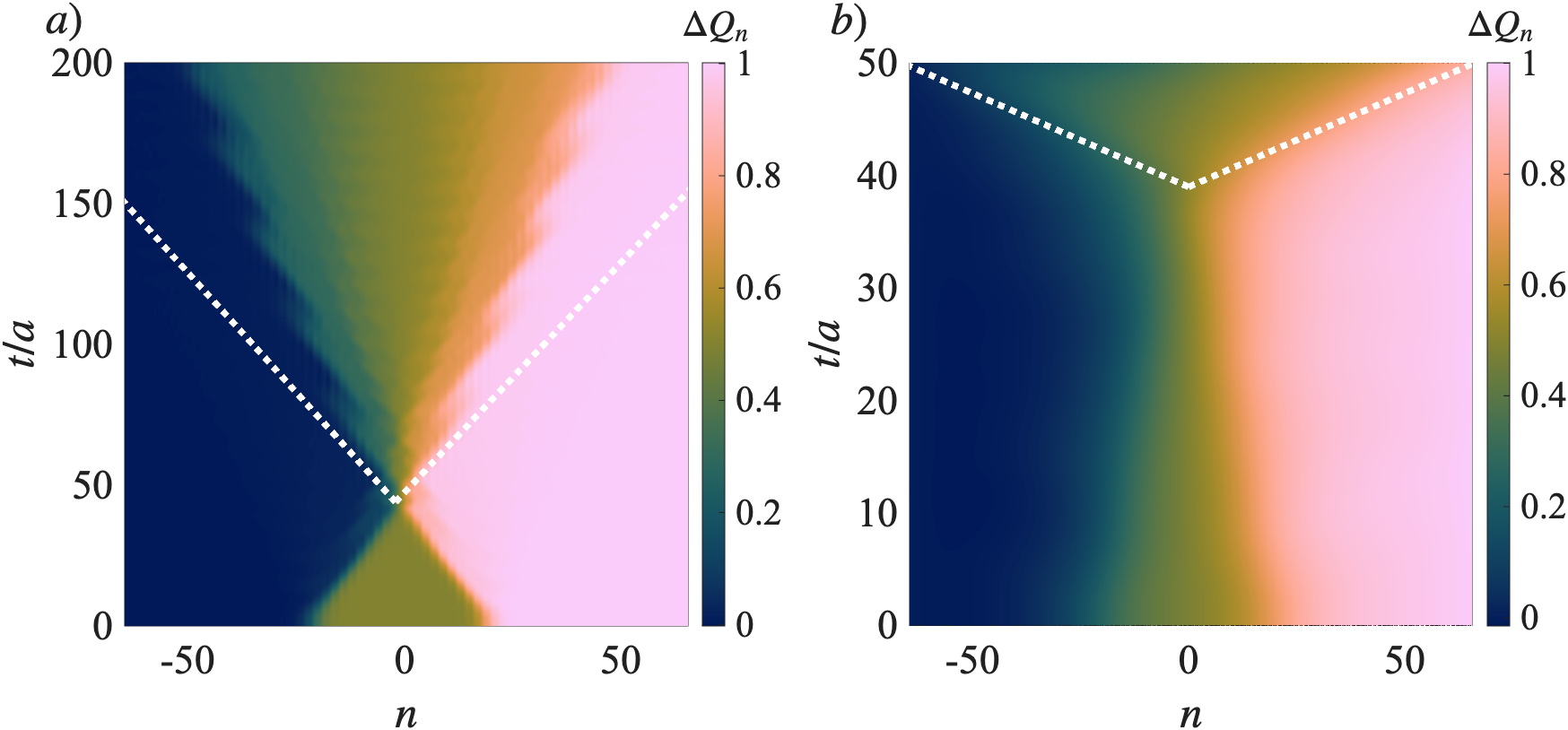}
    \caption{\textbf{Comparison of fractional charge dynamics for slow and fast fermions.} Accumulated charge for a) slow ($Ja=0.3$), and b) fast ($Ja=3.0$) fermions. The rest of parameters used are $N=160$, $\Phi_0=5$, $\xi/a=5$, $ga=0.06$, $dt/a=0.1$, $\overline P_0=-2$. In b), the small excitations transported at approximately the maximum group velocity are captured by the accumulated charge $\Delta Q_{n} $, even though the domain walls in the scalar fermion condensate still approximately follow soliton dynamics.
    }
    \label{fig_collision_charge_slow_fast}
\end{figure}

Let us now focus on the accumulated charge~\eqref{eq:accumulated charge}, addressing how the fractional charges are distributed within the kink and anti-kink, and how they get redistributed as a consequence of the collision.  We show in Fig. \ref{fig_collision_charge_slow_fast} that the accumulated charge has two jumps of $1/2$ as one moves across the kink and then the antikink, leading to a total excess of charge $\Delta Q_N=1$. In the incoming region, this charge is sharply localized into two lumps of $1/2$ that are dragged by the solitons towards each other, and then redistributed into two outgoing fronts that separate from each other after the kink and antikink bounce off each other (see Fig.~\ref{fig_collision_charge_slow_fast}a)).

We have found that the resulting dynamics of the fractional charges after the collision depend on the fermionic tunneling strength $J$ in an unprecedented manner. On the basis of the Born-Oppenheimer approximation, one would naively expect that in the large-$J$ limit, the fermions adapt instantaneously and follow the solitons after the collision. However, this collision is not elastic, and the energy typically gets redistributed into shape modes of the soliton, ripples of the scalar-field energy, and, crucially for the current discussion, fermionic excitations described at the mean-field level by a gapped energy band of bandwidth $J$. As a consequence, for large $J$, it is possible to excite fermionic particle-anti-particle pairs that travel along opposite fronts and propagate much faster than the incoming ones. This situation is depicted in Fig.~\ref{fig_collision_charge_slow_fast}b), where the accumulated charge density $\Delta Q_{n}$ clearly shows steps of 1/2 localized at the fast propagating fermionic wavefronts. This case is even more dramatic as, for these specific parameters,  the kink and antikink in the scalar sector are left in a bound state after the collision. We thus see that the half-charges, which are always confined to the kink or antikink in the single-soliton regime forming a composite excitation, can get released after a collision and completely decouple from the soliton dynamics.

\subsection{Smearing of  collisions by quantum fluctuations}

Let us now introduce quantum fluctuations in the bosonic sector, and generalise our TWA-fGS approach to the case of soliton collisions, which requires defining the initial Wigner function for the scalar sector and the corresponding correlation matrix~\eqref{correlation_matrix} for the fermionic Gaussian state. In analogy to the single-soliton case~\eqref{eq:quadratic_expansion_soliton}, the former is obtained by diagonalizing the small-amplitude expansion of the scalar-field Hamiltonian, but this time using the kink-antikink profile in the resulting elasticity matrix $K(\boldsymbol{\Phi}_{\rm K})\mapsto K(\boldsymbol{\Phi}_{{\rm K}\overline{\rm K}})$~\eqref{eq:el_matrix}. We note that the two individual  Goldstone modes around each soliton are almost degenerate for sufficiently large distances. Hence, the lowest-lying eigenmodes of this elasticity matrix are their symmetric (even) and antisymmetric (odd) superpositions  $\mathcal{M}^{\rm b,(e)}_{n,0}, \mathcal{M}^{\rm b,(o)}_{n,1}$. We can then write an initial Wigner distribution describing the kink-antikink pair as
\begin{equation}
\begin{aligned}
W_0 \propto \; 
\ee^{ -\left(\frac{Q_0^2}{2\sigma_{Q_0}^2}+\frac{(P_0-\overline{P}_0)^2}{2\sigma_{P_0}^2} \right)}\ee^{ -\left(\frac{Q_1^2}{2\sigma_{Q_1}^2}+\frac{P_1^2}{2\sigma_{P_1}^2} \right)}
\prod_{\nu=2}^{N-1} \ee^{ - \left( \frac{Q_\nu^2}{2\sigma_{Q_\nu}^2}  +\frac{P_{\nu}^2}{2\sigma_{P_\nu}^2} \right)},
\end{aligned}
\end{equation}
where we give a non-zero momentum to the even Goldstone mode, such that these solitons approach each other eventually leading to a collision. 
As commented in the single-soliton case, an initial state projection onto the Goldstone mode will induce trajectories in which the kinks can travel in both directions, leading to enlarged ripples within a symmetric light cone. To get closer to the classical regime of the previous section, studying controlled soliton collisions within the framework of the TWA, we can set the fluctuations of the Goldstone modes $\{\sigma^2_{Q_\nu},\sigma^2_{P_\nu},\}_{\nu=0,1}$ close to zero. The quantum fluctuations will thus be introduced by the remaining modes. We explicitly set the occupations of the odd ($\nu=1$) mode to zero, on top of adding a mean momentum of $\overline \Pi_0$ to the initial state occupation of the even Goldstone mode, thus including zero-point fluctuations to the classical scalar field dynamics of Section \ref{SAS_collisions_classical}.
The choice of this initial state could be justified by postselection of colliding trajectories.

In panels e)-f) of Figs. \ref{fig_collision_combined} and \ref{fig_collision_TWA_domain_walls}, we show the results of incorporating leading quantum fluctuations to the collisional dynamics previously studied classically, using the TWA-fGS framework.
The most noticeable effects appear after the solitons collide, where the final expectation values of the scalar field are, in general, smeared out by the quantum fluctuations.
However, the almost-elastic and deep-inelastic collision regimes can be clearly distinguished for weak and strong interactions. The intermediate regime, where the classical solitons show a second collision after being initially separated, exhibits a mixture of both regimes. Here, some soliton trajectories can escape their mutual potential well with the additional kinetic energy from zero-point fluctuations.
Still, we observe that bion formation mediated by the Yukawa coupling persists even when quantum fluctuations are considered, as in Fig.~\ref{fig_collision_combined}-h). The corresponding fermion scalar condensate of Fig.~\ref{fig_collision_TWA_domain_walls}-h) shows how the Dirac field reacts to these quantum fluctuations during the collision. We see that the non-zero values of this condensate are mainly localised within the bion excitation of the scalar field. In general (see other panels of Fig.~\ref{fig_collision_TWA_domain_walls}), when considering the effects of ground state quantum fluctuations on the scalar condensate, we observe the same effect of smeared boundaries of the wavefronts, while still observing a clear correspondence between these domain walls and the scalar field expectation values. 

Let us close this subsection by noting that we have numerically found that, when only one of the fermionic zero-modes is occupied, the interaction strength required to capture the solitons in the potential well of Fig.~\ref{potential_two_solitons} needs to be increased.


\section{\bf Conclusions and Outlook}
\label{sec:conc_out}

We have studied the real-time dynamics of a trapped-ion regularization of the Jackiw-Rebbi model, where the scalar field appears after coarse-graining the quantized vibrations around a zigzag Coulomb crystal, and can thus present quantum fluctuations beyond the typical classical limit in which it is considered a fixed background field. This coarse graining is valid in the vicinity of a structural transition in which quartic interactions become relevant, and one can design a quantum simulator of a self-interacting scalar field that is additionally coupled to a Dirac field encoded in a pair of internal electronic states via a Jordan-Wigner transformation. We have discussed how using laser-induced couplings, one can exploit a specific form of phonon-mediated interactions and spin-dependent forces that emulate the relativistic fermion-boson dynamics of a Yukawa field theory in $D=1+1$ dimensions. Leveraging on previous observations of topological solitons in this platform, we have argued that this quantum simulator offers a promising platform to also observe the binding of fractional fermion charges, exploring the real-time dynamics of the composite system to observe phenomena that arise from the inclusion of quantum fluctuations and back-reaction.

We have presented an analysis of these possible dynamical effects that goes beyond the above crude classical approximation of the soliton,   and also beyond an adiabatic Born-Oppenheimer treatment that assumes that all back-reaction can be understood by assuming that the  Dirac fermions adapt instantaneously to changes of the soliton background. By including quantum fluctuations through a truncated Wigner approximation combined with fermionic Gaussian states, we find that an initially localized kink typically spreads with time, leading to diffusive behavior. As the Yukawa coupling increases, we find that the back-reaction of the fermions onto the soliton inhibits this diffusion, and that the soliton with a bound fractional charge remains localized in a minimum of the Peierls-Nabarro potential. Our simulations demonstrate that for sufficiently strong coupling, the fractional fermion charge is robustly bound to the soliton, with the charge density profile remaining spatially locked to the kink's center.

We have also explored the scattering of a pair of fractional charges bound to a soliton and anti-soliton configuration. In the adiabatic limit, we have reported how the Born-Oppenheimer approximation predicts a scattering potential with both closed and open channels. We have seen that, depending on the Yukawa coupling strength, classical predictions of these collisions go from an almost elastic regime where internal breathing-like ripples also get excited, to highly inelastic regimes in which the soliton-anti-soliton pair forms a composite confined excitation known as a bion that oscillates periodically, leading to further subsequent collisions. When including quantum fluctuations and back-reaction using our methods, we have found remnants of these classical scattering regimes, in which the fluctuations tend to smooth the colliding trajectories. Still, we find that the elastic versus inelastic regimes can be clearly differentiated, and that the appearance of bions survives quantization. 
This rich interplay of quantum fluctuations and back-reaction can be explored through observables that are directly accessible in trapped-ion experiments through fluorescence, sideband spectroscopy, and in-situ imaging of the zigzag order parameter.

Future work could extend beyond the above semiclassical approximation, tracking genuine quantum correlations and entanglement growth between fermionic and bosonic sectors, as well as non-Gaussian effects. Monitoring the time evolution of entanglement entropy, or bipartite fermion–boson entanglement across the kink, would shed light on localization, decoherence, and thermalization in quantum field theories. Exploring stronger couplings and long-range interactions would open the door to simulating richer models such as the Gross--Neveu or massive Thirring theories, including phenomena like dynamical mass generation. Overall, this work establishes a framework for the quantum simulation of interacting relativistic field theories with trapped ions, showing how fractionalization, topology, and back-reaction can be probed in real time, and paving the way toward quantum simulations of more complex relativistic field theories.


\appendix
\section*{Acknowledgments}

A.K. P.V. and A.B. acknowledge support from  the European Union’s Horizon Europe research
and innovation programme under grant agreement No
101114305 (”MILLENION-SGA1” EU Project), from PID2021-127726NB-I00 and PID2024-161474NB-I00 (MCIU/AEI/FEDER,UE), and from QUITEMAD-CM TEC-2024/COM-84. We also acknowledge support from the Grant IFT Centro de Excelencia Severo Ochoa CEX2020-001007-S funded by
MCIN/AEI/10.13039/501100011033, and from the CSIC Research Platform on Quantum Technologies PTI-001.
This work is partially funded by the European Commission–NextGenerationEU, through Momentum CSIC Programme: Develop Your Digital Talent. We acknowledge HPC support from Emilio Ambite, staff hired under the Generation D initiative, promoted by Red.es, an organisation attached to the Spanish Ministry for Digital Transformation and the Civil Service, for the attraction and retention of talent through grants and training contracts, financed by the Recovery, Transformation and Resilience Plan through the EU’s Next Generation funds.


\bibliography{biblio.bib}

\end{document}